\documentclass[twocolumn]{aastex631}

\usepackage[caption=false]{subfig}
\usepackage{amsmath}
\newcommand{\msun}{{M_{\odot}}}
\newcommand{\mstar}{{M_{\ast}}}
\newcommand{\ser}{S\'ersic }
\usepackage[utf8]{inputenc}

\shorttitle{Reconstructing Star Formation Histories of High-Redshift Galaxies}
\shortauthors{Mosleh et al.}

\begin{document}

\title{Reconstructing Star Formation Histories of High-Redshift Galaxies: A Comparison of Resolved Parametric and Non-Parametric Models}

\correspondingauthor{Moein Mosleh}
\email{moein.mosleh@shirazu.ac.ir}

\author[0000-0002-4111-2266]{Moein Mosleh}
\affiliation{Biruni Observatory, College of Science, Shiraz University, Shiraz 71946-84795, Iran}
\affiliation{Department of Physics, College of Science, Shiraz University, Shiraz 71946-84795, Iran}

\author[0009-0002-3624-4640]{Mohammad Riahi-Zamin}
\affiliation{Biruni Observatory, College of Science, Shiraz University, Shiraz 71946-84795, Iran}
\affiliation{Department of Physics, College of Science, Shiraz University, Shiraz 71946-84795, Iran}

\author[0000-0002-8224-4505]{Sandro Tacchella}
\affiliation{Kavli Institute for Cosmology, University of Cambridge, Madingley Road, Cambridge, CB3 0HA, UK}
\affiliation{Cavendish Laboratory, University of Cambridge, 19 JJ Thomson Avenue, Cambridge, CB3 0HE, UK}

\begin{abstract}

We investigate the optimal approach for recovering the star formation histories (SFHs) and spatial distribution of stellar mass in high-redshift galaxies ($z\sim 2-5$), focusing on the impact of assumed SFH models on derived galaxy properties. Utilizing pixel-by-pixel spectral energy distribution (SED) fitting of multi-band photometry, we explore various parametric SFH models (including exponentially declining ($\tau$), delayed-$\tau$, lognormal, and double-power law) alongside spatially resolved non-parametric methods. We first analyze the models using simulated galaxies and then apply them to observed galaxies for validation and as proof of concept, with additional comparisons to results from unresolved SED fitting. Our findings demonstrate that pixel-by-pixel analysis with parametric models is particularly robust in recovering the true SFHs of simulated galaxies, with the double-power law model outperforming others, including non-parametric methods. This model excels in detecting recent starbursts within the last 500 Myr and capturing the stochastic nature of star formation. Conversely, unresolved photometry with simplistic parametric models tends to produce biased estimates of key galaxy properties, particularly underestimating early star formation. Non-parametric methods, resolved or unresolved, typically yield older mass-weighted ages. Biases in early-time SFRs, likely introduced by prior assumptions, further complicate these models. We conclude that the double-power law model, applied in a pixel-by-pixel framework, offers the most reliable recovery of SFHs and produces robust stellar mass maps. Resolved methods simplify modeling dust and metallicity, enhancing parameter interpretability and underscoring the value of flexible parametric models in spatially resolved analyses.

\end{abstract}



\section{Introduction} 

A critical assumption in determining the physical parameters of galaxies from their spectral energy distributions (SEDs) is the treatment of their star-formation histories (SFHs) \citep{walcher2011, conroy2013}. Several studies have demonstrated that key parameters such as stellar mass and age are susceptible to the assumed SFH model \citep{pforr2012, dominguez2015, lower2020, pacifici2023, harvey2024, narayanan2024}. In addition, a key goal in reconstructing star-formation histories (SFHs) is tracing the earliest phases of star formation, particularly in high-redshift galaxies. This is crucial for differentiating SFH models and has profound implications for galaxy formation models \citep{shen2023}. 

The literature presents various parametric, non-parametric, a combination of models \citep[e.g.,][]{belli2019, wang2024}, or cosmological simulation-motivated \citep[e.g.,][]{pacifici2016} approaches, yet selecting a model that accurately reflects realistic SFHs remains challenging \citep{carnall2019, leja2019a}. Common parametric models include exponentially declining \citep{tinsley1972}, delayed-exponentially declining \citep{lee2010}, double power-law \citep{behroozi2013}, and log-normal \citep{gladders2013, diemer2017} SFHs. Despite their utility, these models often struggle to represent the stochastic nature of SFHs. Additionally, parametric models tend to be inadequate for capturing rapid variations in the star-formation rate (SFR), such as those associated with starburst episodes \citep{gladders2013, simha2014}. To address this, parametric models are sometimes supplemented with bursts to capture better the complex SFR evolution over a galaxy's assembly history \citep[e.g.,][]{kauffmann2003, french2018}. However, the inclusion of additional parameters can introduce further complexity to the SED fitting process. Furthermore, the early star formation phase, which we refer to as the `cosmic dawn age' (CDA), can be quantified as the lookback time when a small fraction of the total stellar mass was formed. Here, we define CDA as the epoch when 5\% of the total stellar mass has accumulated ($t_5$). This early formation age is often underestimated in parametric SFH models, leading to biases in inferred galaxy evolution histories.

In recent years, non-parametric SFHs have gained popularity as a more flexible alternative to reflect the complexity of galaxy evolution \citep{ocvrik2006, leja2017, morishita2019, iyer2019}. Advances in techniques such as nested sampling have made these methods more practical, though non-parametric approaches are not without their own challenges \citep{leja2019a, tacchella2022, whitler2023}. These include the selection of priors, the definition of age bins, the degree of SFH continuity, and the computational demands of such models. Specifically, the choice of bin size can critically affect the ability to resolve recent stochastic changes in SFR \citep{suess2022}, as well as the prior on adjacent bins \citep{wan2024}, and the overall base of the SFH \cite{turner2024}. Besides that, computational limitations often constrain the number of bins, restricting the resolution of non-parametric SFHs.

In addition to uncertainties related to SFHs, other SED modeling parameters such as dust distribution and metallicity evolution introduce further complications to the SED modeling. Observations in the mid- and far-infrared reveal that dust distribution is often non-uniform within galaxies, complicating the modeling of dust attenuation \citep{calzetti2000, charlot2000}. Complex dust models, which tend to introduce additional parameters, are frequently employed to account for these variations \citep{dacunha2008}.

Therefore, despite advances in SED modeling, these complexities can hamper efforts to recover accurate estimates of galaxies' physical parameters and their evolution, particularly when dealing with the total fluxes of galaxies \citep{conroy2010, tortorelli2024}. One way to mitigate such difficulties is through spatially resolved or pixel-by-pixel analysis, which reduces parameter space by applying simplified assumptions to distinct regions within a galaxy \citep{abraham1999, conti2003, zibetti2009, wuyts2012, hemmati2014, smith2018, abdurrouf2021}. This method allows for more straightforward dust attenuation models and provides a clearer interpretation of mixed stellar populations across different regions, leading to more accurate SED fits and consequently more reliable estimates of physical parameters, such as stellar masses and SFRs, across different regions of a galaxy \citep{ji2023, lofaro2024, polletta2024}. Additionally, different regions within a galaxy often exhibit distinct assembly histories, which contribute to variations in stellar ages and metal enrichment timescales. As a result, the overall SFH of a galaxy is an aggregate of the star-formation processes in its various regions, and resolving SFHs spatially can provide a more comprehensive picture of a galaxy's evolution \citep[e.g.,][]{tacchella2016, Nelson2021, abdurrouf2023, Bellstedt2024}.

One key result of pixel-by-pixel SED modeling is that the total stellar masses derived from this approach often differ from those obtained using integrated, unresolved fluxes \citep[e.g.,][]{zibetti2009, sorba2015, mosleh2020} \citep[but also see][]{martinze2017, song2023}. This discrepancy has been attributed to factors such as the underlying assumptions about the SFH and the outshining effect \citep{papovich2001}, where young, luminous stars dominate the light but contribute less to the overall stellar mass \citep{sorba2018, gimenez2024}. However, the accuracy of pixel-by-pixel analyses, including estimates of specific star formation rates (sSFR) and stellar ages, still depends heavily on the chosen SFH models. As shown by \citet{jain2024} using spatially resolved analysis, the assumptions based on declining SFH models can shift the peak of star formation to different epochs, affecting both the recovered stellar masses and the correct sSFR for recent times. Therefore, it is crucial to systematically test various SFH models in the context of pixel-by-pixel analysis to identify any discrepancies and ensure that the derived physical parameters, such as stellar masses, SFR, and ages, are robust across different assumptions. This testing is essential to improving the reliability of spatially resolved studies of galaxy evolution.

Moreover, the ability to recover details of the SFH, particularly during different epochs, is crucial \citep{faucher2018, tacchella2020, narayanan2024}. Features such as starbursts or sudden quenching events (cessation of star formation) are challenging to capture, and we are particularly interested in how well these features can be recovered using resolved SED fitting compared to non-parametric methods, which are often limited by the flexibility of their age binning. Accurately resolving these events can have significant implications for interpreting the evolutionary history of galaxies \citep{dressler2024}. Additionally, to better estimate the CDA, we need to explore the variety of SFH models and assess how accurately different approaches capture the early phases of star formation in order to refine our understanding of galaxy evolution.

Therefore, in this work, we aim to examine the recovery of the SFH and other physical parameters (e.g., stellar mass, SFR) with high spatial resolution, by fitting the SEDs of individual pixels using both parametric and non-parametric SFH models. This approach allows us to assess the efficacy of different SFH models, particularly in comparison to non-parametric methods based on total (or unresolved) fluxes, in capturing the true SFH and recovering the total stellar mass of galaxies. To rigorously test our ability to recover SFHs, we created mock galaxies with realistic mass distributions and complex SFHs, comparable to those produced by hydrodynamical simulations. These mock galaxies allow us to evaluate how well we can recover physical properties and SFHs using various assumptions. To this end, we utilize mock galaxies with known SFHs, resembling those observed at $2 \lesssim z \lesssim 5$ by the James Webb Space Telescope (JWST) and the Hubble Space Telescope (HST). By applying both parametric and non-parametric methods to these mock galaxies, we assess which models best recover the input SFHs and physical parameters. As a consistency check, we apply methodologies developed and validated on mock galaxies to observed galaxies within the same redshift range. This allows us to test the robustness of the methods in the presence of real observational challenges, such as measurement uncertainties, real dust effects, and instrumental limitations. By validating our approach on real data, we assess its practical effectiveness and demonstrate its applicability to future observational studies, enhancing the study's impact. Beyond testing SFH models, this analysis offers an opportunity to explore optimal approaches for constructing reliable 2D stellar mass maps, a crucial tool for understanding how different regions of galaxies assemble and evolve.

In addition, comparing the performance of parametric and non-parametric models at earlier epochs is critical, as the SFHs during these periods influence the estimation of stellar masses \citep{vanmierlo2023, gimenez2023}. For instance, previous studies have shown that non-parametric models tend to yield larger stellar masses than parametric ones \citep{leja2019b}. We aim to test this further in the context of pixel-by-pixel analysis by applying both parametric and non-parametric SFH models to determine the impact of each on the derived stellar masses. This comparison will help clarify how differences in SFH modeling--especially regarding the treatment of early and late epochs--affect the accuracy of stellar mass estimates and other key physical parameters in spatially resolved studies. Throughout this work, we adopt the following cosmological parameters: $\Omega_m = 0.3$, $\Omega_\Lambda = 0.7$, and $H_0 = 70\ \mathrm{km\ s^{-1}\ Mpc^{-1}}$.\\

\section{Data \& Methodology} 

In this section, we present the data sources and methodologies employed to build and analyze a sample of high-redshift ($z\sim2-5$) mock galaxies, as well as to conduct a comparable analysis of observed galaxies at similar redshifts. To simulate realistic conditions and examine observed galaxies in depth, we utilized deep imaging data from the JWST and HST. These datasets enable us to create synthetic multi-filter images of mock galaxies that mimic real observational data, providing a basis for consistent comparison. Additionally, we used the catalog derived from these images to acquire general properties of the observed galaxies.

We begin this section with an overview of the imaging datasets and catalogs used, followed by a description of the observed galaxy sample. Based on these, we define the sample of mock galaxies, outlining the processes used to build their properties, including their star formation histories (SFHs) and spatially resolved maps of stellar mass, age, dust, and metallicity. Finally, we detail the SED fitting procedures applied to both real and simulated galaxies, distinguishing between integrated and spatially resolved (pixel-by-pixel) approaches.

\subsection{Data}
The spatially resolved analysis conducted in this study utilizes recent deep imaging and photometric observations of the GOODS-South field \citep{giavalisco2004} by the James Webb Space Telescope (JWST) and Hubble Space Telescope (HST). We selected the GOODS-South field due to its extensive wavelength coverage, spanning from the rest-frame ultraviolet (UV) to the near-infrared (NIR). This broad coverage is achieved through a combination of 23 different filters. Specifically, the data includes observations from 9 filters (F090W, F115W, F150W, F200W, F277W, F335M, F356W, F410M, F444W) by the JWST/NIRCam as part of the JWST Advanced Deep Extragalactic Survey (JADES) program \citep{rieke2023, eisenstein2023}, which offers high-resolution imaging and deep photometric data crucial for studying distant galaxies. Additionally, 5 filters (F182M, F210M, F430M, F460M, and F480M) were acquired from the JWST Extragalactic Medium-band Survey \citep[JEMS;][]{williams2023}, enhancing the spectral resolution which can help to disentangle the contributions from different stellar populations and the emission lines. Furthermore, the dataset incorporates 9 filters from the Cosmic Assembly Near-infrared Deep Extragalactic Legacy Survey (CANDELS) and the 3D-HST survey \citep{faber2011, koekemoer2011, grogin2011, skelton2014, momcheva2017}, which provides deep and wide-field imaging across rest-frame UV and optical using the 9 filters observed by ACS (F435W, F606W, F775W, F850LP, F814W) and WFC3 (F105W, F125W, F140W and F160W) filters.

To perform pixel-by-pixel SED fitting, we first needed uniformly reduced and calibrated images obtained by different telescopes and through different filters. Therefore, we used the photometric and imaging data provided by the DAWN JWST Archive (DJA) which are reduced using \texttt{GRIZLI} pipeline \citep{brammer2023}, as detailed in \cite{valentino2023}. The drizzled mosaic images were all background subtracted and astrometrically aligned. The images from the NIRCam short-wavelength channel (SW) and long-wavelength channel (LW) were drizzled to a pixel scale of 20 and 40 mas pixel$^{-1}$, respectively. We further resampled the SW bands to a common scale of 40 mas pixel$^{-1}$.

In the second step, we homogenized the resolutions of images at different wavelengths to match the NIRCam/F480M filter, which has the longest wavelength. For this purpose, we first built effective point spread functions (ePSFs) from isolated, unsaturated, and bright stars in each image, using the method from \cite{andersonking2000}. As shown by \cite{ji2023} and \cite{weaver2024}, using practical PSFs is more robust than using simple \texttt{WEBBPSF} models. We then used the \texttt{PYPHER} package \citep{boucaud2016} to find the convolution kernels by employing Wiener filtering to homogenize the ePSFs to the resolution of the F480M filter. These steps ensured that the observational data were prepared consistently and accurately, allowing for reliable SED fitting and analysis.

\subsection{Observed Galaxy Sample}

In this paper, we identify 12 observed galaxies from the DJA v7.0 catalog of the GOODS-South field, with secure spectroscopic redshifts in the range of $z \sim 1.9 - 4.2$ and stellar masses $\log (\mstar/\msun) > 9$. These galaxies were selected based on having observations in more than 21 filters, thus placing them within the very deep region of the GOODS-South field. The positions, redshifts, and stellar masses of these galaxies are presented in Table \ref{tb1}. The redshift range $2 \lesssim z \lesssim 5$ is particularly interesting, as concerns about the reliability of stellar mass estimates increase at high redshift.  At these redshifts, the Universe is relatively young, and even short periods of intense star formation can have a significant impact on a galaxy's stellar mass. Additionally, this redshift range provides good wavelength coverage for rest-frame near-infrared observations, which are essential for robust stellar mass estimates.

\begin{figure}[h]
    \centering
    \includegraphics[width=0.9\linewidth]{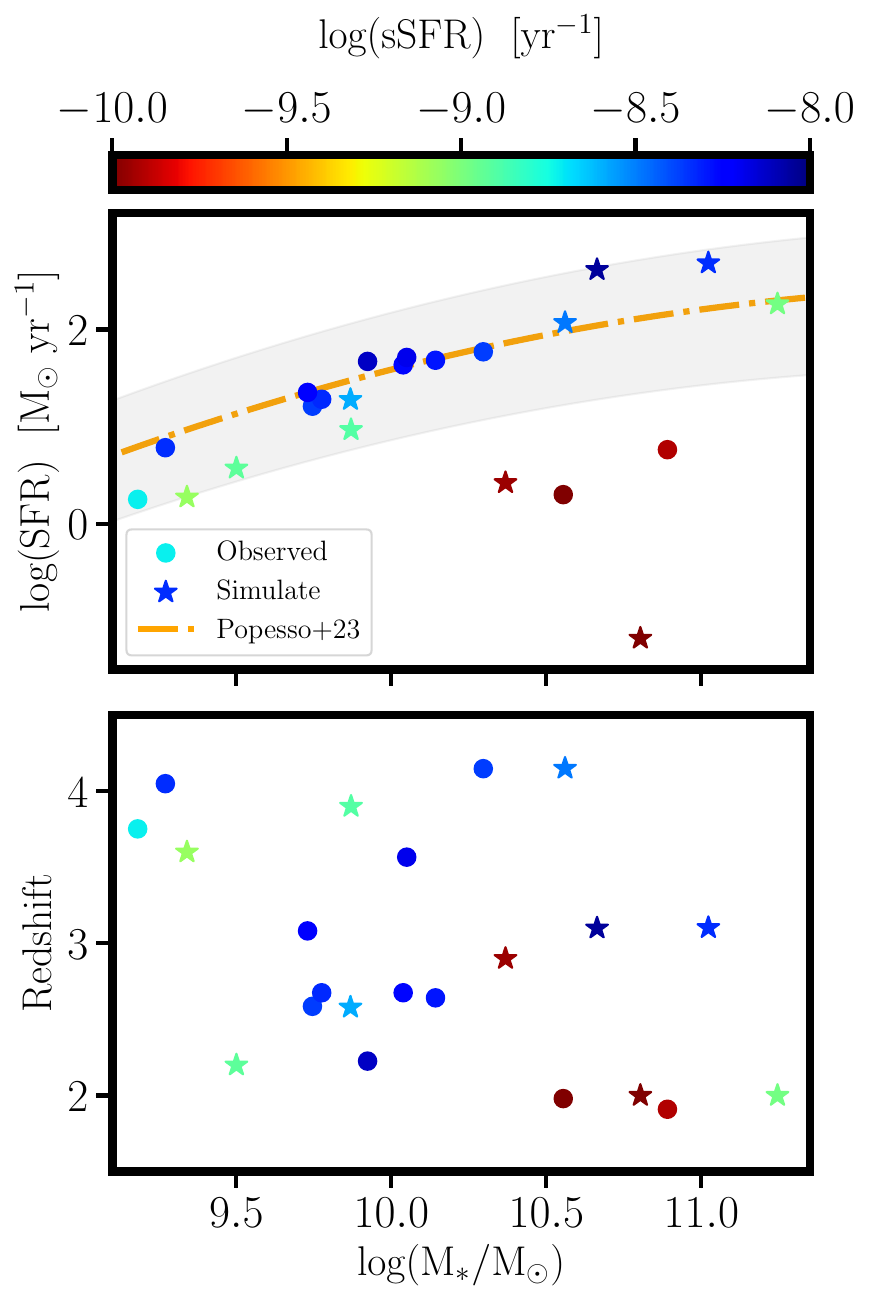}
    \caption{\textbf{Top panel}: The SFR-Mass distribution of the 12 observed (solid circles) and 10 simulated galaxies (star symbols), studied in this paper. The objects are color-coded according to their sSFR. The dot-dashed line presents the main-sequence relation for star-forming galaxies at $z\sim3.5$ from \cite{popesso2023} and the gray region depicts the $0.5$ dex scatter about the main-sequence relation at the redshift range of $z=2-5$. \textbf{Bottom panel:} Redshift distribution of the objects shown in the top panel as a function of mass. (see Sections 2.2 and 2.3 for more details.)}
    \label{fig1}
\end{figure}

The distribution of these galaxies along the main-sequence relation is shown in the top panel of Figure \ref{fig1}, with solid circles color-coded according to their specific star-formation rates (sSFR). The bottom panel depicts their redshift distributions as a function of stellar mass. As illustrated in the plot, the majority of these galaxies (83\%) are star-forming, closely following the SFR-Mass relation at redshift $z \sim 3.5$ from \cite{popesso2023}. The remaining two galaxies in this sample (17\%) are quiescent, with $\log(\text{sSFR}/\mathrm{yr}^{-1}) < -10$, at $z \sim 2$. Thus, this sample ensures a diverse range of star-formation histories (SFHs) can be studied.

We note that the total stellar masses and star-formation rates (SFRs) reported in this study are estimated from the integrated quantities obtained through the pixel-by-pixel analysis described in Section 2.4. In addition, the stellar masses used in this paper are calculated from the integrated star-formation histories, i.e. including living stellar masses, the remnants, and the mass return to the ISM.

\begin{deluxetable}{ccccc}
\tablenum{1}
\tablecaption{Properties of the 12 observed galaxies, including their coordinates, redshifts, and estimated stellar masses. \label{tb1}}
\tablewidth{0pt}
\tablehead{
\colhead{DJA ID} & \colhead{RA} & \colhead{DEC} &
\colhead{$z_{spec}$} & \colhead{$\text{log}(\mstar/\msun)$}
}
\startdata
17225 &	53.1431 & -27.8155 & 4.14 & 10.14 \\
24722 &	53.1563 & -27.7990 & 2.67 & 9.61 \\
24924 &	53.1576 & -27.7990 & 2.67 & 9.89 \\
25340 &	53.1630 & -27.7977 & 1.98 & 10.34 \\
25648 &	53.1588 & -27.7972 & 1.91 & 10.67 \\ 
25850 &	53.1738 & -27.7964 & 3.56 & 9.90 \\ 
26980 &	53.1526 & -27.7939 & 3.08 & 9.59 \\
27106 &	53.1758 & -27.7934 & 4.04 & 9.12 \\
27864 &	53.1606 & -27.7918 & 2.64 & 9.97 \\
29915 &	53.1433 & -27.7868 & 3.75 & 8.97 \\
33326 &	53.1614 & -27.7817 & 2.58 & 9.58 \\
37222 &	53.1544 & -27.7715 & 2.22 & 9.79 \\
\enddata
\end{deluxetable}

\begin{deluxetable}{cccc}
\tablenum{2}
\tablecaption{Properties of 10 mock galaxies generated for this study, including their stellar masses, star formation rates (SFRs), and redshifts, to analyze SFH recovery. \label{tb2}}
\tablewidth{0pt}
\tablehead{
\colhead{ID} & \colhead{$\text{log}(\mstar/\msun)$} & \colhead{$\text{log}(\mathrm{SFR}) [\msun yr^{-1}]$} & 
\colhead{$z$} 
}
\startdata
1 & 11.02   & 2.68   & 3.1 \\
2 & 10.37 & 0.42   & 2.9 \\ 
3 & 10.80 & -1.18  & 2   \\ 
4 & 9.87  & 1.28   & 2.6 \\ 
5 & 10.56 & 2.07   & 4.1 \\ 
6 & 10.66 & 2.61   & 3.1 \\ 
7 & 11.24   & 2.26   & 2   \\ 
8 & 9.34  & 0.27   & 3.6 \\
9 & 9.50  & 0.57   & 2.2 \\ 
10 & 9.86 & 0.96   & 3.9 \\ 
\enddata
\end{deluxetable}

\begin{figure*}[ht!]
\includegraphics[width=0.96\textwidth]{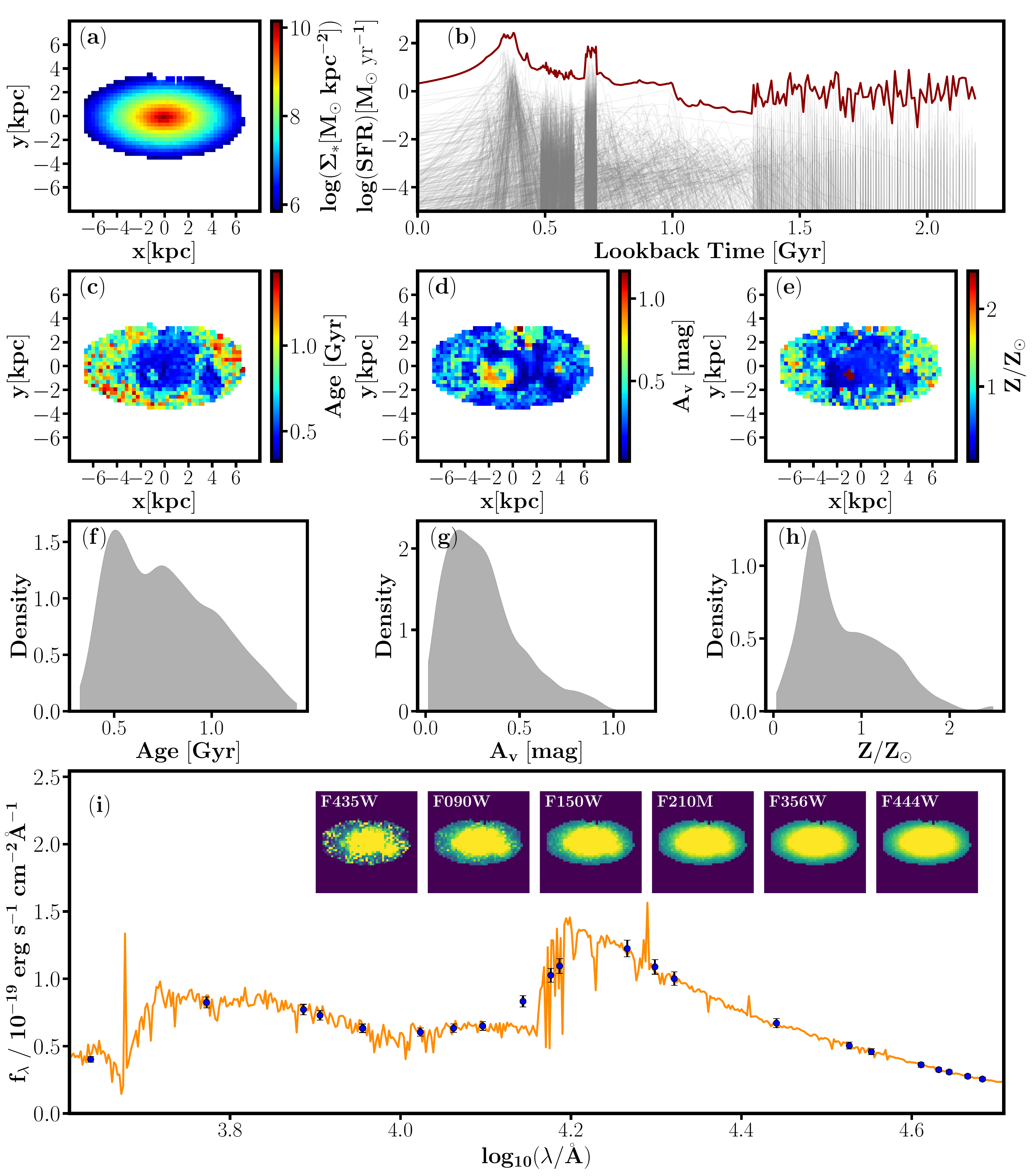}
\caption{Example of the stellar mass surface density, age, dust, and metallicity distributions for a simulated galaxy (ID = 2 in Table \ref{tb2}). Panel (b) illustrates the star formation histories (SFHs) of individual pixels (gray lines) alongside the galaxy's total SFH (red line). The SFHs for the mock galaxies are designed to represent a wide range of conditions observed in the universe. Panel (i) presents the mock spectrum of the galaxy with overlaid photometric points, along with postage stamp images in selected filters, providing insight into its spectral properties and multi-wavelength appearance. See Section 2.3 for details on the generation of these maps and SFHs.}
\label{fig2}
\end{figure*} 

\subsection{Simulated Galaxy Sample}

To evaluate the robustness of our spatially resolved SFH modeling, we constructed mock galaxies with controlled distributions of stellar age, metallicity, and dust attenuation. This approach allows us to systematically test the recovery of these properties under idealized conditions, ensuring that any biases in the modeling process can be directly traced to specific assumptions. The mock galaxies were tailored to match the spatial resolution, noise properties, and observational characteristics of our dataset, providing a realistic yet controlled framework for our analysis.

We created 10 mock galaxies with assigned stellar masses, SFRs, and redshifts comparable to those of the observed galaxies presented in Section 2.2 (see Table \ref{tb2}). The star symbols in Figure \ref{fig1} illustrate the distributions of these galaxies across the SFR-mass and redshift-mass relations. Similar to the observed sample, approximately 80\% of these galaxies are star-forming, while the remaining 20\% are quiescent.

The 2D mass maps of these mock galaxies were generated using \ser models \citep{sersic1963}, with half-mass radii ranging from $0.7-1.2$ kpc and \ser indices in the range of \textbf{$n \sim 1-1.5$}. These parameters, along with varying ellipticity and orientations, were chosen to resemble the small-scale structures of galaxies at redshifts of $\sim2-4$ \citep[e.g.,][]{allen2024, jia2024}. The normalized 2D \ser images were then multiplied by the total stellar masses of the simulated galaxies to create realistic mass distributions. Analogous to the observed galaxies, the mass maps were sampled with a pixel scale of 0.04 arcseconds per pixel. The boundaries of these maps were defined to ensure that the integrated masses from all pixels were consistent with the total initial masses to within 0.1\% accuracy. An example of a stellar mass map of a mock galaxy (ID=2) is shown in panel (a) of Figure \ref{fig2}.

Next, we individually assigned a stellar population model to each pixel in the mass map. Constructing these models requires assumptions about SFH, metallicity, dust content, and the contribution of nebular emission lines. However, the 2D distributions of certain parameters, such as stellar population ages and dust, are not random in real galaxies. This prevents us from assigning pixel values purely at random. Instead, we used 2D maps of stellar population parameters derived from our own observed galaxies as the initial guess. To introduce realistic variations, we then perturbed the value of each pixel within the 16th to 84th percentile range of its own posterior distribution, thereby generating new 2D maps of parameters. This method was specifically used for the distributions of metallicity ($Z/Z_{\odot}$), dust attenuation ($A_V$), and ionization parameters (U) of the nebular emission lines. The distribution of dust attenuation and metallicity for the second simulated galaxy in our sample are shown in panels (d) and (e) of Figure \ref{fig2}, along with their range of values in panels (g) and (h) of the same plot. These distributions illustrate the variations in chemical composition and dust content that can occur within different regions of a galaxy. 

To realistically model the stochastic nature of star-formation histories (SFHs) in high-redshift galaxies, we developed mock galaxies using two primary SFH templates for each pixel: a modified double-power law model and, in a smaller subset of cases, the post-starburst model from \cite{wild2020}. The double-power law SFH serves as the dominant model, allowing for both rising and declining SFRs, while the PSB model was included to increase the diversity of SFHs and capture stochastic star formation variations. These models were further enhanced by incorporating random bursts, ensuring that they encompass a realistic variety of star formation events (see panel b of Figure \ref{fig2}).
 
Initially, the parameters of each model were estimated from the 2D distributions of observed galaxies and then randomly perturbed in the same way as described for other parameters. These parameters include the rising and falling slopes and the characteristic time ($\alpha$, $\beta$, and $\tau$) of the double-power law SFH model: 

\begin{equation}
\mathrm{SFR}(t) \propto \big{[}(\frac{t}{\tau})^{\alpha} + (\frac{t}{\tau})^{-\beta} \big{]}^{-1}
\end{equation}

The post-starburst model combines two distinct components: an exponentially declining star formation history (SFH) representing older stellar populations' formation in the early universe and a double-power-law model for the subsequent starburst event. Key parameters include the time of the initial onset of star formation, $t_{form}$, the age of the burst, $t_{burst}$, representing the epoch when the starburst occurred, and $f_{burst}$, indicating the fraction of stellar mass formed during the burst. Additionally, the model incorporates the parameters $\alpha$ and $\beta$ from the double-power-law, which describe the rising and falling phases of the burst, respectively, as well as the characteristic timescale, $\tau$, of the exponentially declining component:

\begin{multline}
\mathrm{SFR}(t) \propto \frac{1-f_{burst}}{\int_{t_{form}}^{t_{burst}} \mathrm{SFR}_{e} dt} \times \mathrm{SFR}_{e} + \\ 
\frac{f_{burst}}{\int_{t_{burst}}^{0} \mathrm{SFR}_{burst} dt} \times \mathrm{SFR}_{burst}
\end{multline}

where SFR$_{e}$ represented as exponentially declining model and SFR$_{burst}$ is double power law model.

Additionally, a log-normal SFH model:

\begin{equation}
\mathrm{SFR}(t) \propto \frac{1}{t}\exp\big{[}-\frac{(\ln(t)\ -\ T_0)^2}{2\tau^2}\big{]} 
\end{equation}

was also assumed for one-third of the pixels of the last simulated galaxies (ID=10) to avoid biasing the final results. In this model, $\tau$ and $T_0$ are free parameters but lack intuitive interpretation \citep{carnall2019}. Following \citet{diemer2017}, we reparameterized these in terms of $t_{max}$, the peak time of star formation, and $\mathrm{{FWHM}}$, the full width at half maximum of the SFH, making the model more interpretable. The parameters are randomly selected within the ranges $t_{max} \sim 0.1-0.2$ Gyr (lookback time) and $\mathrm{FWHM}\sim0.01-0.1$ Gyr. 

To better capture the stochastic nature of star formation and mimic real observed galaxies, we introduced randomly distributed starbursts over the galaxy's lifetime. These bursts were modeled as delta functions, with their stellar mass fraction drawn from a uniform prior. However, to prevent SFHs from being overly smooth and to better resemble observed variations, some bursts were intentionally clustered by replicating them two or three times at similar lookback times. This introduces stronger biases in SFHs, making them more realistic. The gray lines in panel (b) of Figure \ref{fig2} represent the assumed SFHs of the pixels for the mock galaxy, and the solid red line shows the total SFH. The final SFH of each simulated galaxy is the sum of the SFHs of all its pixels.

We then used \texttt{BAGPIPES} (Bayesian Analysis of Galaxies for Physical Inference and Parameter Estimation, \citep{carnall2018}) to generate the corresponding fluxes for each JWST and HST filter. This step involved creating detailed mock SEDs by assuming stellar population models of \cite{BC2003}, a \cite{kroupa2001} IMF, and the inclusion of emission lines using \texttt{CLOUDY} \citep{ferland1998, ferland2013} and assuming \cite{byler2017} method, in addition to the SFHs, metallicity and dust content described above for each pixel of the mock galaxy. The \texttt{BAGPIPES} then integrated these SEDs over the transmission curves of the 23 filters to obtain synthetic photometric fluxes. These fluxes were used to create mock images of the galaxies for each filter separately. By simulating the images for each filter, we approximated the observational conditions, including resolution. To account for observational uncertainties, we assigned a fixed 5\% flux uncertainty, which was used in the subsequent SED fitting analysis. To further illustrate the properties of the simulated galaxies, Figure \ref{fig2} includes the mock spectrum of one representative galaxy along with its corresponding photometry. Additionally, we show postage stamp images in a few filters to provide a visual representation of their appearance across different wavelengths.

\subsection{SED fitting}
 
The spatially resolved stellar mass maps are created by performing SED fitting on individual pixels. First, postage stamps of $32'' \times 32''$ are created from the PSF-matched images of observed galaxies in all filters. The segmentation (or mask) map provided by DJA v7.0 is used to identify the regions and borders of each observed galaxy. The fluxes of each pixel in different filters and their associated errors, which are estimated from the noise-equalized region around each object, are extracted separately.

The stellar mass maps and the general SFHs of galaxies are obtained by performing SED fitting on individual pixels of galaxies based on different SFH assumptions. We used the SED modeling code \texttt{BAGPIPES}, considering the \cite{BC2003} stellar population evolution models with ages less than the age of the universe at the object's redshift. Nebular emission lines were included, with the assumed ionization parameter (\textit{U}) uniformly distributed between $-3 < \log(U) < -1$, as modeled by \texttt{CLOUDY}. We used the \cite{calzetti2000} attenuation curve with flat priors for $A_V$ in the range $[0, 4]$, and a uniform metallicity $Z/Z_{\odot}$ in the range $[0, 2.5]$. The \cite{kroupa2001} initial mass function was also considered, with the prior distribution of the stellar mass formed $\log(\mstar/\msun) \in [1, 14]$. The assumed models and related priors are summarized in Table \ref{tb3}.

SED fitting on a pixel-by-pixel basis is performed for each mock or observed galaxy using five different SFH assumptions. The tested SFH models include four parametric models: exponentially declining or $\tau$ model (EXP), delayed-exponentially declining or Delay-$\tau$ (DLY), double power-law (DPL), and log-normal (LGN), as well as a non-parametric approach (NPM). Table \ref{tb4} summarizes the list of parameters and priors.

The results of SED fitting using total (or unresolved) fluxes are also obtained for additional comparisons. The SFH assumptions are the same as those used in the pixel-by-pixel methods. Additionally, we used \texttt{Prospector} \citep{leja2017, johnson2021} and assumed continuity (PSP: CNT, $\nu=2$ an $\sigma=0.3$) and bursty (PSP: BRT, $\nu=2$ an $\sigma=1$) non-parametric methods \citep{tacchella2022} with the same priors described in the Tables \ref{tb3} and  \ref{tb4}. This approach allows for comparing pixel-by-pixel analysis and popular unresolved non-parametric methods, especially using a different code. The summary of models and priors of parameters assumed while using \texttt{Prospector} is listed in Table \ref{tb4}. Note that the NPM acronym is related to the pixel-by-pixel analysis based on the non-parametric continuity model using \texttt{BAGPIPES}, and those with PSP acronyms are related to the results based on unresolved fluxes using \texttt{Prospector}. Moreover, unresolved models are labeled in Figures using the `$\dagger$' sign. 

\begin{deluxetable*}{ccc}
\tablenum{3}
\tablecaption{Summary of key assumptions and prior distributions applied to the various models used in the SED fitting process. \label{tb3}}
\tablewidth{0pt}
\tablehead{
\colhead{Parameter} & \colhead{prior} & \colhead{Description} 
}
\startdata
SPS Model & Bruzual \& Charlot (2003) & Stellar population synthesis model \\
IMF & Kroupa (2001) & Stellar Initial Mass Function \\
Dust Law parametrization & Calzetti et al. (2000) & Dust law \\
$A_V$ & Uniform: (0, 4) & $V$-band attenuation \\ 
SFH & DPL; LGN; DPL; EXP; Continuity (CNT); Bursty (BRT) & Star formation history \\ 
$\log_{10}(M/M_\odot)$ & uniform: (1, 14) & Formed mass \\ 
$Z_{*}/Z_\odot$ & uniform: (0,2.5) for Bagpipes, (0.1,1.5) for Prospector & Stellar metallicity \\
$\log_{10} U$ & uniform: (-3, -1) & Ionization Parameter \\
$\log_{10} (Z_{gas}/Z_{\odot})$ & uniform: (-2.5, 0.5) & Gas-phase metallicity (Prospector only)\\
\enddata
\end{deluxetable*}

\begin{deluxetable*}{cccc}
\tablenum{4}
\tablecaption{Prior distributions and parameter ranges assumed for different SFH models in the SED fitting process. NPM refers to the non-parametric SFH model used in BAGPIPES, while PSP refers to the non-parametric SFH model used in Prospector. \label{tb4}}
\tablewidth{0pt}
\tablehead{
\colhead{SFH Model} & \colhead{Parameter} & \colhead{prior} & \colhead{Description} 
}
\startdata
Lognormal & $t_{max}$ & uniform: (0, 14) & Age of Universe at peak SFR [Gyr]\\
& FWHM & uniform: (0.01, 2) & FWHM of SFH [Gyr]\\
\\
Double Power Law & $\alpha$ & log-uniform: (0.01,1000) & Falling slope \\
& $\beta$ & log-uniform: (0.01,1000) & Rising slope \\ 
& $\tau$ & uniform: (0, 14) & Turnover [Gyr] \\ 
\\
Delayed-$\tau$ & $\tau$ & uniform: (0.01, 10) & e-folding timescale \\ 
& Age & uniform: (0.1,15) & Time since SF began [Gyr] \\
\\
$\tau$ & $\tau$ & uniform: (0.01, 10) & e-folding timescale \\ 
& Age & uniform: (0.1,15) & Time since SF began [Gyr] \\
\\
Continuity (NPM, PSP) & $N_{bins}$ & 8 & Number of age bins \\
& SFR Ratio & Student\textquotesingle s-t: $\nu$ = 2, $\sigma$ = 0.3 & Ratio of the SFRs in adjacent  \\
 & & & bins of the non-parametric SFH \\
 \\
Bursty (PSP) & $N_{bins}$ & 8 & Number of age bins \\
& SFR Ratio & Student\textquotesingle s-t: $\nu$ = 2, $\sigma$ = 1 & Ratio of the SFRs in adjacent  \\
 & & & bins of the non-parametric SFH \\
\enddata
\end{deluxetable*}

\begin{figure*}
    \centering
    \includegraphics[width=0.49\linewidth]{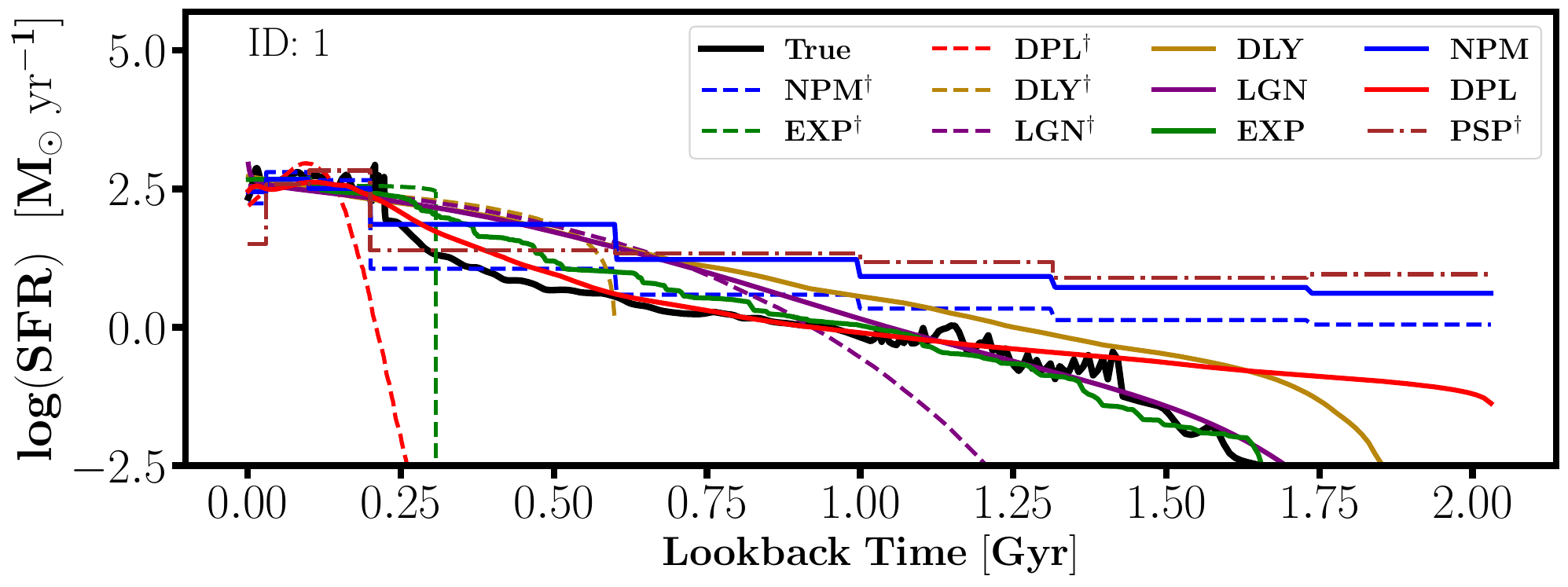}
    \includegraphics[width=0.49\linewidth]{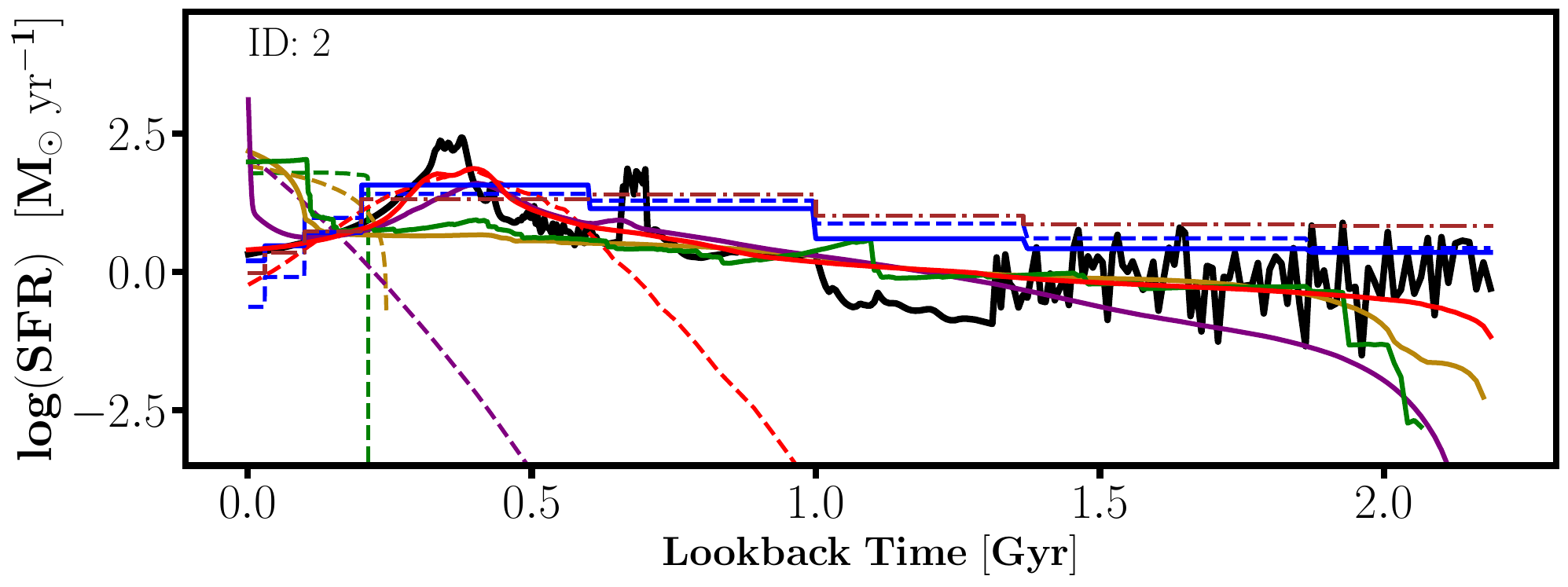}
    \includegraphics[width=0.49\linewidth]{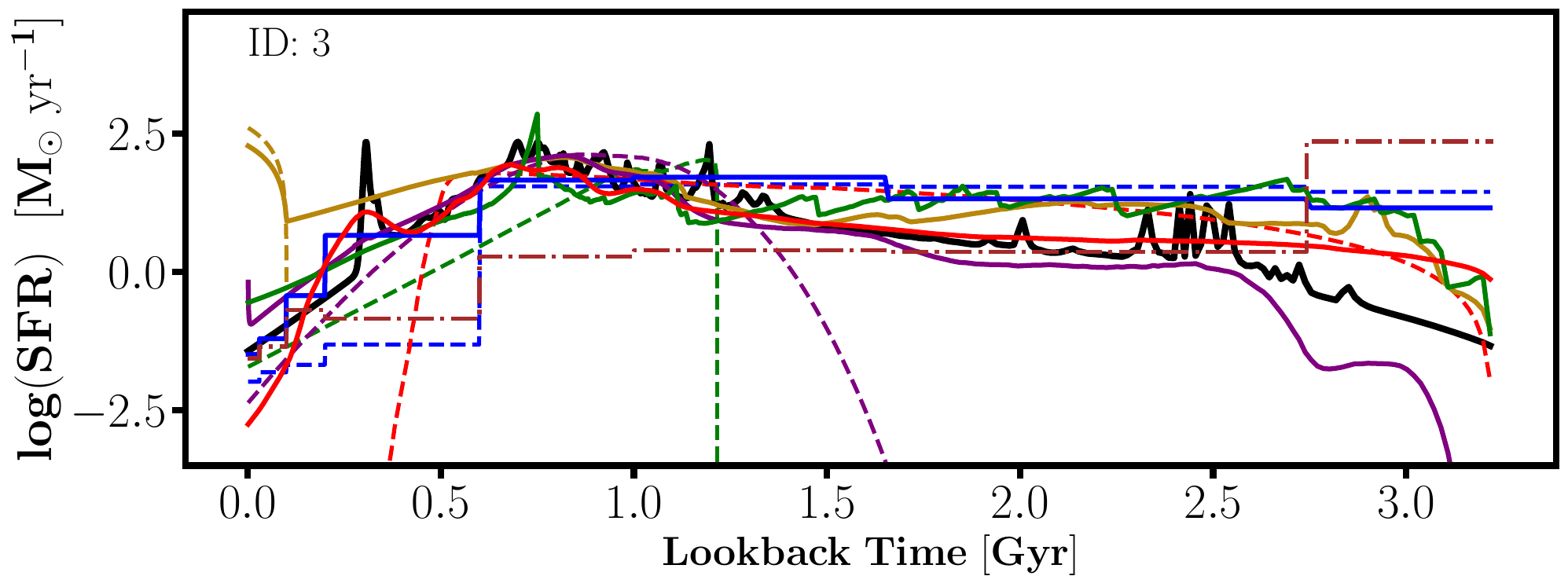}
    \includegraphics[width=0.49\linewidth]{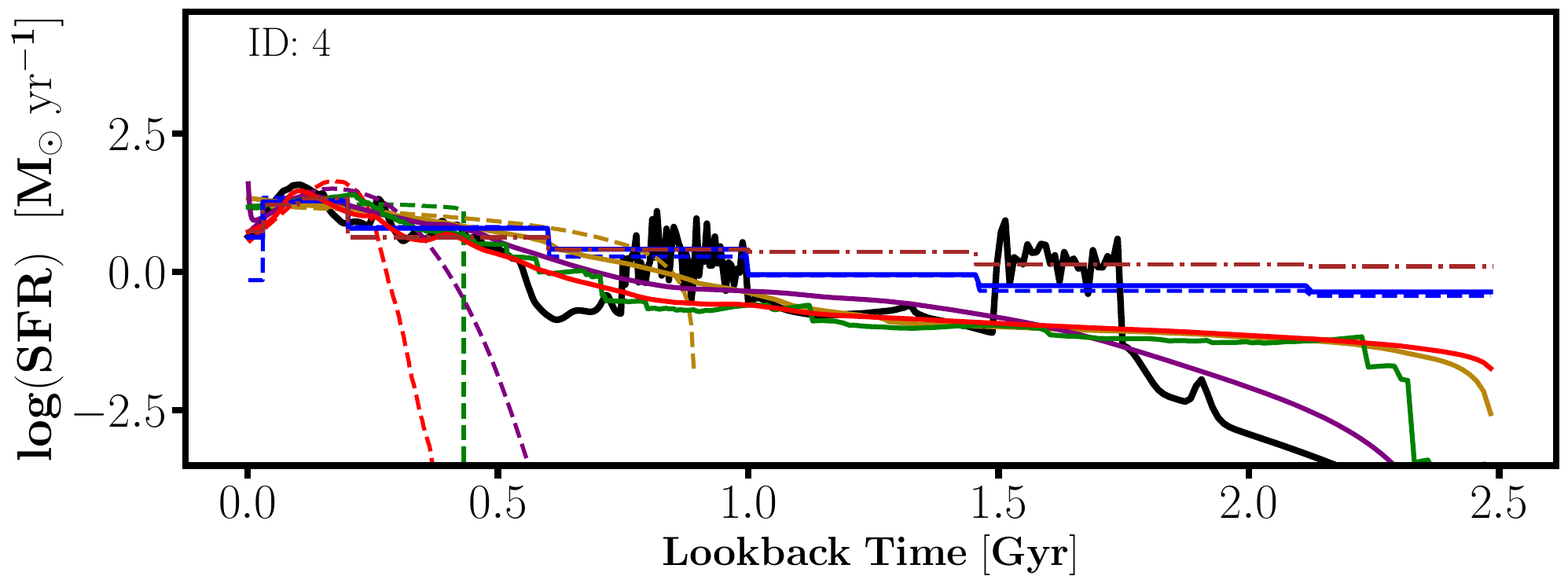}
    \includegraphics[width=0.49\linewidth]{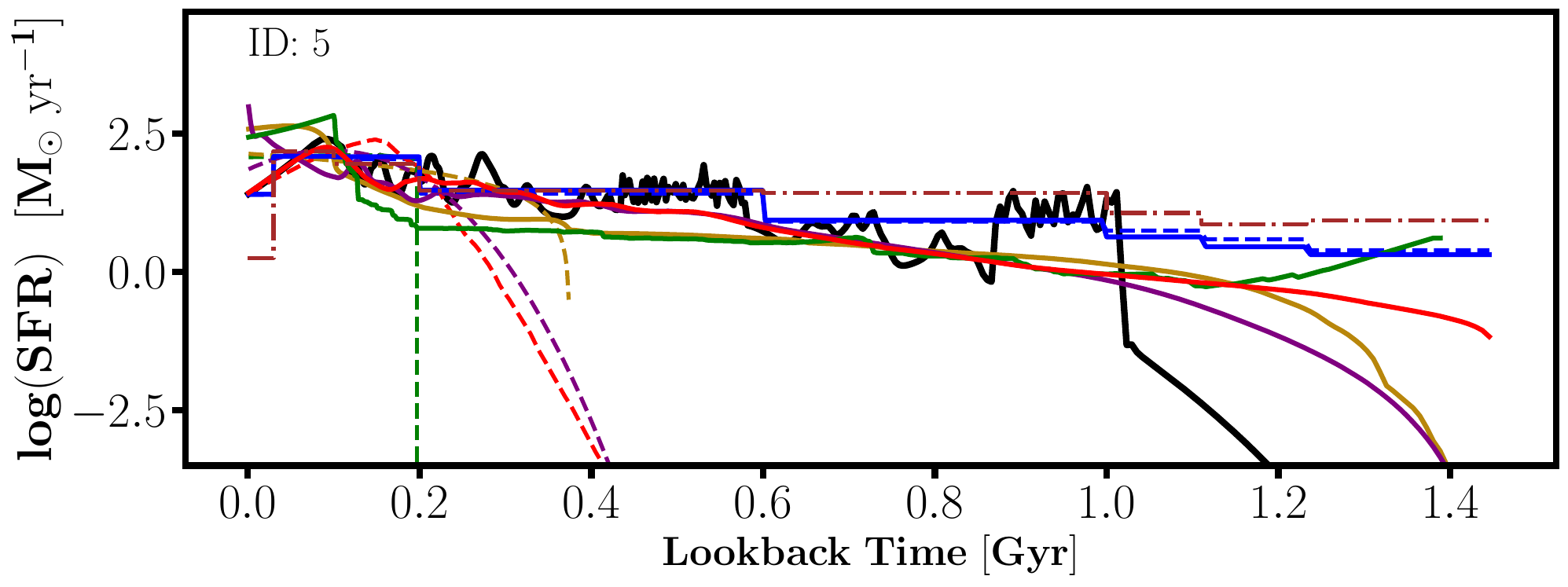}
    \includegraphics[width=0.49\linewidth]{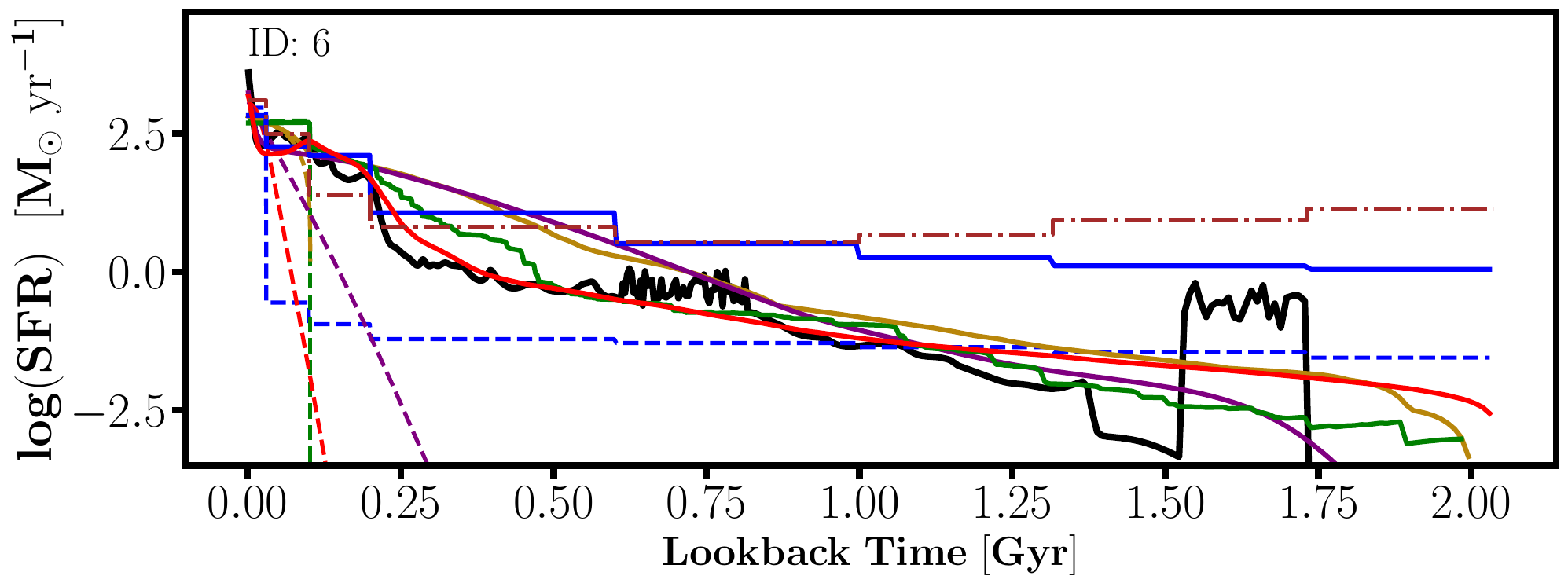}
    \includegraphics[width=0.49\linewidth]{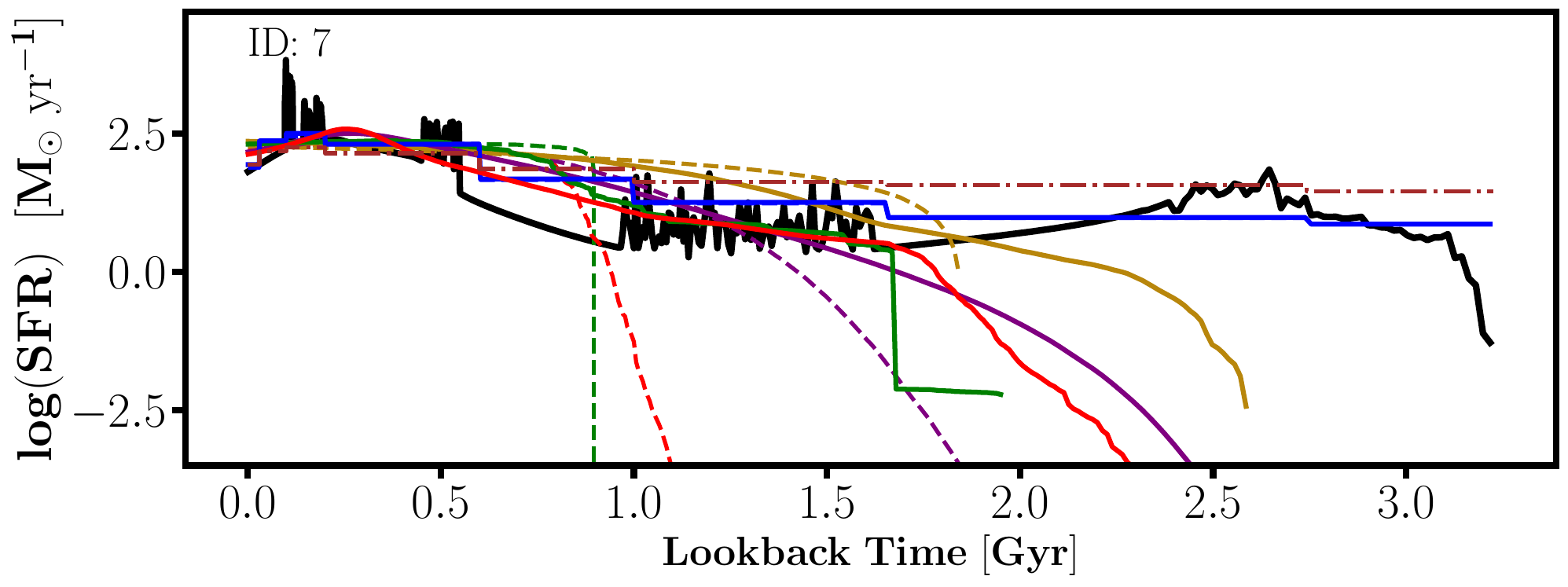}
    \includegraphics[width=0.49\linewidth]{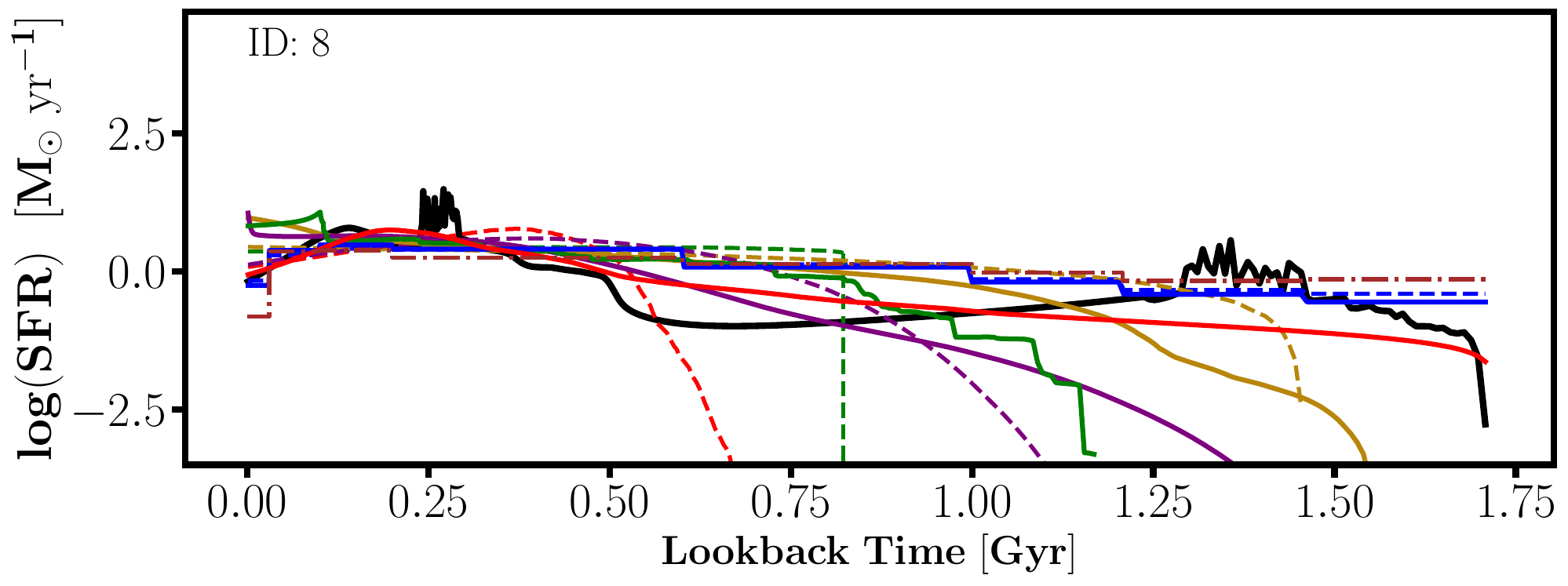}
    \includegraphics[width=0.49\linewidth]{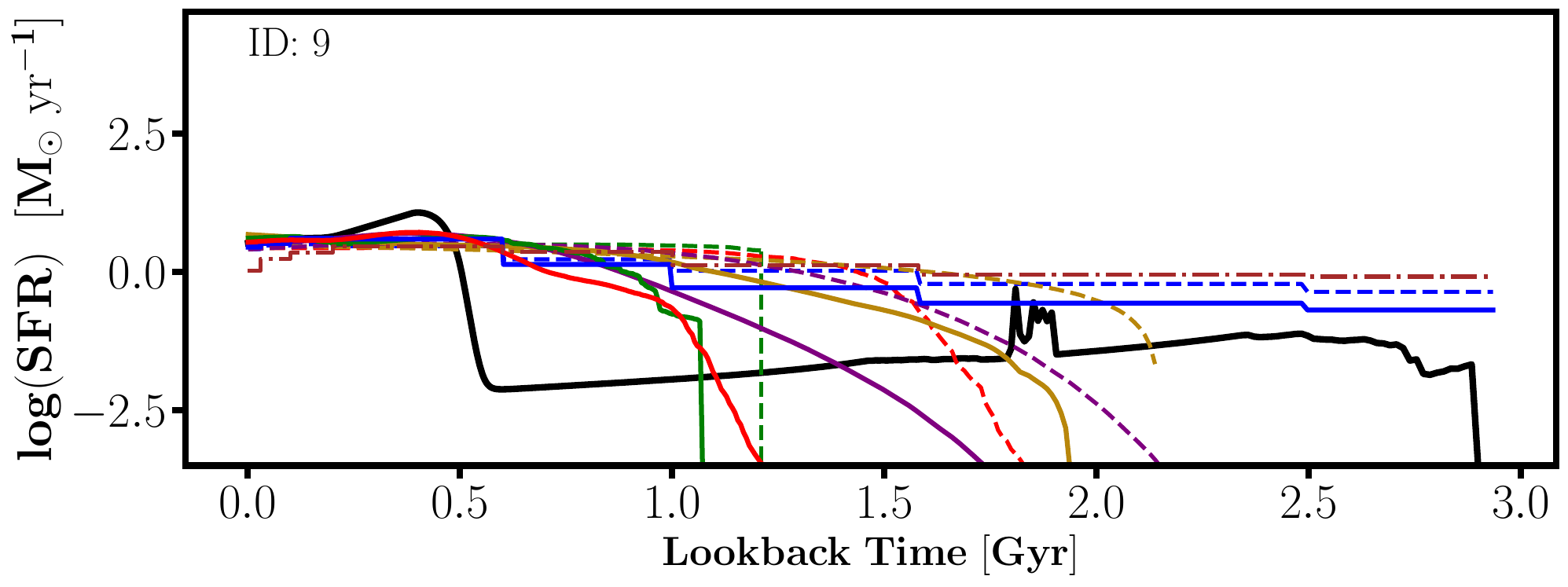}
    \includegraphics[width=0.49\linewidth]{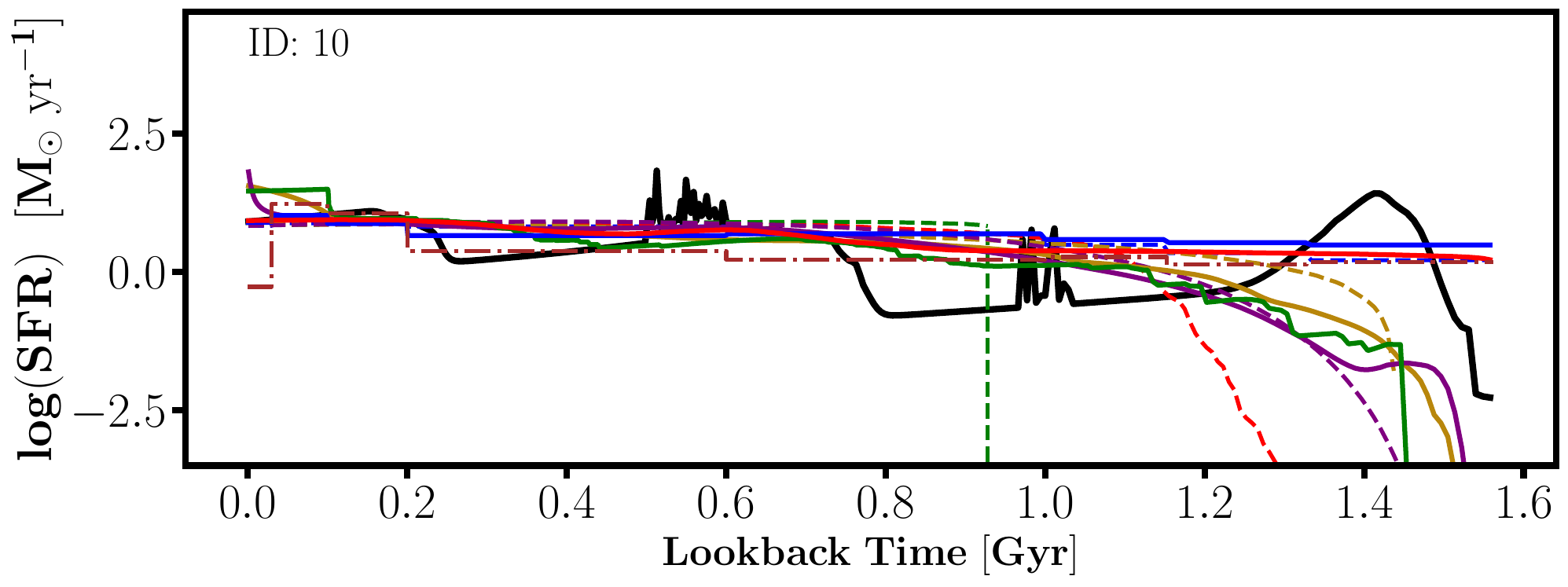}
    \caption{Star formation histories (SFHs) of the mock galaxies. The black lines represent the input (true) models, while the dashed and solid lines show the results from unresolved and resolved analyses, respectively. Unresolved parametric models generally struggle to accurately reproduce the true SFHs. In contrast, the resolved parametric models, particularly the DPL and LGN models, demonstrate greater flexibility and are more consistent with the true SFHs. }
    \label{fig3}
\end{figure*}

\begin{deluxetable}{ccccc}
\tablenum{5}
\tablecaption{Median deviation of key parameters from the simulated model for each SFH model. The continuity non-parametric models based on unresolved fluxes using Bagpipes and Prospector are denoted as NPM$\dagger$ and PSP$\dagger$, respectively. \label{tb5}}
\tablewidth{0pt}
\tablehead{
\colhead{Model} & \colhead{$\Delta \mathrm{\log}(\mstar/\msun)$} & \colhead{$\Delta \mathrm{log(SFR)}$} & \colhead{$\Delta t_{50}$} & \colhead{$\Delta t_{5}$}\\
\colhead{} & \colhead{} & \colhead{[$\msun$ yr$^{-1}$]} & \colhead{[Gyr]} & \colhead{[Gyr]}
}
\startdata
EXP & 0.03 $\pm$ 0.04 &  0.29 $\pm$ 0.41 & 0.01 $\pm$ 0.13 &  0.47 $\pm$ 0.75 \\
DLY & 0.04 $\pm$ 0.07 &  0.24 $\pm$ 0.38 & 0.00 $\pm$ 0.12 &  0.39 $\pm$ 0.71 \\
DPL & 0.00 $\pm$ 0.00 & -0.07 $\pm$ 0.07 & 0.01 $\pm$ 0.02 &  0.17 $\pm$ 0.40\\
LGN & 0.00 $\pm$ 0.04 &  0.06 $\pm$ 0.26 & 0.03 $\pm$ 0.04 &  0.49 $\pm$ 0.48\\
NPM & 0.04 $\pm$ 0.02 & -0.06 $\pm$ 0.06 & 0.12 $\pm$ 0.08 &  -0.19 $\pm$ 0.37\\
NPM$^{\dagger}$ & 0.03 $\pm$ 0.02 & -0.09 $\pm$ 0.07 & 0.12 $\pm$ 0.20 & -0.06 $\pm$ 0.28 \\
PSP$^{\dagger}$ & 0.11 $\pm$ 0.11 & -0.15 $\pm$ 0.13 & 0.28 $\pm$ 0.35 &  -0.56 $\pm$ 0.64 \\
\enddata
\end{deluxetable}

\begin{deluxetable}{cccc}
\tablenum{6}
\tablecaption{Median deviation of the star formation history from the simulated model for each SFH model. `LateTime' corresponds to the most recent SFH, while `EarlyTime' refers to the first 500 Myr of star formation.\label{tb6}}
\tablewidth{0pt}
\tablehead{\colhead{Model} & \colhead{$\Delta$SFH} [dex] & \colhead{$\Delta$SFH$\mathrm{_{Late Time}}$} & \colhead{$\Delta$SFH$\mathrm{_{Early Time}}$}}
\startdata
EXP & -0.02 $\pm$ 1.02 &  0.07 $\pm$ 0.47 & -0.23 $\pm$ 2.81 \\
DLY & 0.27 $\pm$ 0.92 &  0.10 $\pm$ 0.51 & -0.22 $\pm$ 2.54 \\
DPL & 0.00 $\pm$ 0.63 & 0.00 $\pm$ 0.24 &  0.54 $\pm$ 2.40 \\
LGN & -0.03 $\pm$ 0.84 &  0.12 $\pm$ 0.36 & -1.25 $\pm$ 2.04 \\
NPM & 0.64 $\pm$ 0.83 & 0.02 $\pm$ 0.31 &  1.07 $\pm$ 1.68 \\
NPM$^{\dagger}$ & 0.49 $\pm$ 0.87 & -0.11 $\pm$ 0.48 &  1.21 $\pm$ 1.95 \\
PSP$^{\dagger}$ & 0.76 $\pm$ 1.11 & -0.09 $\pm$ 0.36 &  1.60 $\pm$ 1.96 \\
\enddata
\end{deluxetable}

\begin{figure}
    \includegraphics[width=1\linewidth]{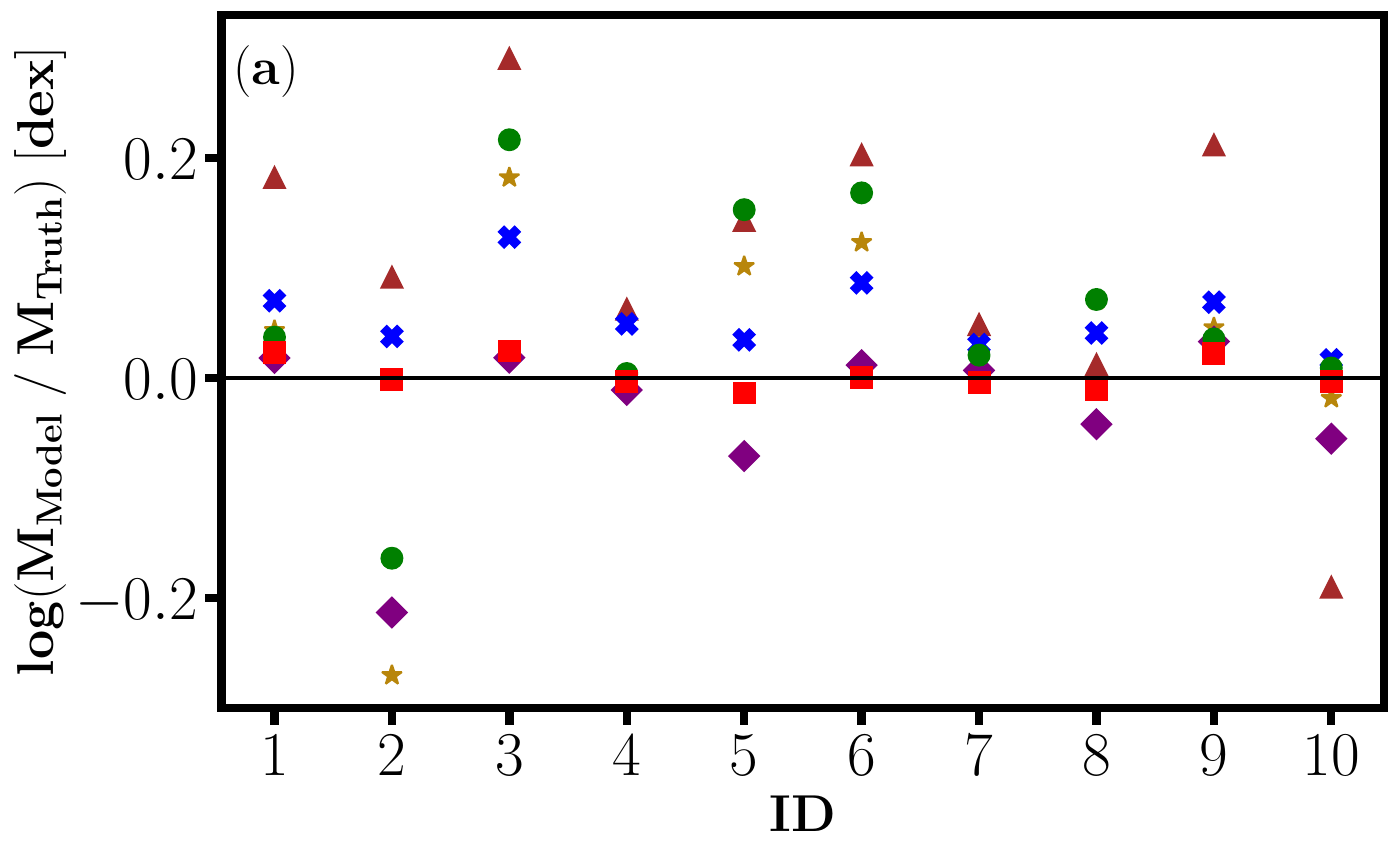}
    \includegraphics[width=1\linewidth]{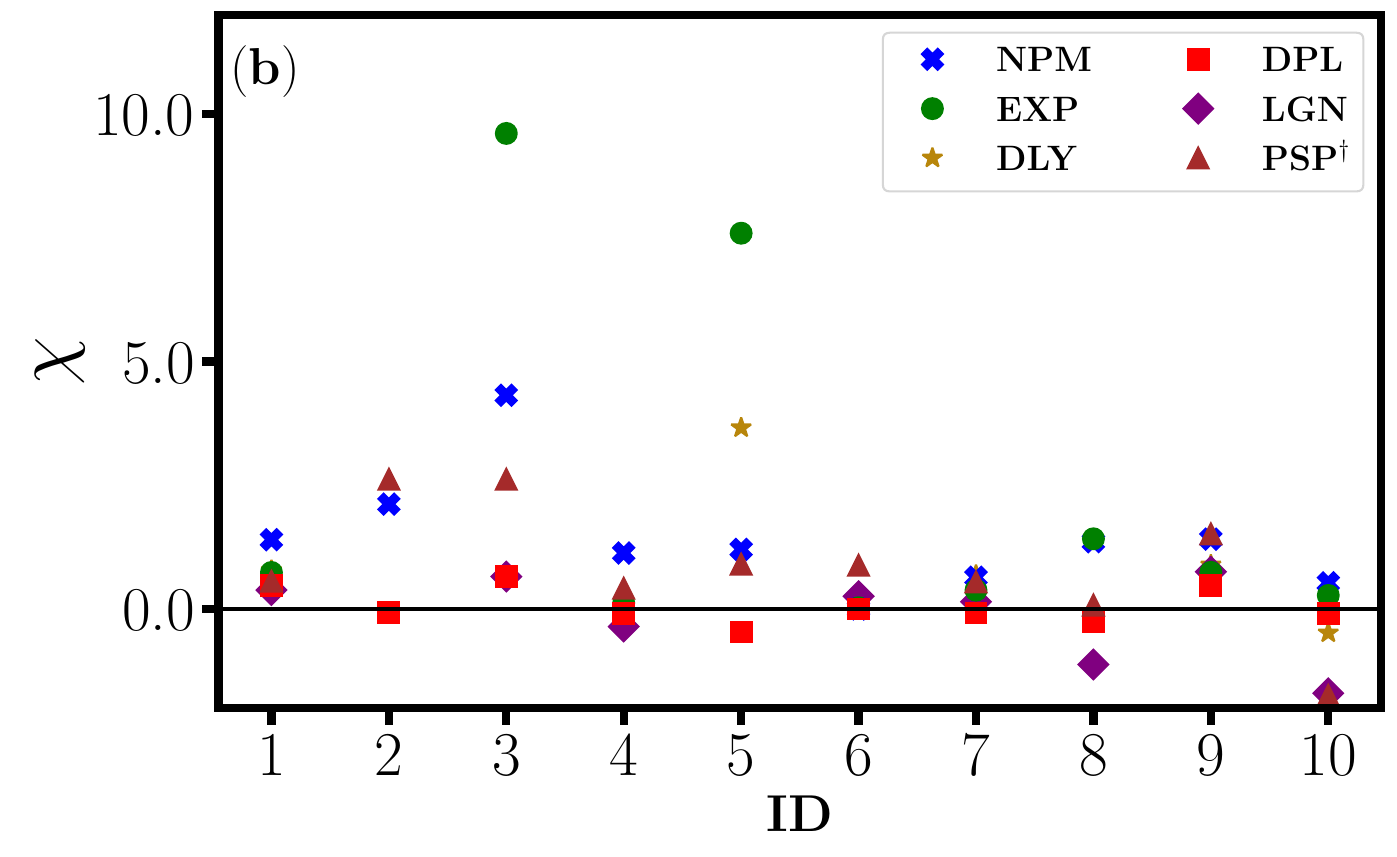}
    \caption{Panel (a) shows relative differences in stellar mass formed from different spatially resolved analyses compared to the true models. In addition, unresolved non-parametric models using Prospector (PSP, dark-red triangles) are compared. The integrated SFHs from the pixel-by-pixel analysis, particularly those using the DPL and LGN parametric models, show the smallest differences among all models. Panel (b) illustrates the normalized residuals ($\chi$= (Model-Truth)/ Model Uncertainty) for each method, highlighting the reliability of the associated uncertainties. The PSP and NPM models exhibit median $\chi$-values close to 1 with moderate scatter, indicating well-calibrated uncertainties. The DPL model shows a small median $\chi$-value with a low scatter, suggesting a slight overestimation of uncertainties. The LGN model also has a small median $\chi$-value but with a larger scatter, implying variability in uncertainty estimates. The EXP and DLY models have relatively low median $\chi$-values but exhibit high scatter, suggesting their uncertainties may be underestimated in some cases. The $\chi$-value range is limited for clarity, with five extreme outliers omitted.}
    
    \label{fig4}
\end{figure}

\begin{figure*}
    \centering
    \includegraphics[width=0.24\linewidth, height=0.24\linewidth]{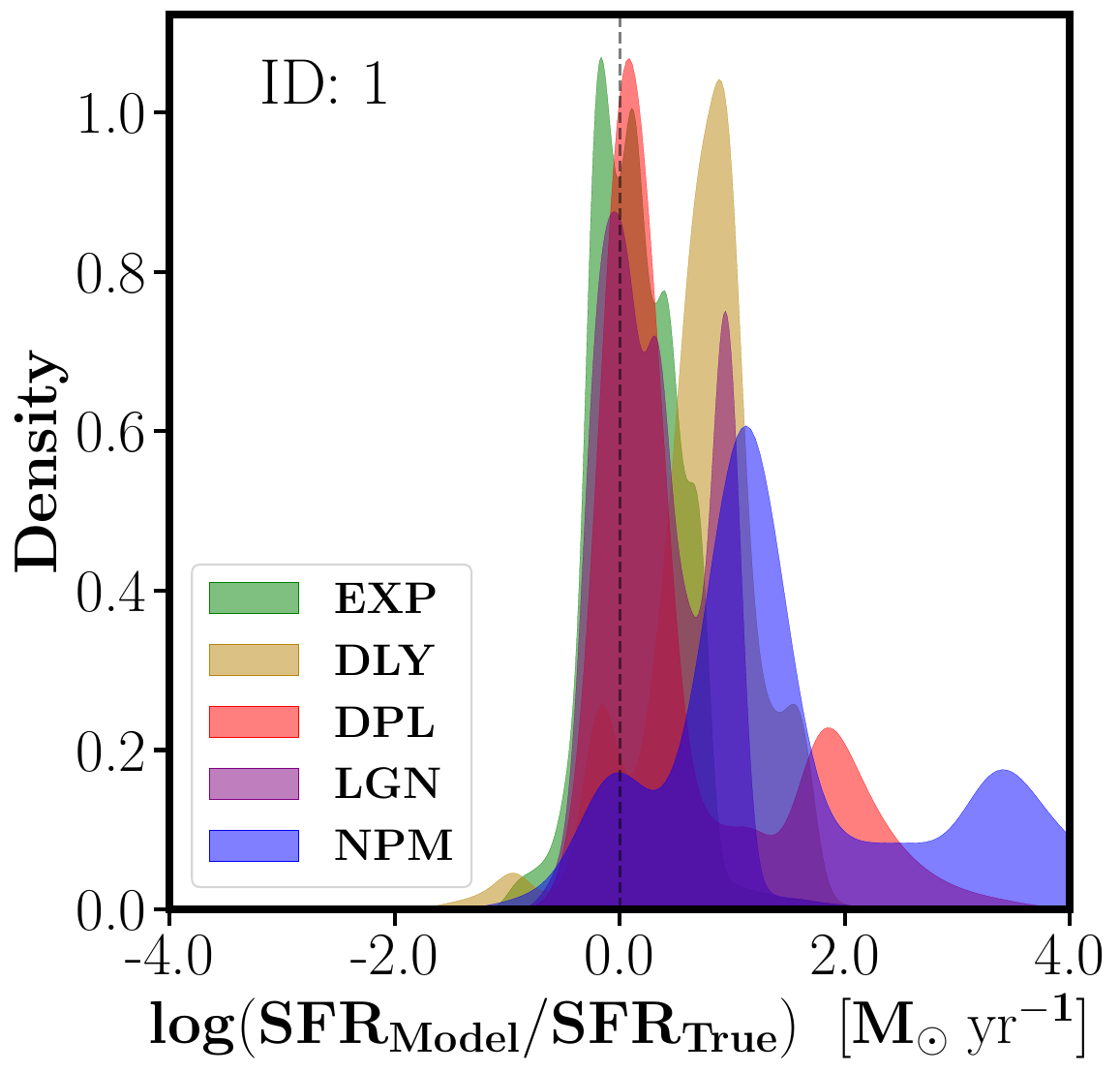}
    \includegraphics[width=0.24\linewidth, height=0.24\linewidth]{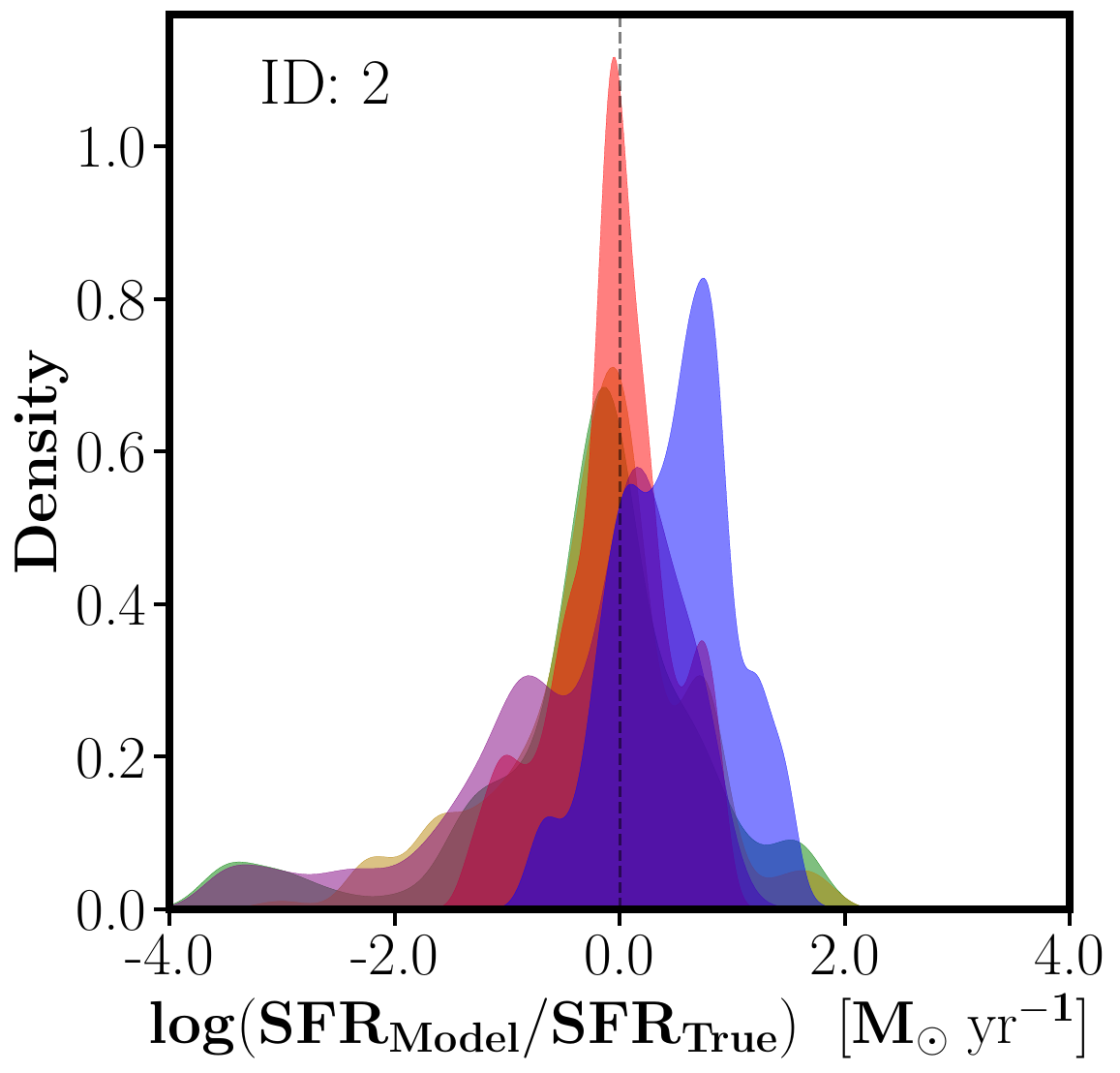}
    \includegraphics[width=0.24\linewidth, height=0.24\linewidth]{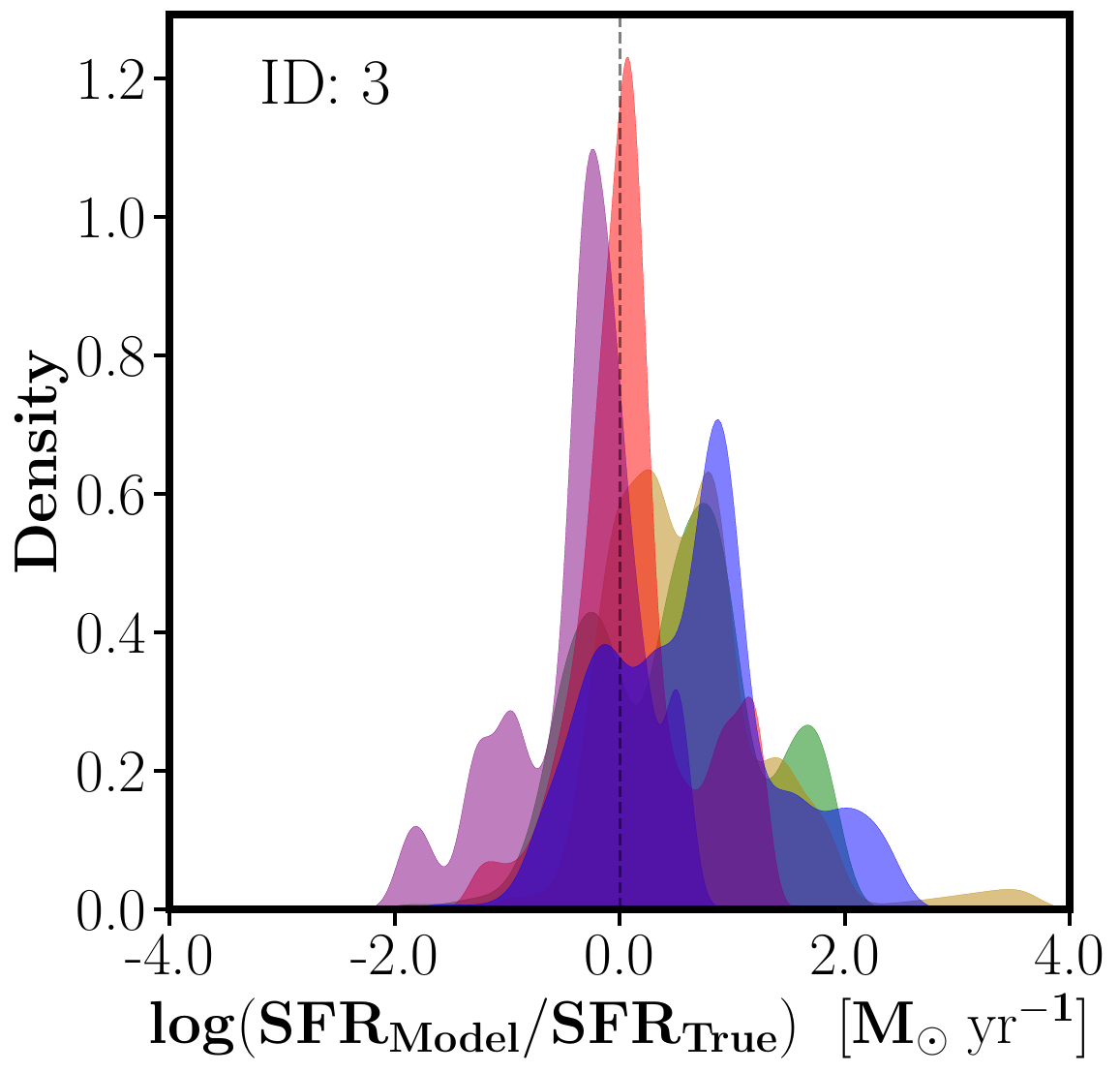}
    \includegraphics[width=0.24\linewidth, height=0.24\linewidth]{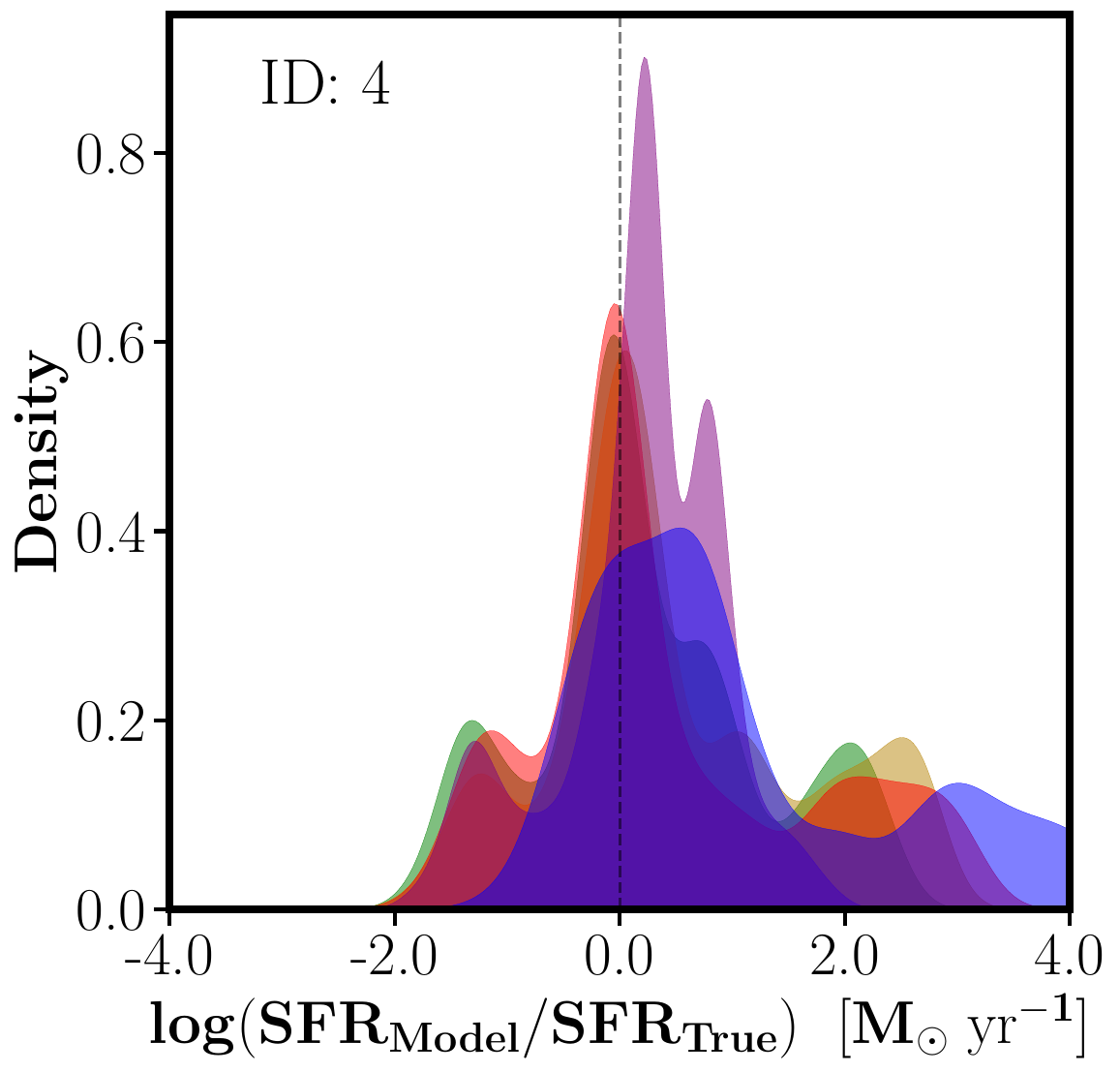}
    \includegraphics[width=0.24\linewidth, height=0.24\linewidth]{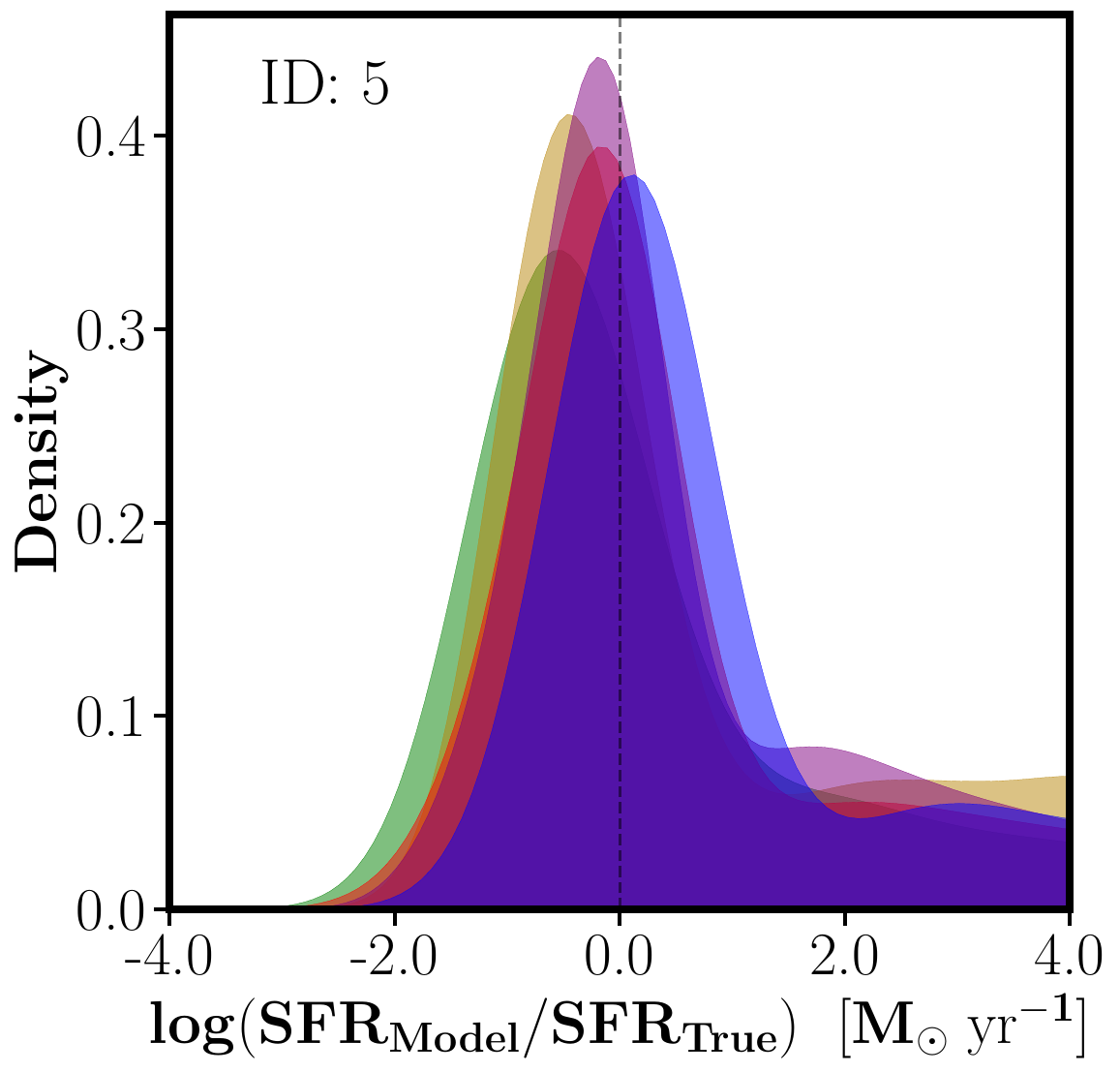}
    \includegraphics[width=0.24\linewidth, height=0.24\linewidth]{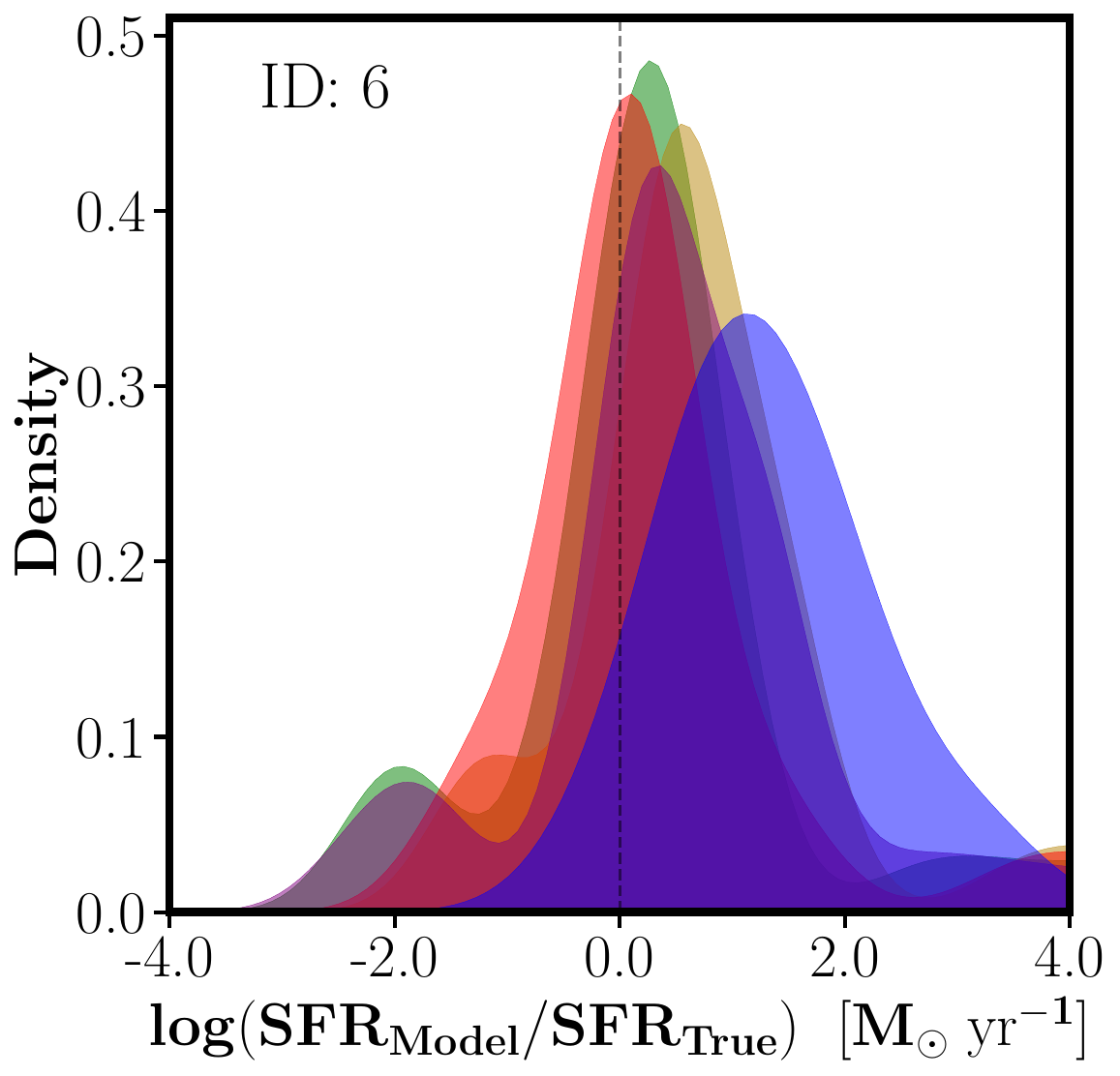}
    \includegraphics[width=0.24\linewidth, height=0.24\linewidth]{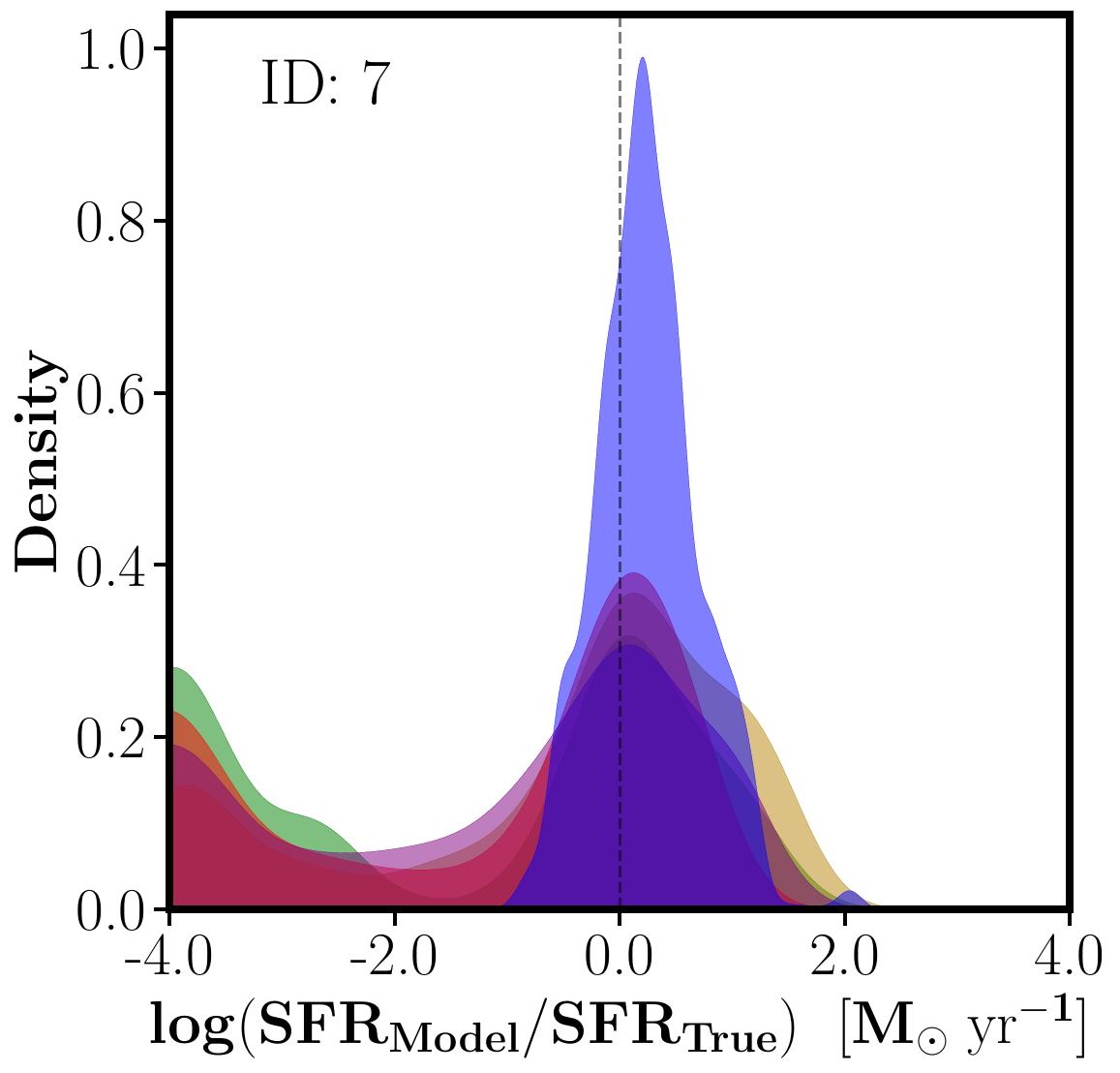}
    \includegraphics[width=0.24\linewidth, height=0.24\linewidth]{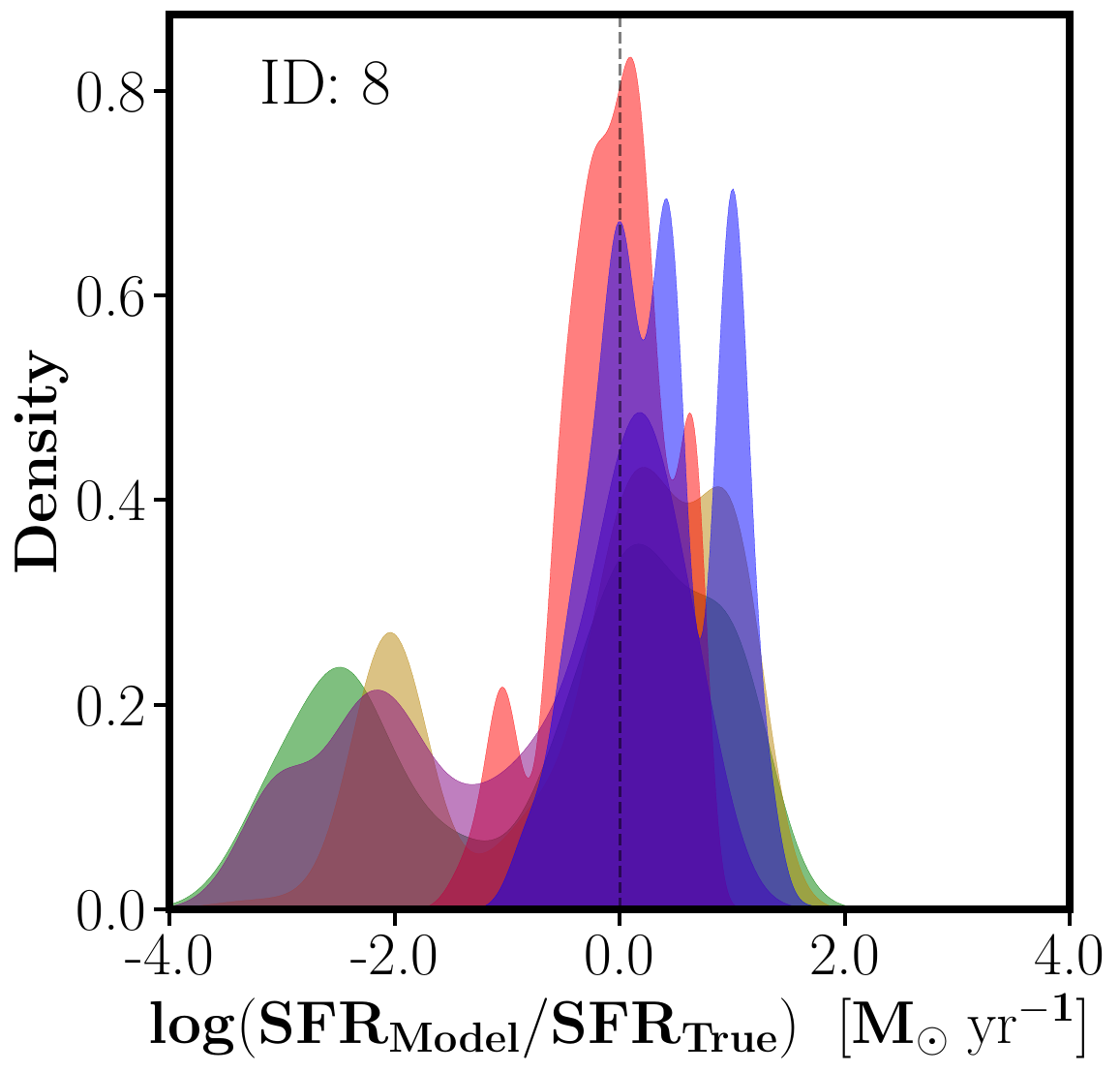}
    \includegraphics[width=0.24\linewidth, height=0.24\linewidth]{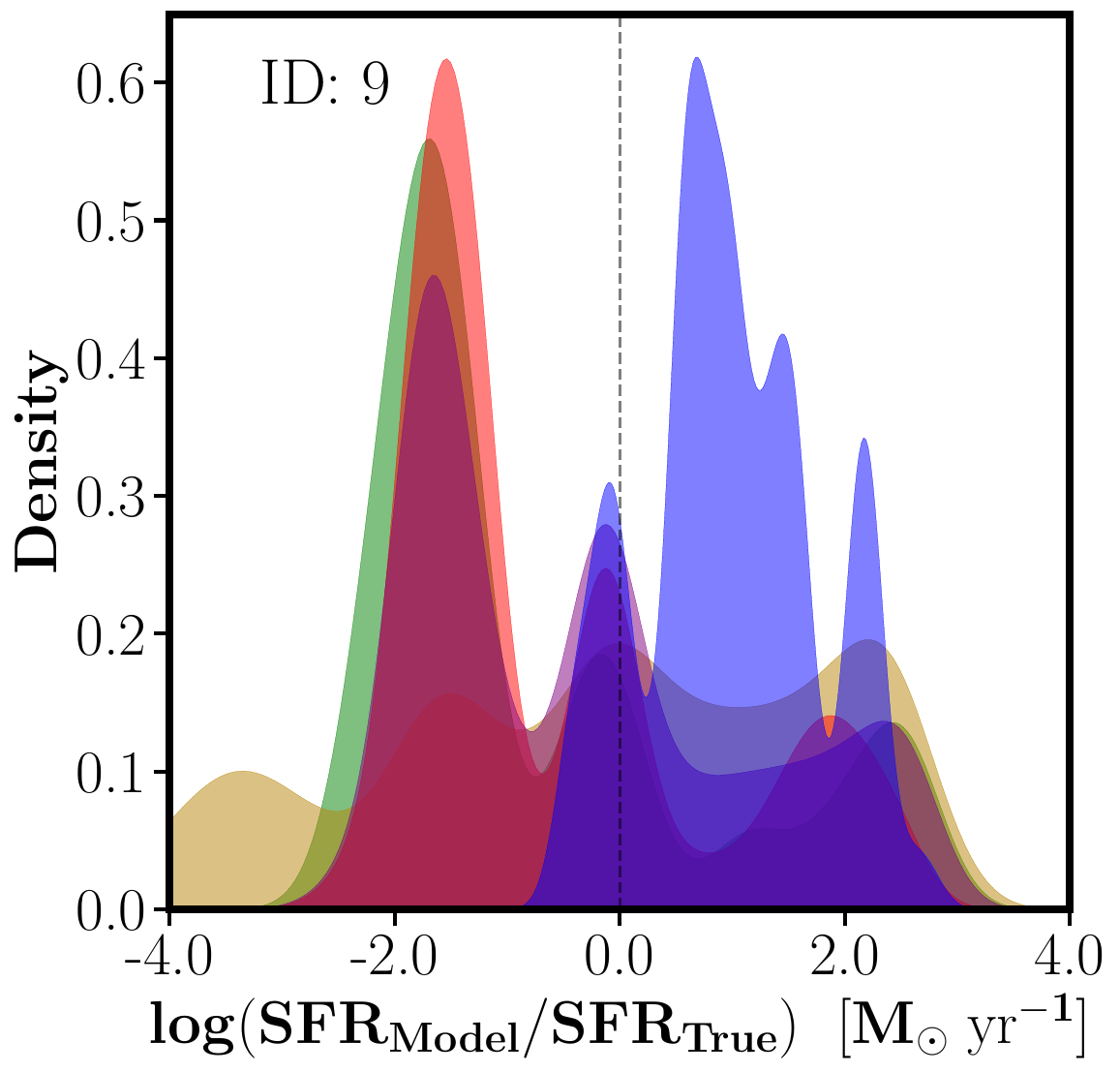}
    \includegraphics[width=0.24\linewidth, height=0.24\linewidth]{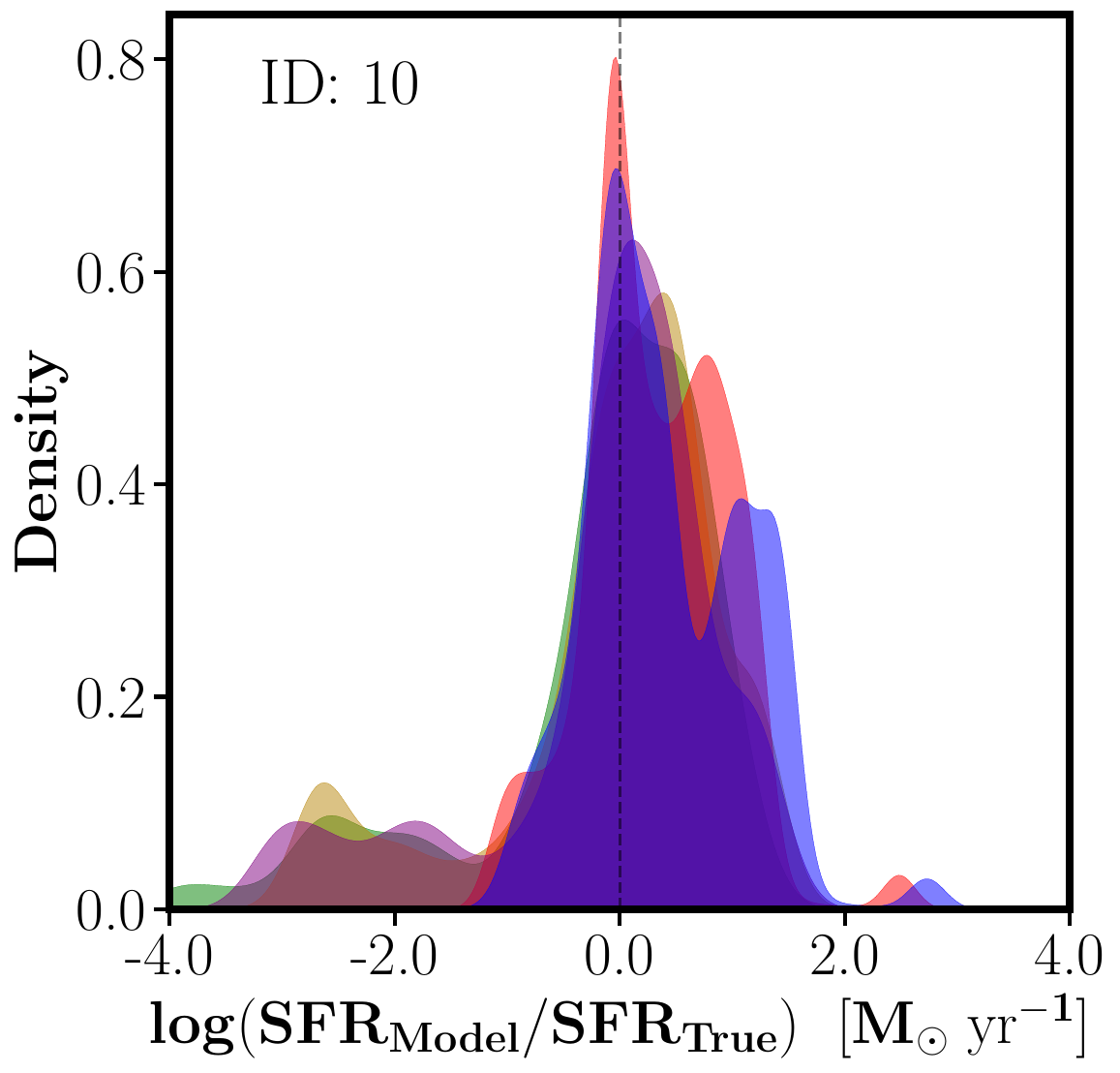}
    \includegraphics[width=0.48\linewidth, height=0.24\linewidth]{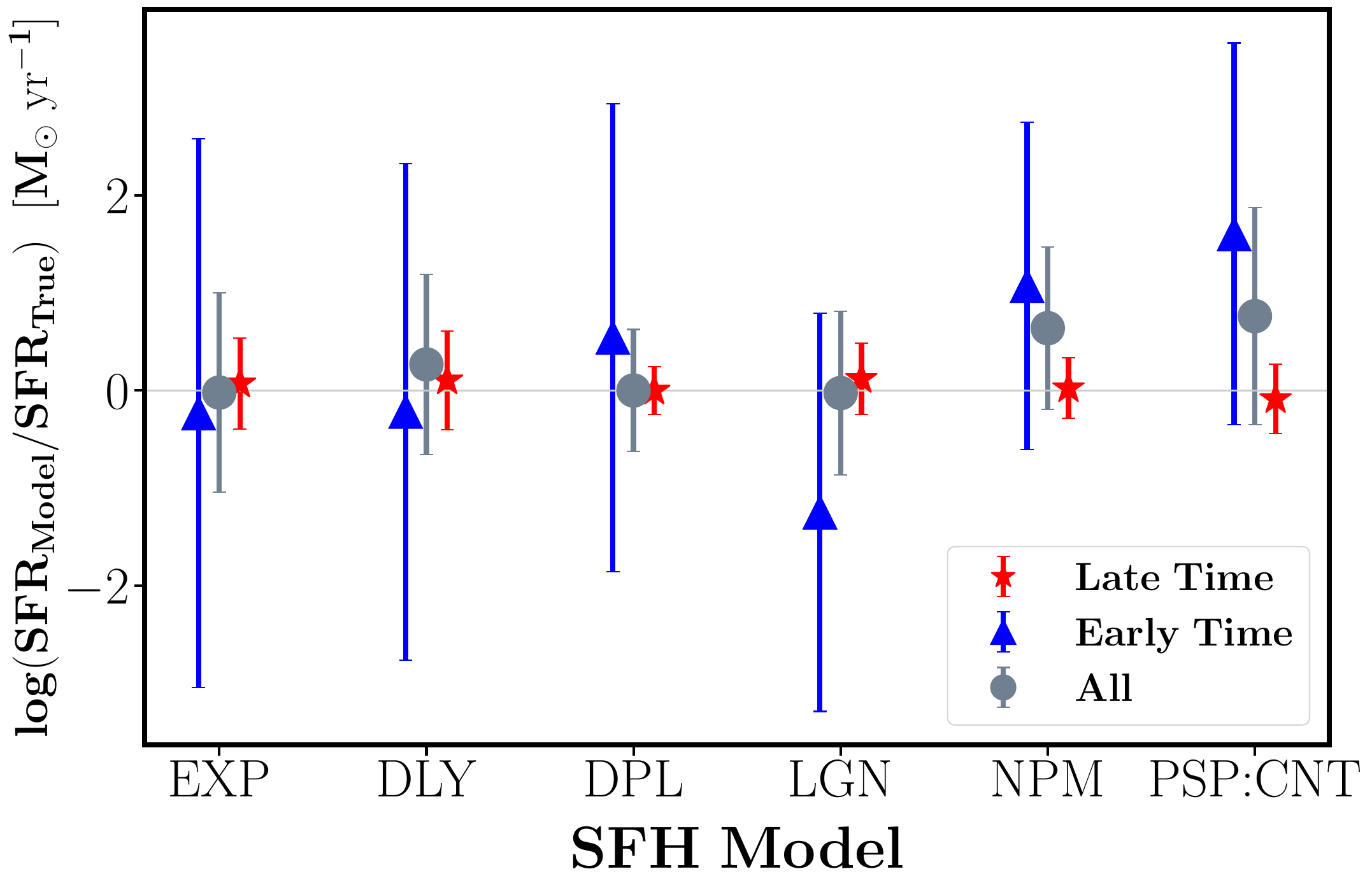}
    \caption{The distributions of the relative differences in SFHs compared to the input models are shown for each object. These distributions are built from SFR values across all time and all pixels, rather than just the recent SFR at the time of observation. For most mock galaxies, the resolved parametric models, particularly the DPL and LGN models, are centered around zero, indicating minimal bias. The width of the distributions, reflecting the variability and scatter of each model, is notably larger for the resolved non-parametric (NPM) models. The tails of the distributions highlight significant deviations, particularly at early formation times. In the bottom right panel, these deviations are quantified for different models, showing the total SFH (gray circles), the SFH over the recent 500 Myr (red stars), and early epochs (blue triangles). The DPL models exhibit the least deviations and scatter across all cases.}
    \label{fig5}
\end{figure*}

\section{Results} 

In this section, we present the results of our analysis for mock galaxies, focusing on the recovery of star formation histories (SFHs) using various assumptions and comparing these with the original SFHs. Additionally, we examine key galaxy properties such as stellar mass, star formation rates (SFRs), and formation ages. We then extend the analysis to an observed sample and discuss the relative deviations in the primary galaxy properties.

\subsection{SFH Recovery of Simulated Galaxies}

Figure \ref{fig3} displays the SFHs of the simulated galaxies derived from our spatially resolved and unresolved analysis. The black lines in each panel show the true SFHs from the input models, while the unresolved and pixel-by-pixel SFHs derived from different models are represented by color-coded dashed and solid lines, respectively. To avoid clutter, we have omitted the scatter for each model, focusing instead on the key trends. It is important to note that the input SFH models of mock galaxies were intentionally varied to represent a range of scenarios, including rising and falling SFRs as well as both short- and long-term starbursts. This diversity ensures that our results are robust across different types of galaxies, such as starburst, quiescent, or post-starburst systems.

As shown in Figure \ref{fig3}, the parametric SFHs derived from unresolved fluxes fail to accurately recover the original SFHs, particularly in the early stages of star formation and in capturing short-timescale variations in SFR. This limitation is consistent with previous studies, where unresolved SFHs are often affected by the outshining effect \citep[e.g.,][]{jain2024, gimenez2024}. Although the differences in stellar masses formed using these models compared to the true values remain small and within the uncertainties (with a scatter of $\pm 0.07$ dex), the inferred cosmic dawn age (CDA), quantified by $t_5$ (the lookback time when 5\% of the total stellar mass was formed), can be underestimated by up to $\sim 1-1.5$ Gyr. We define $\Delta t_5$ as the difference between the true and recovered $t_5$, further highlighting the biases in parametric SFHs based on unresolved fluxes. The results from the continuity non-parametric models using unresolved fluxes (PSP$\dagger$ as brown dot-dashed lines and NPM$\dagger$ as dashed blue lines in Figure \ref{fig3}) perform better in recovering the overall shape of the SFHs. Despite this improvement, the stellar masses in the PSP$\dagger$ model remain slightly overestimated, with a median offset of $0.11\pm0.11$ dex (Table \ref{tb5}). Additionally, the CDA in this model is overestimated by $0.56\pm0.64$ Gyr, which remains significant for galaxies at these redshifts.

However, SFH models based on pixel-by-pixel analysis more accurately capture the early stages of galaxy formation and reproduce the general behavior of past SFRs, irrespective of the assumed model. This highlights the effectiveness of spatially resolved analysis in recovering the true stellar mass assembly history of galaxies. For example, the minimum difference between the recovered and true early formation ages ($t_{5}$) is $0.17\pm0.40$ and $-0.19\pm0.37$ Gyr (in the case of the DPL and NPM models respectively) indicating a good match between the resolved models and the true galaxy ages. These findings are consistent with recent studies showing that pixel-by-pixel SED fitting extends SFHs to earlier epochs \citep[e.g.,][]{gimenez2024}. We note that for most simulated galaxies, the resolved models, in particular DPL one, outperform the unresolved version. However, in simulated galaxy 9, the unresolved DPL appears to perform slightly better than the resolved DPL due to its extremely low SFR for most of its history ($\log(SFR) < 0$, reaching $\sim -2.5$ later) followed by a sharp burst ($\Delta \log(SFR) \sim 3$) in the last 0.5 Gyr. The unresolved fit smooths over this extreme burst, whereas the resolved fit, struggles to fully recover its timing and amplitude.

To further quantify how well each model reproduces the overall SFH, panel a of Figure \ref{fig4} presents the relative differences in stellar mass between the original and estimated values for each mock galaxy, with median values listed in Table \ref{tb5}. The total stellar masses derived from the resolved analyses exhibit minimal bias, with a typical $\Delta$ formed stellar mass of $0.01\pm0.04$ dex across the models.

In addition to evaluating the ability of different methods to reproduce the true stellar masses and SFHs, it is equally important to assess whether the associated uncertainties are reliable. To address this, we calculated the normalized residuals ($\chi$ = (Model-Truth)/ Model Uncertainty) for each model and mock galaxy, as shown in panel b of Figure 4. Our analysis reveals that the DPL model exhibits the most reliable uncertainty estimates, with a nearly unbiased median and low scatter ($\chi=-0.070\pm0.34$), suggesting slightly overestimated uncertainties. The LGN model shows a similarly small median deviation but with a larger scatter ($\chi=-0.10\pm4.21$), indicating greater variability in its uncertainty estimates. The unresolved Prospector (PSP) and non-parametric pixel-by-pixel (NPM) models show reasonable $\chi$-values of $0.72\pm1.20$ and $1.28\pm1.11$, respectively. In contrast, the EXP and DLY models, while having small median $\chi$-values, exhibit significantly larger scatter ($0.56\pm4.68$ and $0.73\pm6.99$), indicating that their uncertainties are likely less reliable. This high scatter may arise from their rigid SFH assumptions, which do not fully capture the complexity of star formation histories.

In general, the DPL and LGN models show the smallest systematic differences in stellar masses, SFRs, and formation ages ($t_{50}$, where 50\% of the stellar masses are formed, see e.g., \cite{tacchella2022a}). Spatially resolved non-parametric (NPM) models, which use continuity SFHs, also align well with the overall shape of the SFHs, though the median difference in $t_{50}$ suggests larger systematic biases compared to resolved parametric models. Both resolved and unresolved non-parametric models (NPMs and PSP) tend to overestimate SFRs in the early stages of galaxy evolution. This trend will be discussed further in discussion section 4, where we explore the potential causes of this behavior.

We also quantify the deviations in SFH shape for each mock object in Figure \ref{fig5} and Table \ref{tb6}, which show the density distribution of relative differences between each model and the true SFHs. These distributions are constructed from SFR values across all time and all pixels, providing a comprehensive view of how well different models recover the full SFH. The spatially resolved parametric models, particularly the DPL and LGN models, exhibit the smallest deviations and scatter, with the median and 1$\sigma$ scatter illustrated in the bottom-right panel (gray circles). For the recent 500 Myr of SFH evolution (red stars), the DPL model continues to outperform others, although the LGN and resolved-NPM models also exhibit small systematic differences. These models are particularly effective at capturing recent changes in SFR, as demonstrated in the mock galaxies 2, 3, and 4, where strong starburst episodes and the following SFR declines are present.

\begin{figure*}
    \centering
    \includegraphics[width=0.49\linewidth]{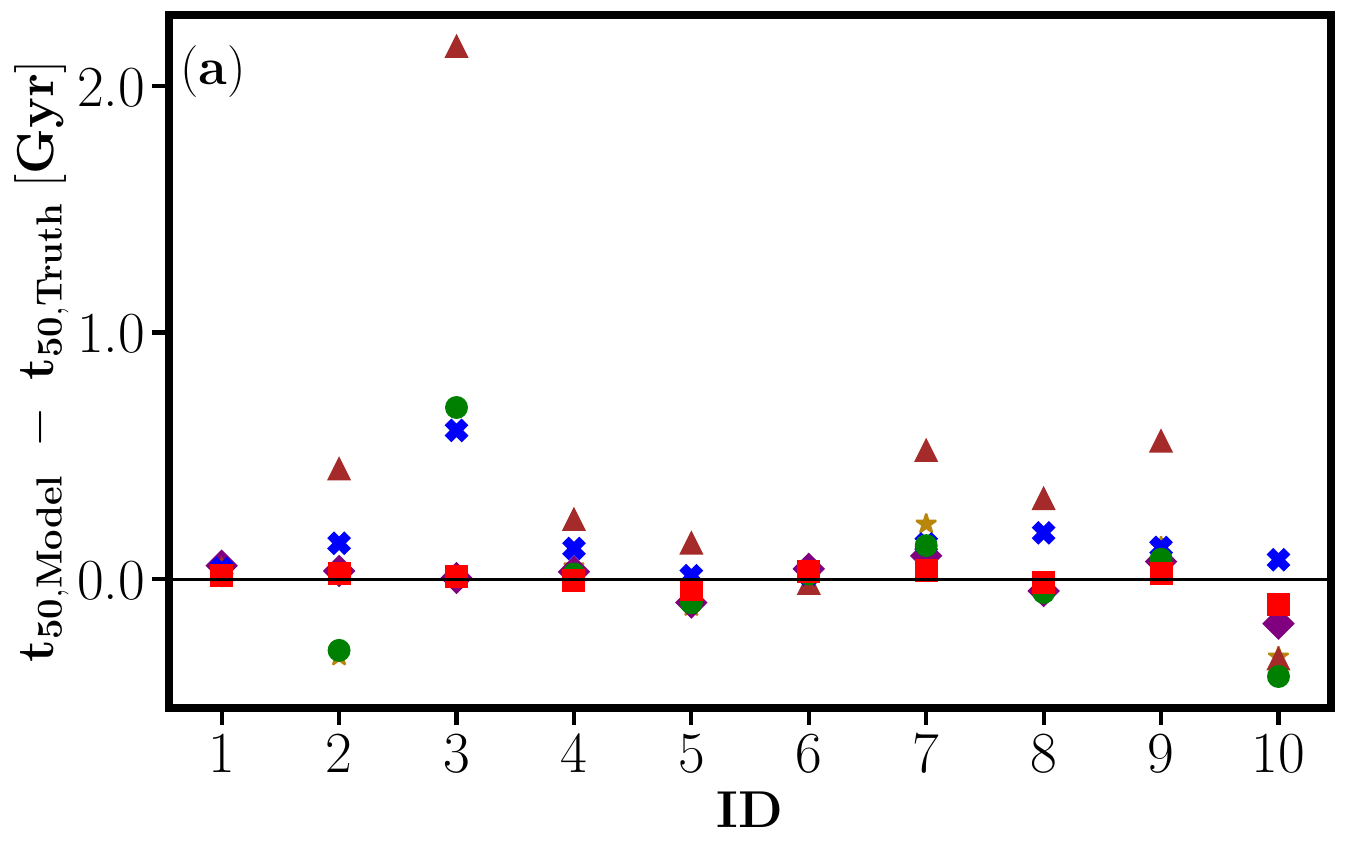}
    \includegraphics[width=0.49\linewidth]{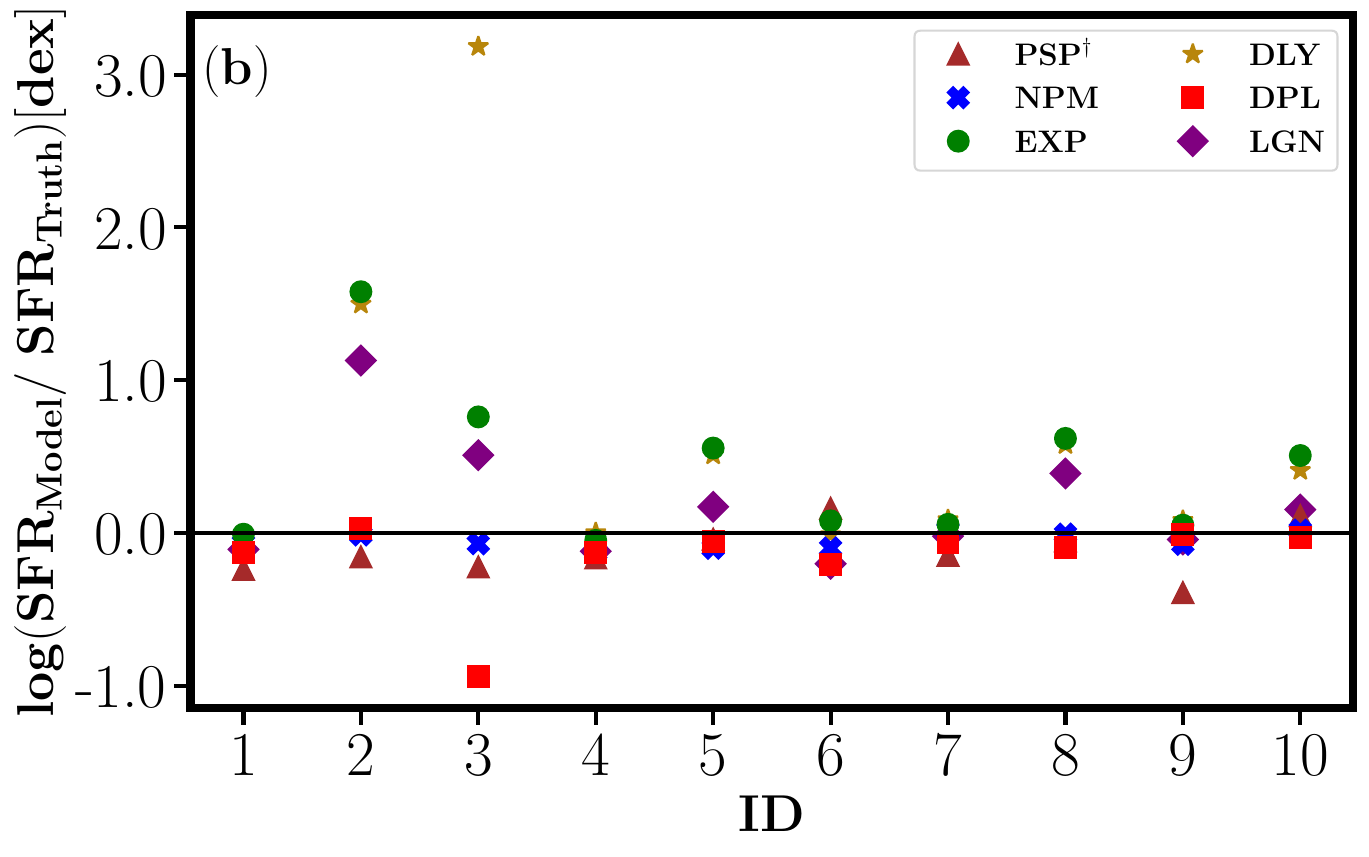}
    \caption{Similar to panel (a) of Figure 4, the relative differences in formation age ($t_{50}$) are shown in the left panel (a), and the recent SFR in the right panel (b), comparing results from different SFH analyses to the true models. The SFH models based on resolved DPL and LGN show good agreement with the input models for both formation ages and estimated SFRs.}
    \label{fig6}
\end{figure*}

Nevertheless, while the resolved DPL and LGN parametric models successfully capture the general trends and transitions in the SFHs, non-parametric methods tend to introduce less smooth transitions, making it more difficult to pinpoint specific features such as the peak of recent starbursts or the rate of SFR decline. These irregularities create uncertainties in precisely determining the timing and intensity of such events. Additionally, non-parametric models often overestimate SFRs at early epochs, as seen in the differences in SFHs during early stages (blue triangles in Figure \ref{fig5}). The DPL, EXP, and DLY models mitigate this bias more effectively (see Table \ref{tb6} for quantitative comparisons). The large scatter in early-stage SFR estimates occurs across models due to the sharp SFR declines at early times, which introduces significant scatter into our analysis. For consistency in our analysis, we assumed a minimum $\log SFR(t) = -3$  when computing deviations and scatter in Figure \ref{fig5} and Table \ref{tb6}, specifically in cases where the SFR dropped sharply to very low values or when some models did not provide meaningful constraints at early times. This prevents artificially large deviations from dominating the scatter measurements but may introduce some bias in the early-time comparisons. Thus, caution is required when interpreting scatter and biases at early stages. The overall effect of over- or underestimation of SFHs is reflected in the estimated formation time ($t_{50}$), as shown in the left panel of Figure \ref{fig6}. Here, the resolved DPL model exhibits the smallest bias, while the unresolved non-parametric PSP model (red triangles) shows substantial systematic deviations.

Overall, parametric models that account for rising and falling star formation rates (e.g., DPL and LGN), when applied in a pixel-by-pixel analysis, show greater flexibility in capturing SFR variations over time, especially in star-forming galaxies. By contrast, models such as the exponentially declining ($\tau$) model exhibit less flexibility in representing such stochastic behavior. The right panel (b) of Figure \ref{fig6} compares the recent (last 100 Myr) SFRs to the input models. DPL, the resolved non-parametric model (NPM), and the unresolved PSP model all consistently estimate recent SFRs. However, some models, such as the resolved LGN, DLY, and EXP models, display notable deviations in recent SFRs, particularly with late SFR upturns in mock galaxies 2 and 3. These deviations may stem from the dominance of certain SFH shapes in specific regions of the galaxies or the inherent model structure \citep[][]{haskell2024}. 

Short-term starbursts, particularly those occurring in the early stages of galaxy formation (within the first $\sim$1 Gyr), are rarely resolved by these models. Incorporating an additional burst component into SFH models could help capture these events, though this would introduce additional parameters requiring further constraints. Nonetheless, models like DPL and LGN, which allow for both rising and falling slopes over time, deliver the closest match to the true galaxy SFHs within a pixel-by-pixel framework.

\subsection{Exploring the Resolved Maps}

Another key goal of this study is to evaluate the validity and accuracy of two-dimensional analyses based on pixel-by-pixel modeling. The spatial distributions of galaxy parameters, such as stellar mass, SFR, and age, provide critical insights into galaxy formation and evolution. Therefore, assessing the accuracy of these resolved maps is essential for understanding how well different SFH models capture the underlying structure of galaxies. In this subsection, we compare the input and recovered 2D distributions of key parameters, focusing on how well different SFH models perform in replicating these maps.

\begin{figure*}
    \centering
    \includegraphics[width=1\linewidth ,height=0.17\linewidth]{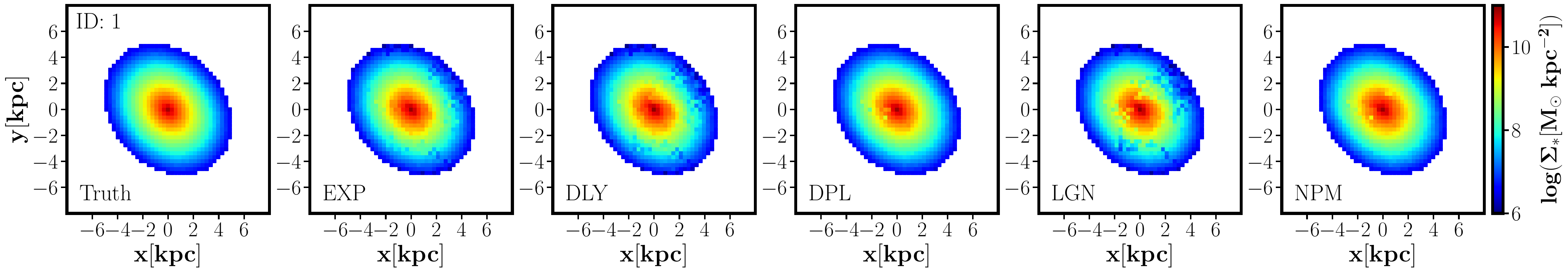}
    \includegraphics[width=1\linewidth ,height=0.17\linewidth]{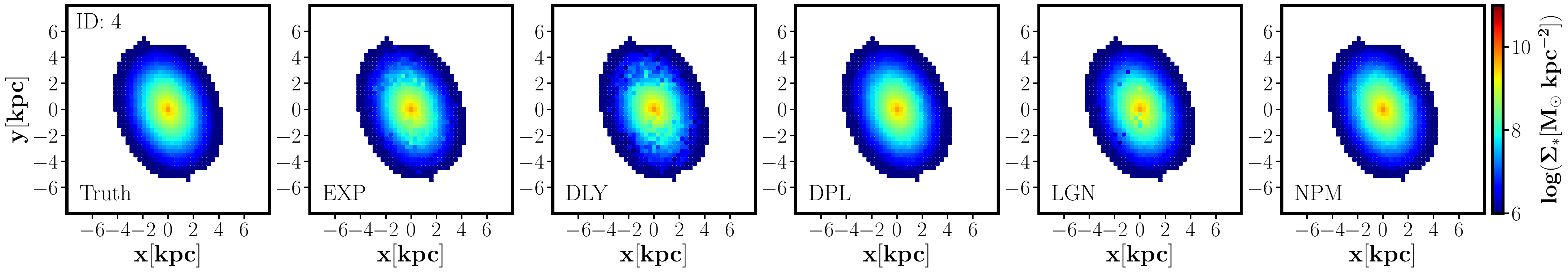}
    \caption{Comparison of stellar mass surface density maps derived from different SFH models with the true input models, shown on the left, for two mock galaxies.}
    \label{fig7}
\end{figure*}

Figure \ref{fig7} presents a comparison between the original simulated stellar mass surface density maps and those recovered from our spatially resolved analysis for two mock galaxies. The left-hand panels display the original, input stellar mass surface density distributions, while the other panels show the maps derived from various SFH models. At first glance, it is evident that all models recover the general structure of the original maps reasonably well. However, certain models stand out for their ability to produce smoother and more coherent distributions. In particular, the DPL, LGN, and NPM models yield less fragmented and more refined stellar mass maps compared to the exponentially declining (EXP) and delayed-exponentially declining (DLY) models, which introduce more scatter and irregularities in the spatial distributions.

While the detailed investigation of mass profiles and half-mass radii is beyond the scope of this paper, our preliminary analysis suggests that the DPL and LGN models also provide the closest match to the input radial mass profiles. This observation further supports the notion that these models are more effective at capturing the overall stellar mass assembly history in galaxies, producing smoother and more realistic maps. Future work will explore the impact of SFH models on derived galaxy profiles in more depth.

\begin{figure*}
    \centering
    \includegraphics[width=1\linewidth ,height=0.17\linewidth]{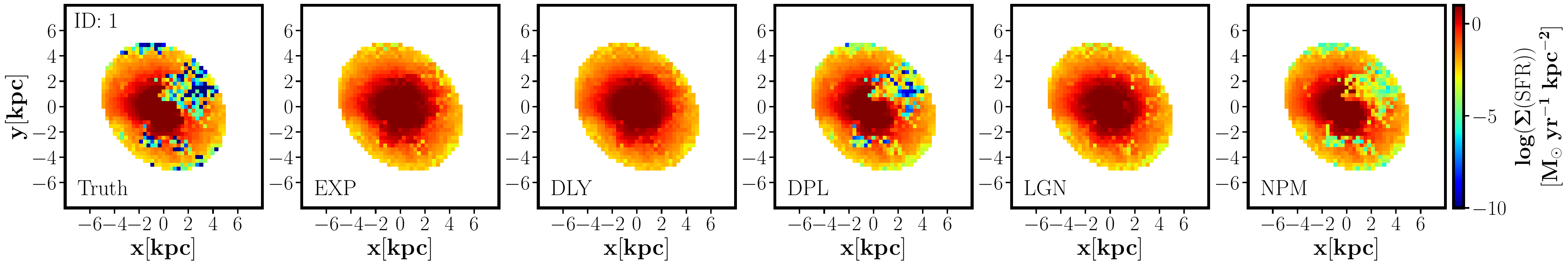}
    \includegraphics[width=1\linewidth ,height=0.17\linewidth]{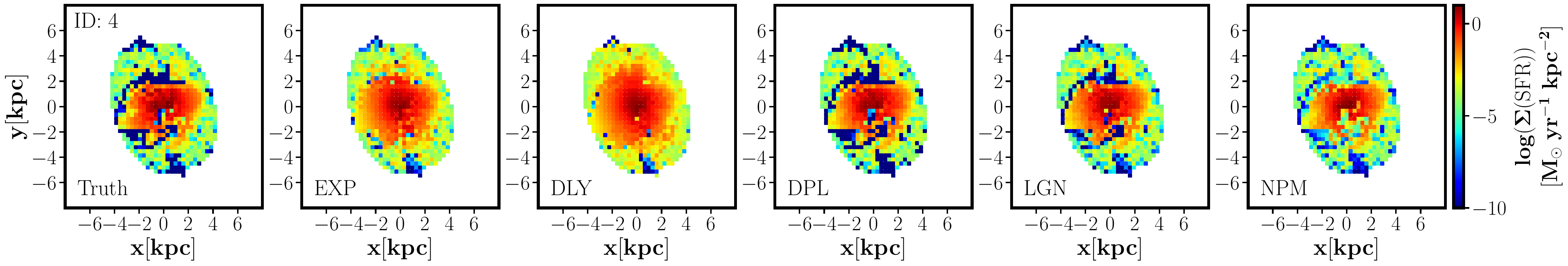}
    \caption{The same as Figure \ref{fig7}, but for the SFR surface density maps.}
    \label{fig8}
\end{figure*}

\begin{figure*}
    \centering
    \includegraphics[width=1\linewidth ,height=0.17\linewidth]{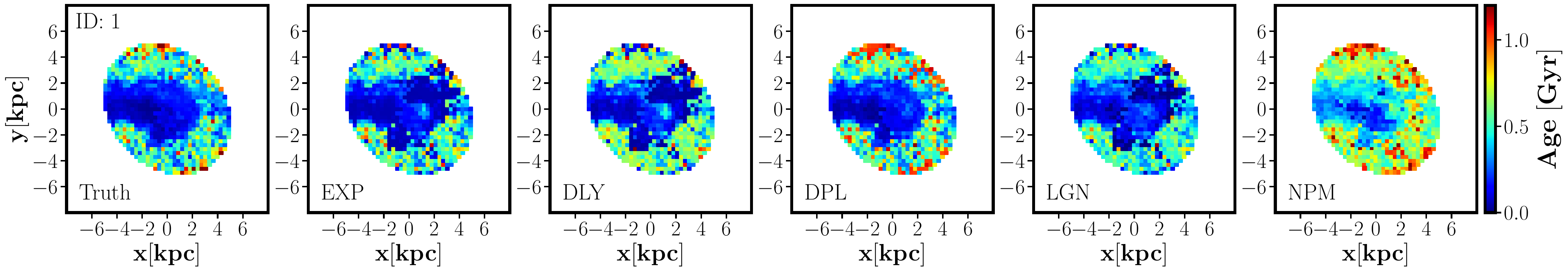}
    \includegraphics[width=1\linewidth ,height=0.17\linewidth]{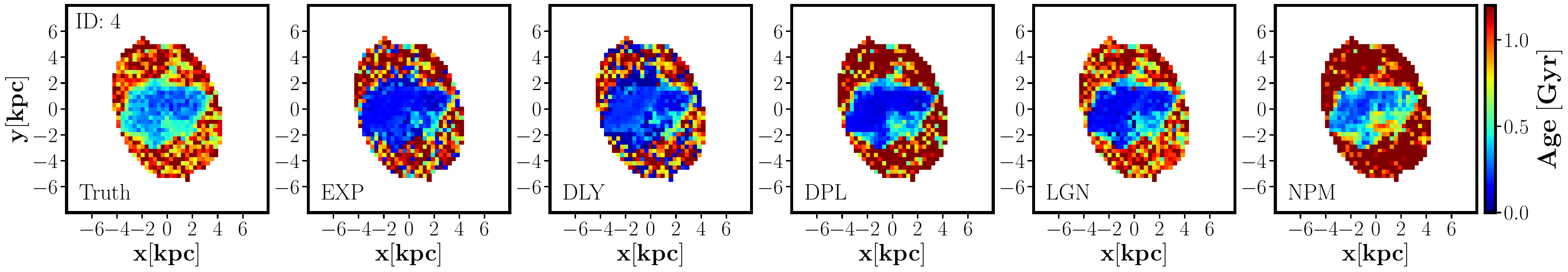}
    \caption{The same as Figure \ref{fig7}, but for the stellar age maps}
    \label{fig9}
\end{figure*}

Figures \ref{fig8} and \ref{fig9} display the corresponding maps for SFR surface density and age distributions, respectively. Upon visual inspection, the SFR and age maps derived from the DPL and LGN models most closely resemble the original simulations in terms of overall structure and spatial trends. These models better recover the broad distribution of star formation and stellar ages, capturing the key features of the true maps. In comparison, the EXP and DLY models exhibit stronger deviations, likely due to their more rigid SFH assumptions.

The resolved non-parametric (NPM) models also show notable performance, especially in the SFR surface density maps, but exhibit some overestimation in the age maps. This tendency towards overestimation is consistent with the trends observed in the overall SFHs, as presented in the previous subsection and Table \ref{tb5}. While the NPM model assumed for this work is more flexible in capturing variations in star formation, its lack of strong constraints on age bins can lead to broader uncertainties in age estimates, particularly for older stellar populations.

Overall, the comparison of the resolved maps highlights the importance of choosing appropriate SFH models for spatially resolved analysis. The DPL and LGN models emerge as the most reliable in reproducing both the structural and temporal properties of galaxies, suggesting that these parametric models strike a good balance between flexibility and constraint, enabling them to better track the complex star formation and mass assembly histories of galaxies.

\begin{figure*}
    \centering
    \includegraphics[width=0.49\linewidth]{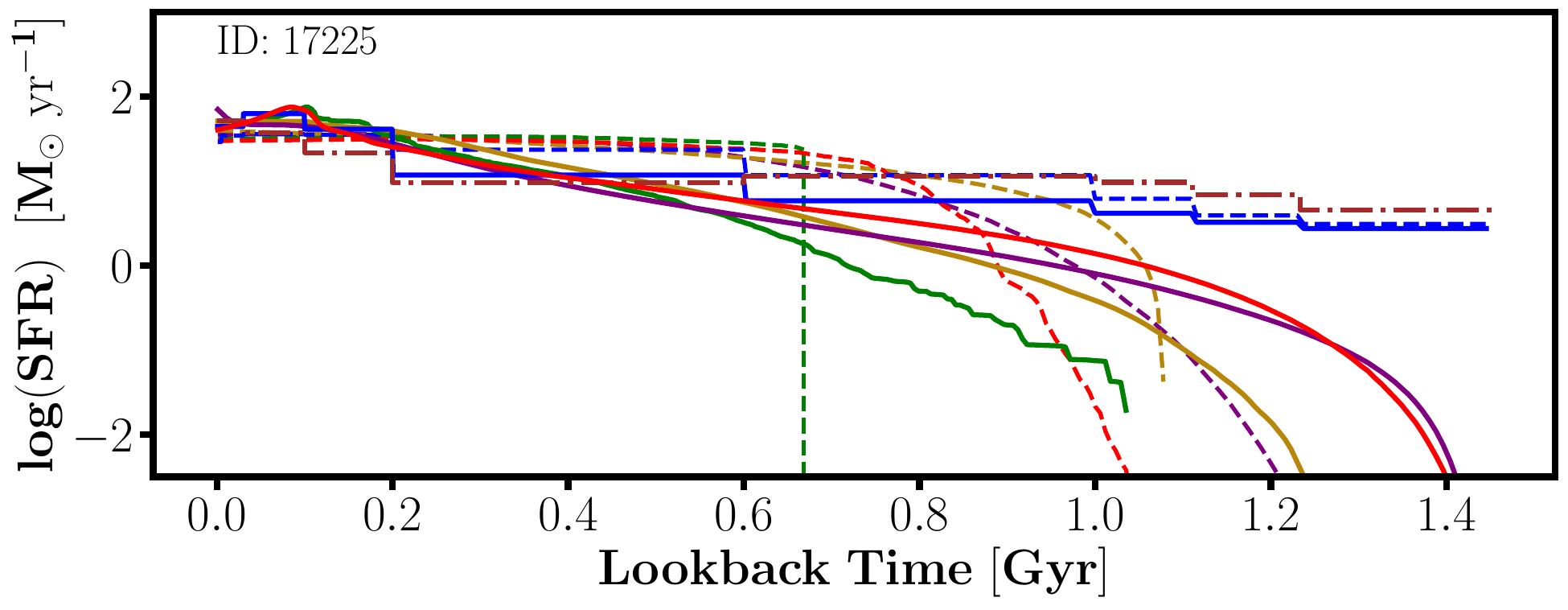}
    \includegraphics[width=0.49\linewidth]{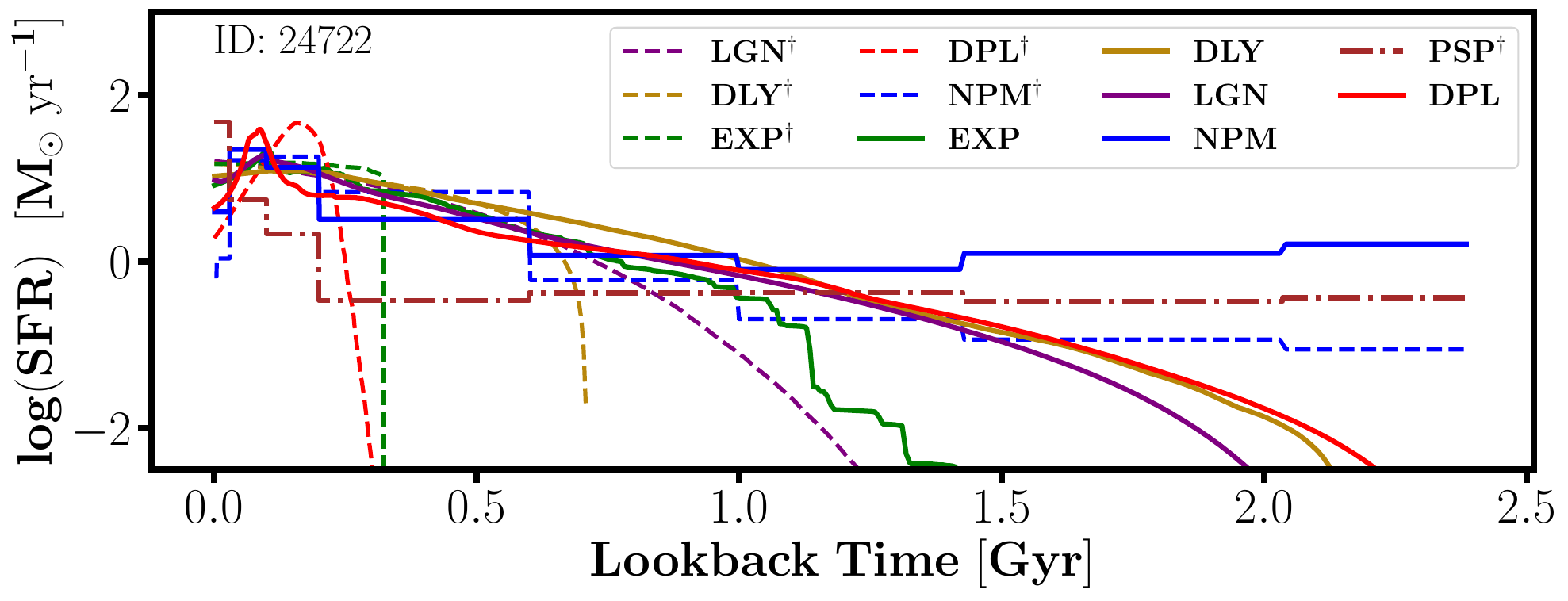}
    \includegraphics[width=0.49\linewidth]{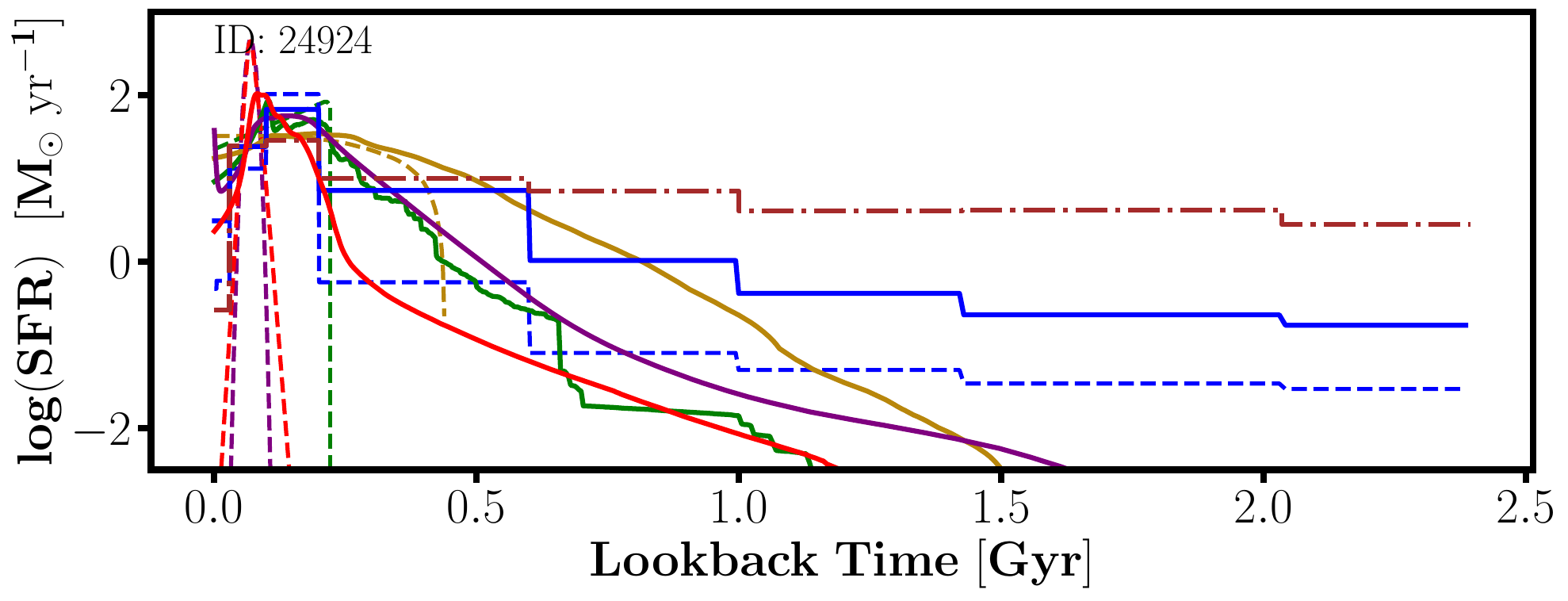}
    \includegraphics[width=0.49\linewidth]{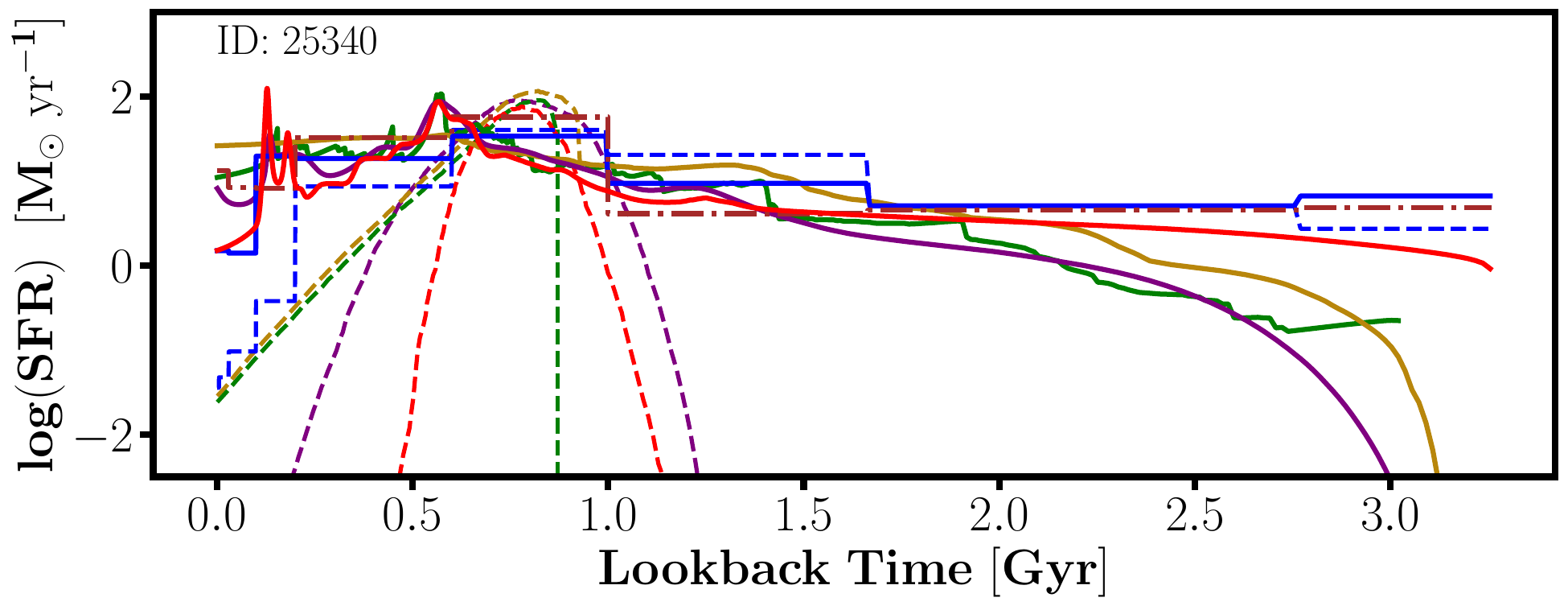}
    \includegraphics[width=0.49\linewidth]{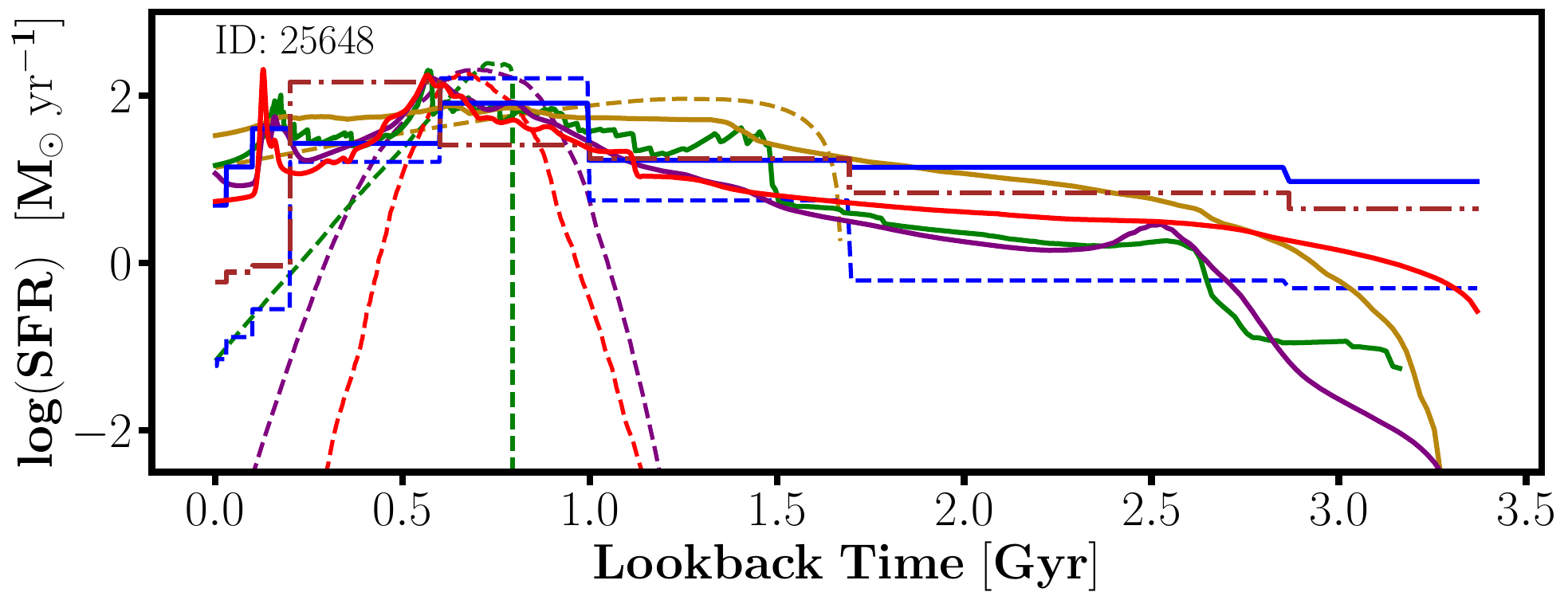}
    \includegraphics[width=0.49\linewidth]{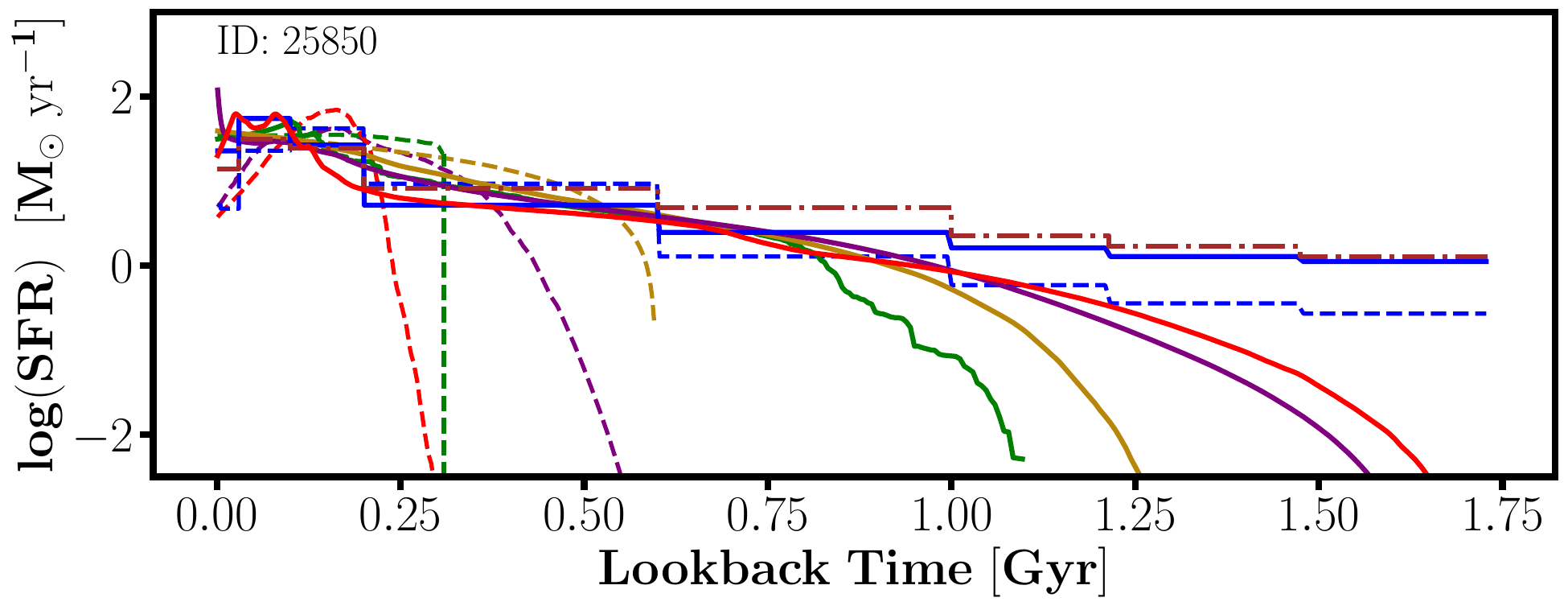}
    \includegraphics[width=0.49\linewidth]{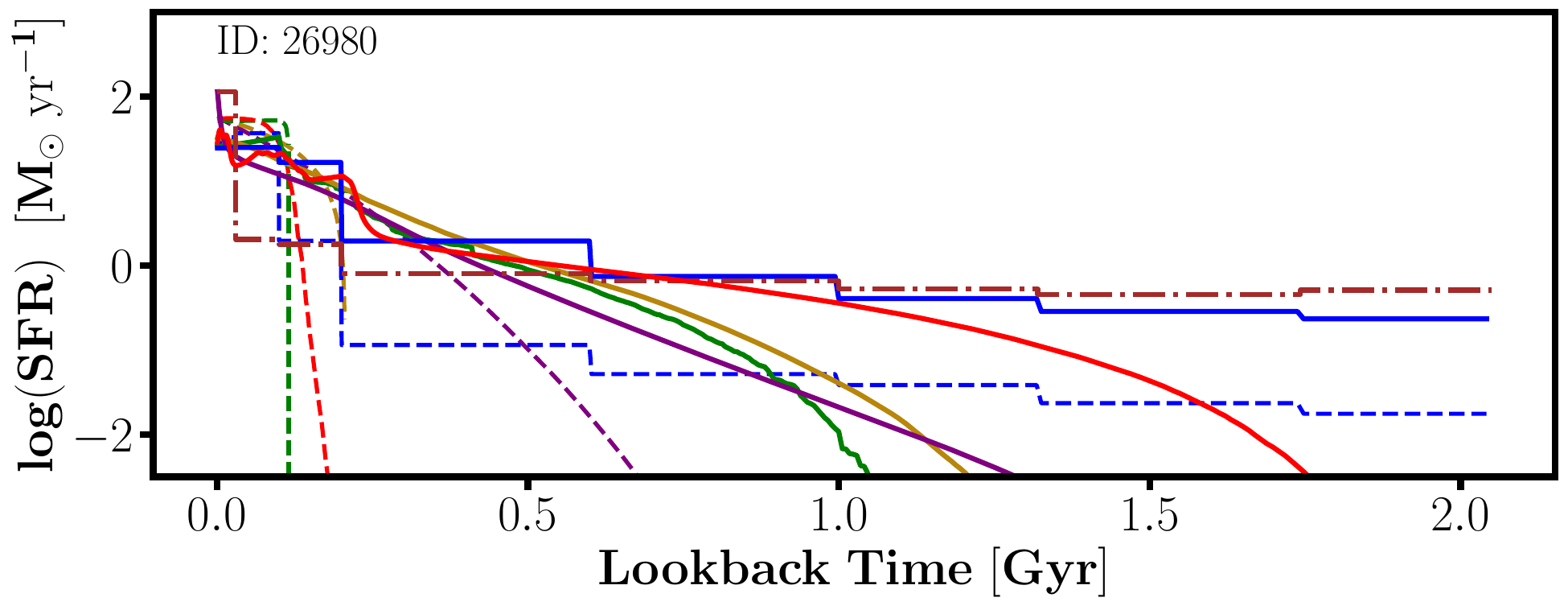}
    \includegraphics[width=0.49\linewidth]{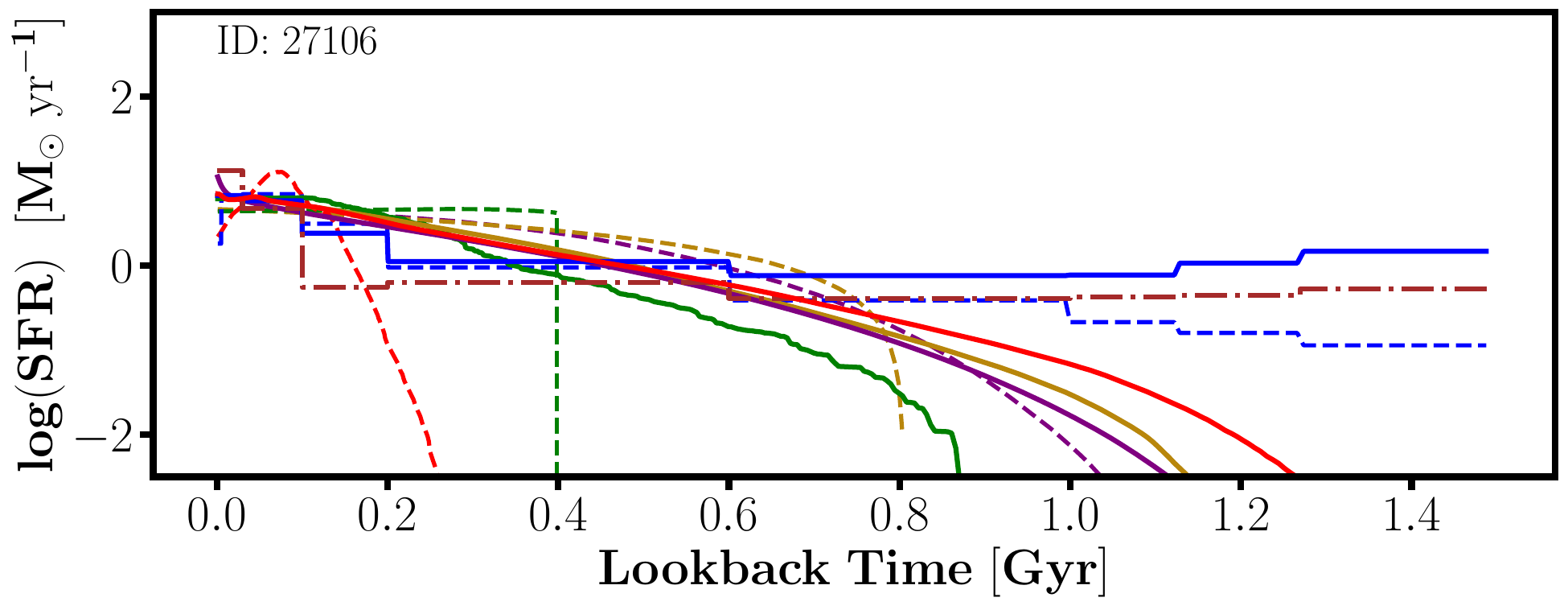}
    \includegraphics[width=0.49\linewidth]{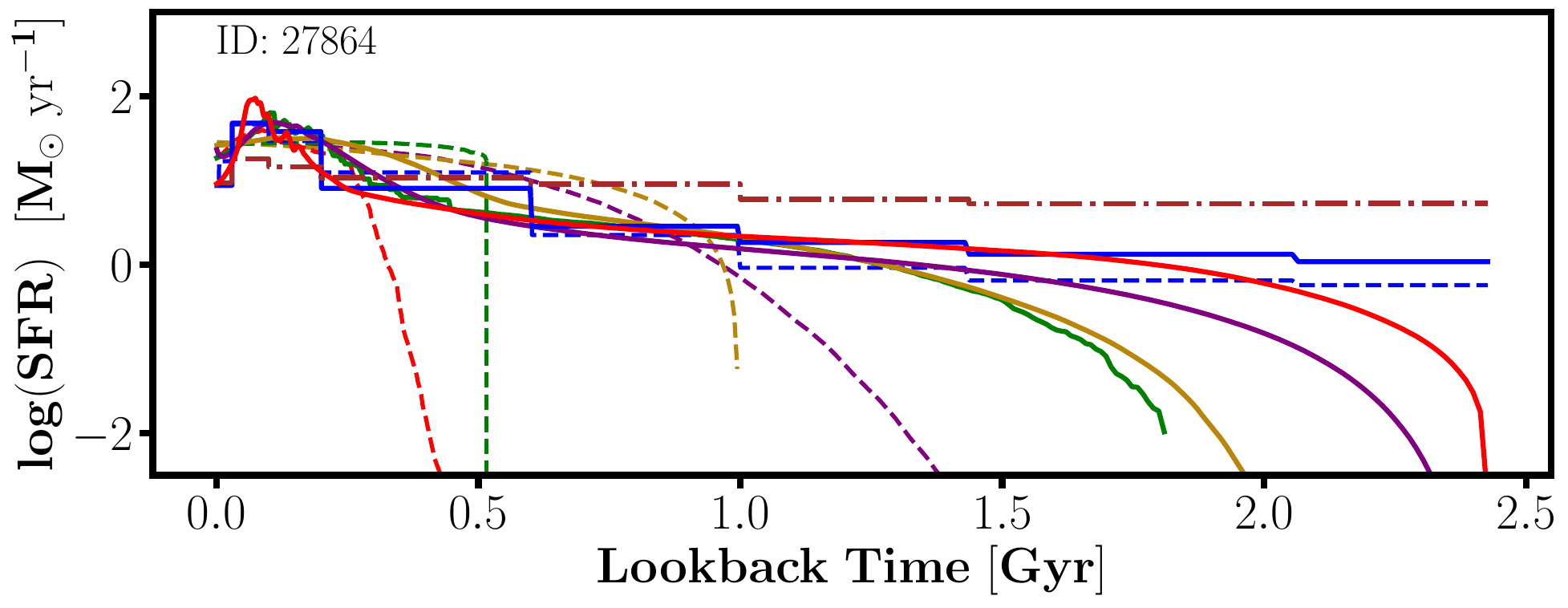}
    \includegraphics[width=0.49\linewidth]{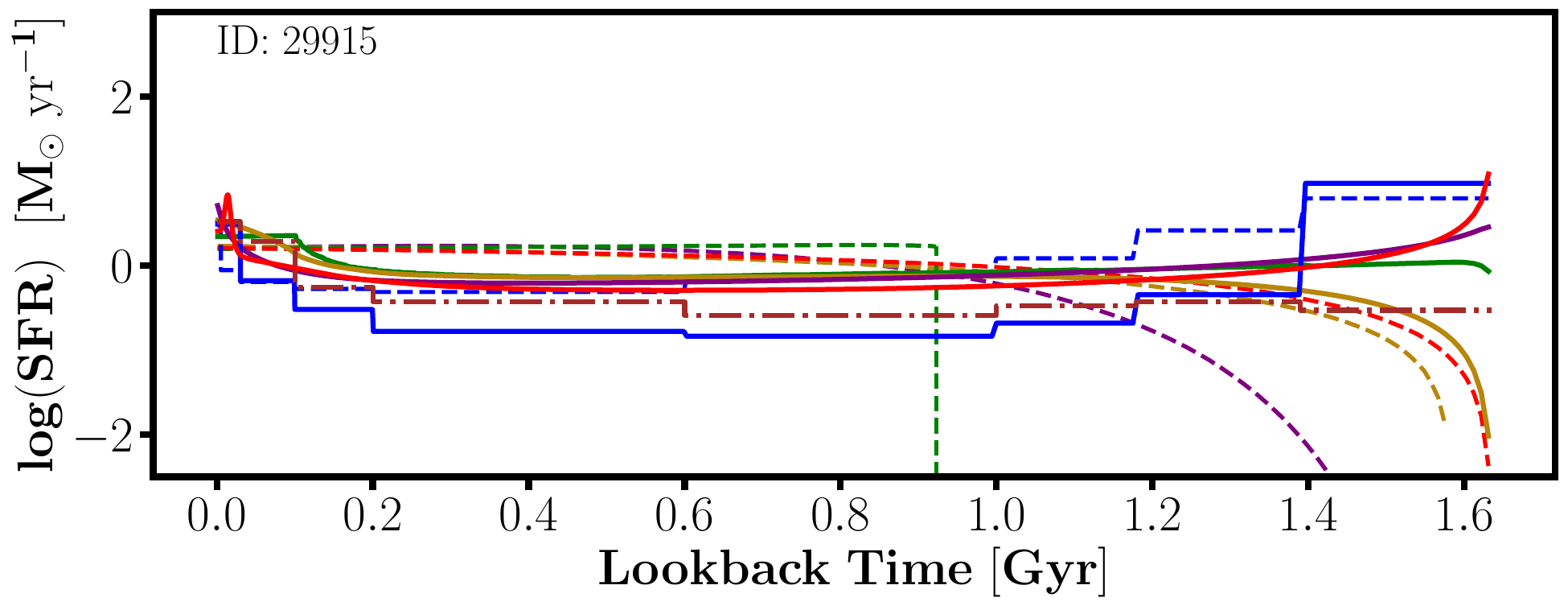}
    \includegraphics[width=0.49\linewidth]{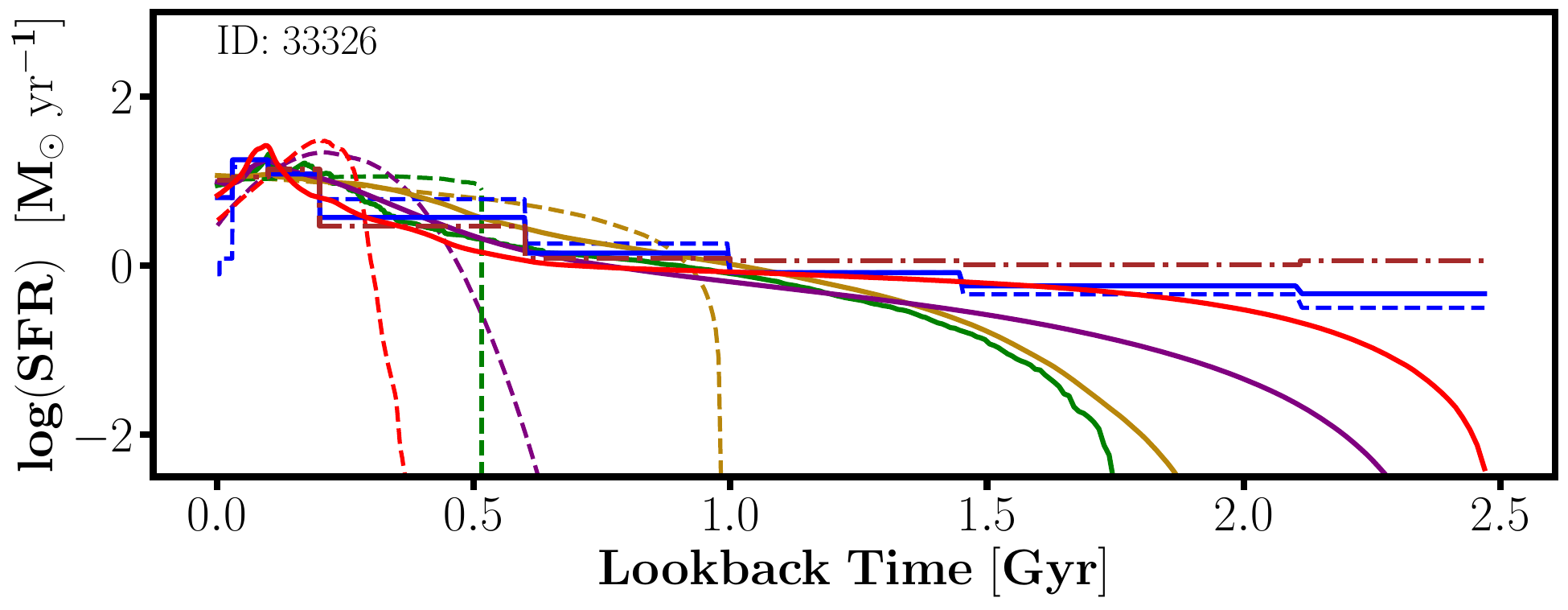}
    \includegraphics[width=0.49\linewidth]{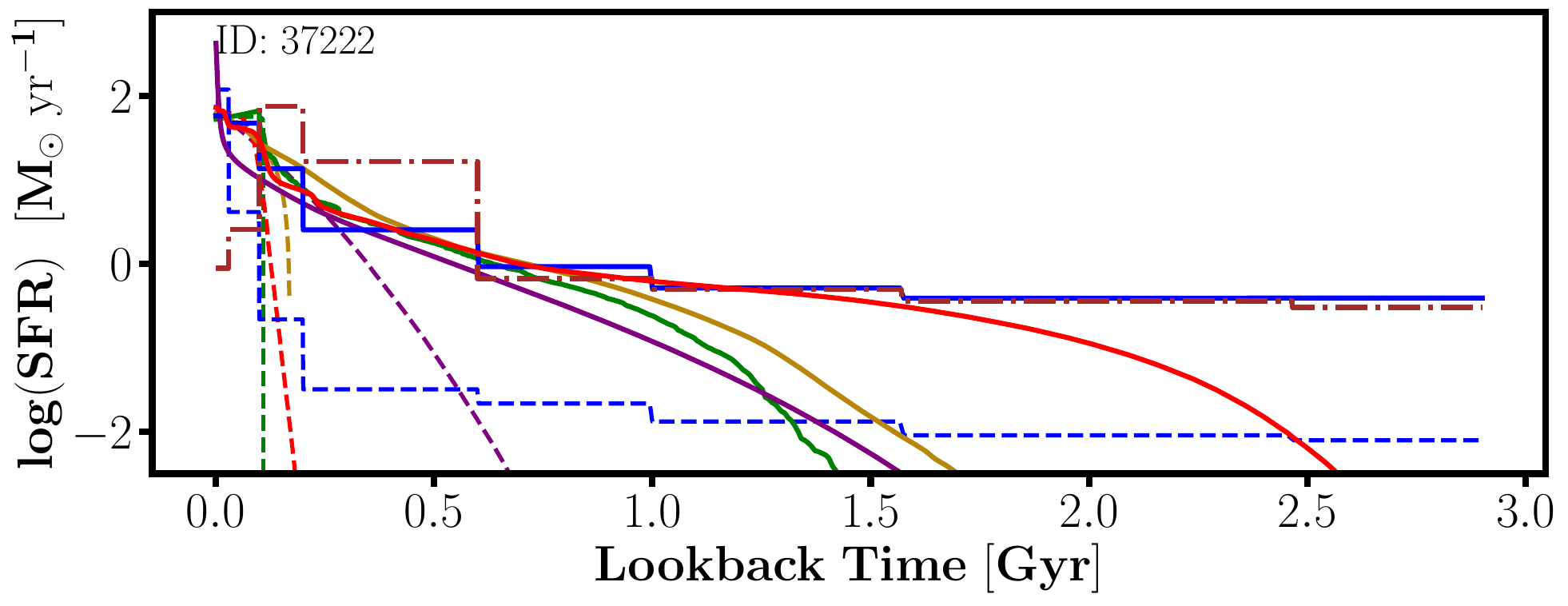}
    \caption{Star formation histories of the 12 observed galaxies. Solid lines represent the resolved models, while dashed lines correspond to the unresolved models. Pixel-by-pixel models, particularly flexible parametric forms like the DPL and LGN, effectively capture variations, including smoother starburst events, compared to non-parametric methods. At cosmic dawn ages, pixel-by-pixel parametric models exhibit distinct behavior compared to both resolved and unresolved non-parametric approaches.}
    \label{fig10}
\end{figure*}

\subsection{Results for Observed Galaxies}

In this subsection, we present the SFHs of our observed galaxy sample, derived from various models, as shown in Figure \ref{fig10}. Similar to the approach used for the simulated galaxies, we first examine the SFHs from the spatially resolved analysis (solid lines) and compare them to those derived from unresolved photometry using different parametric SFH models (dashed lines). As seen in the simulated data, the unresolved parametric SFH models, except the unresolved non-parametric models, tend to systematically underestimate the cosmic dawn ages (CDAs) of the observed galaxies (between $\sim0.5$ to $0.7$ Gyr compared to the resolved DPL models). This underestimation is most pronounced in models that assume exponentially declining or delayed-exponential SFHs, which fail to account for the early epochs of the star formation history. The discrepancies are most evident when compared to the resolved DPL model.

\begin{figure}
    \includegraphics[width=1\linewidth]{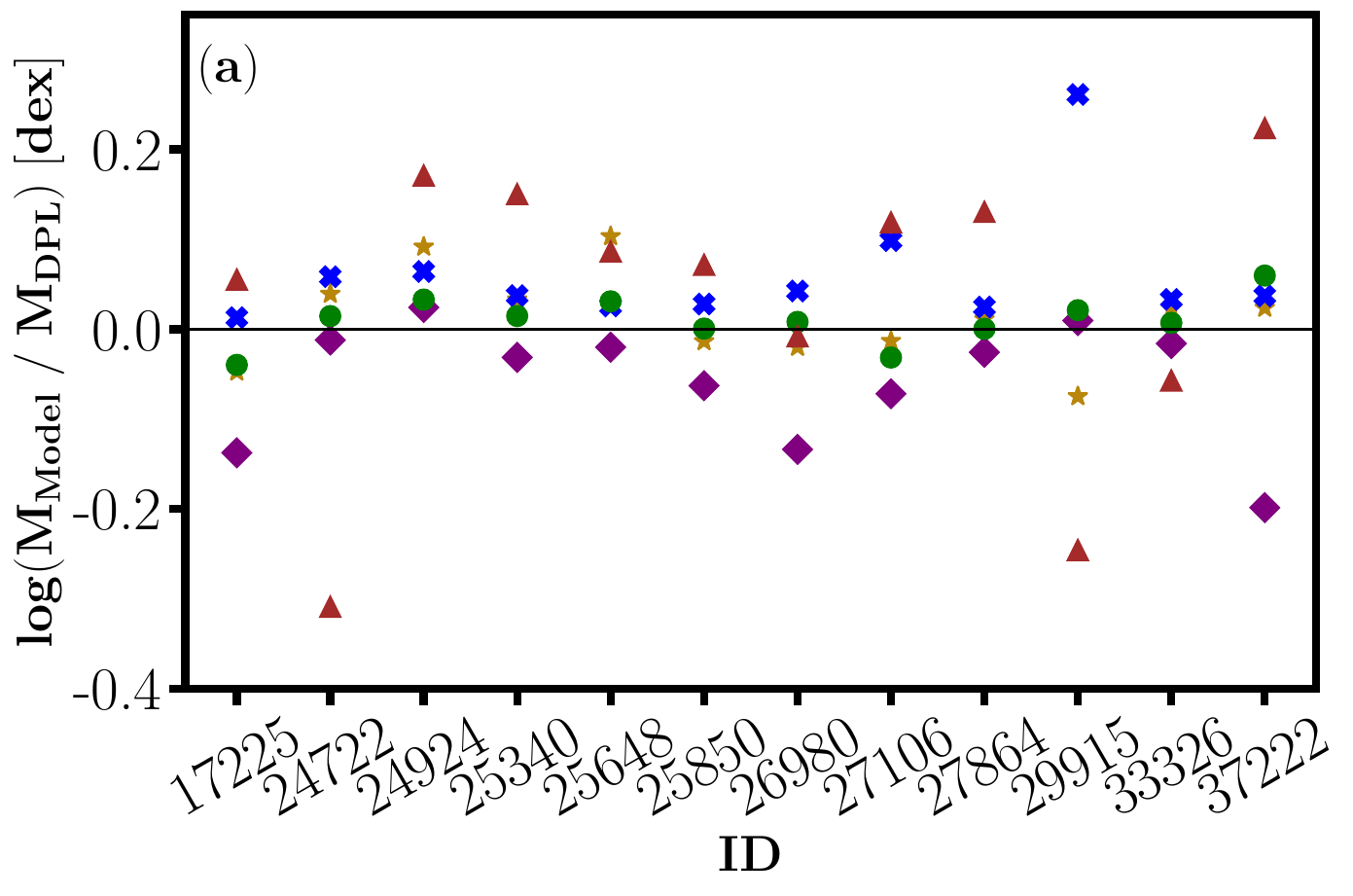}
    \includegraphics[width=1\linewidth]{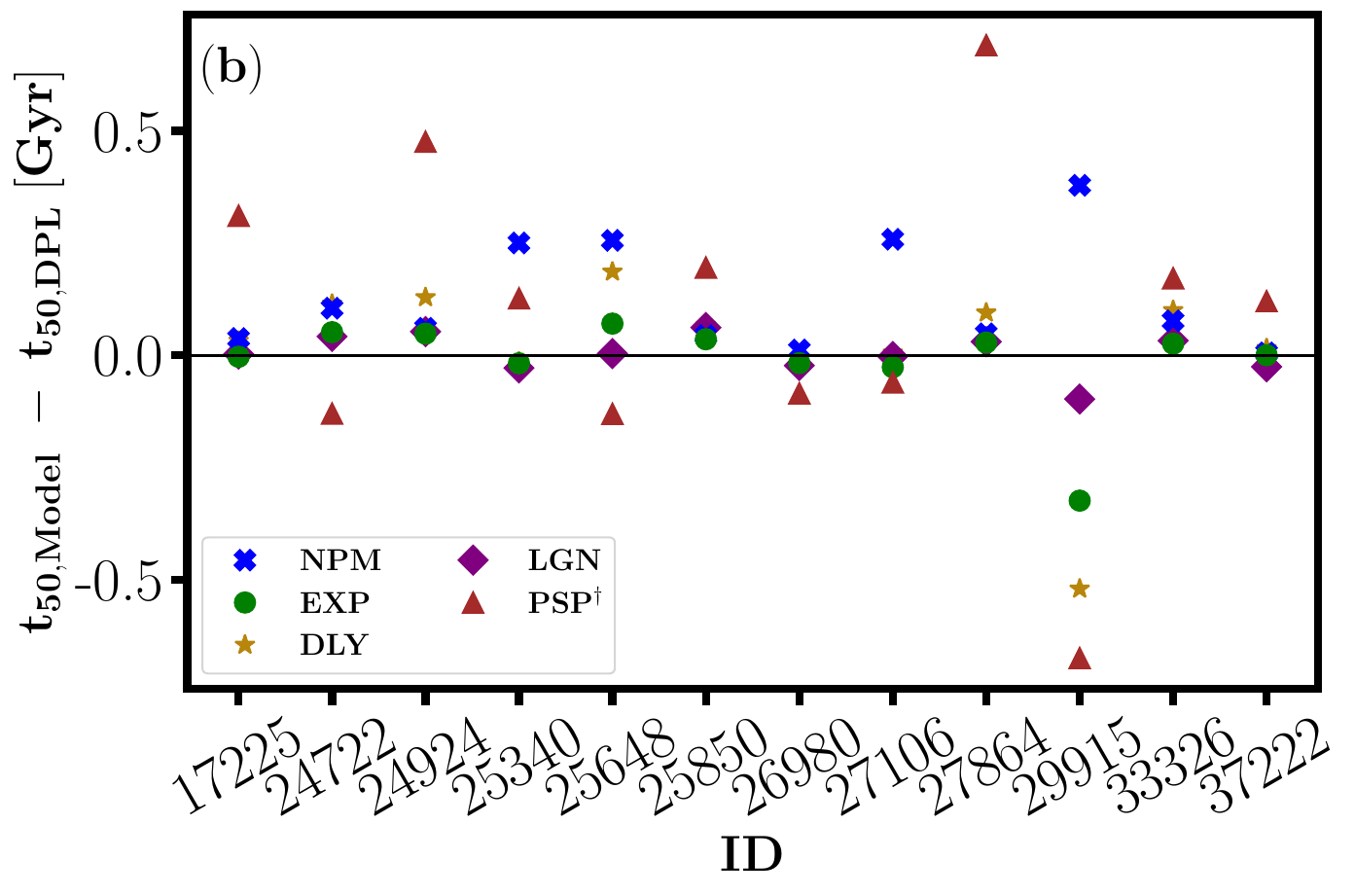}
    \includegraphics[width=1\linewidth]{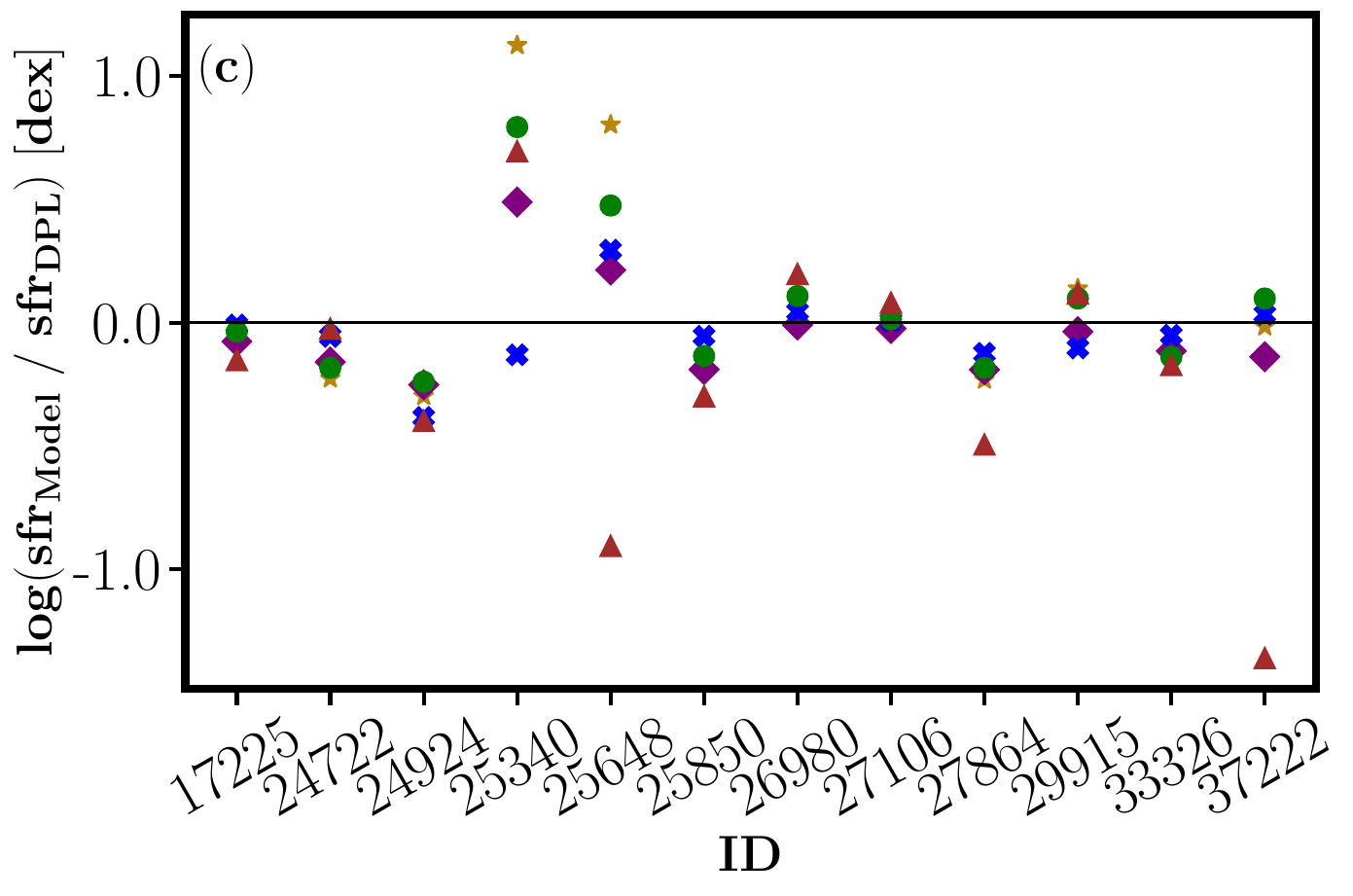}
    \caption{Similar to Figure \ref{fig4}, but showing the relative differences in stellar mass formed from different spatially resolved analyses for the observed galaxies compared to the DPL models. The error bars are omitted for clarity due to the overlap between data points. The median values of the relative differences for each parameter are reported in Table \ref{tb7}.}
    \label{fig11}
\end{figure}

The results across different SFH models become notably more consistent when spatially resolved approaches are applied. We use the resolved DPL model as our fiducial model and compare the total stellar masses, star formation rates (SFRs), and ages across each model. Table \ref{tb7} reports the relative differences for each of these parameters. For most galaxies, the median differences for resolved models are minimal. In panel (a) of Figure \ref{fig11}, the stellar mass differences relative to the DPL model are shown for each galaxy. As illustrated, the scatter for most cases is small, typically less than $0.1$ dex. The largest deviations are found in the unresolved non-parametric (PSP$\dagger$) model, where overestimation of the SFH at early epochs contributes to the observed discrepancies. The relative differences of the CDA ($t_5$) for the PSP$\dagger$ is $-0.60\pm0.28$, which means that the star formation started earlier in this model. This effect is also evident in the formation ages (panel (b) of Figure \ref{fig11}), where deviations of up to $\pm 0.6$ Gyr are observed, which is significant given the age of the universe at these redshifts.

By examining the resolved SFHs for each galaxy (Figure \ref{fig10}), we observe many short-term ($\lesssim 100$ Myr) variations in SFR, revealing bursts of star formation that may otherwise go unnoticed in unresolved analyses. For instance, galaxies 25340 and 25648 exhibit peaks in their recent star formation activity within the last 600-700 Myr, as revealed by the resolved DPL, LGN, and even EXP models. These fluctuations in star formation activity, along with their widths and strengths, can potentially provide valuable insights into the formation history of galaxies. While the non-parametric models (both resolved and unresolved) capture the overall trends in SFHs, they tend to smooth over the finer details, and the exact timing and magnitude of these starbursts are often lost due to the broader age bins assumed in these models. 

The consistency of the recent SFRs derived from different models, as shown in panel (c) of Figure \ref{fig11}, underscores the reliability of the resolved models in estimating current star formation activity, particularly when compared to unresolved parametric models. The spatially resolved approach appears to better preserve the nuances of star formation across the galaxy, especially in the most recent epochs (see Table \ref{tb7}).

\begin{figure*}
    \centering
    \includegraphics[width=1\linewidth,height=0.19\linewidth]{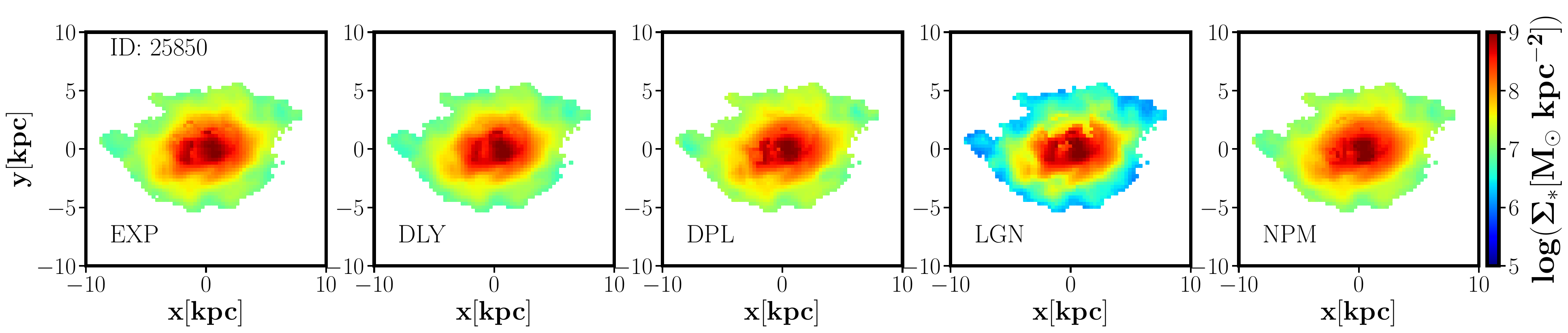}
    \includegraphics[width=1.\linewidth,height=0.19\linewidth]{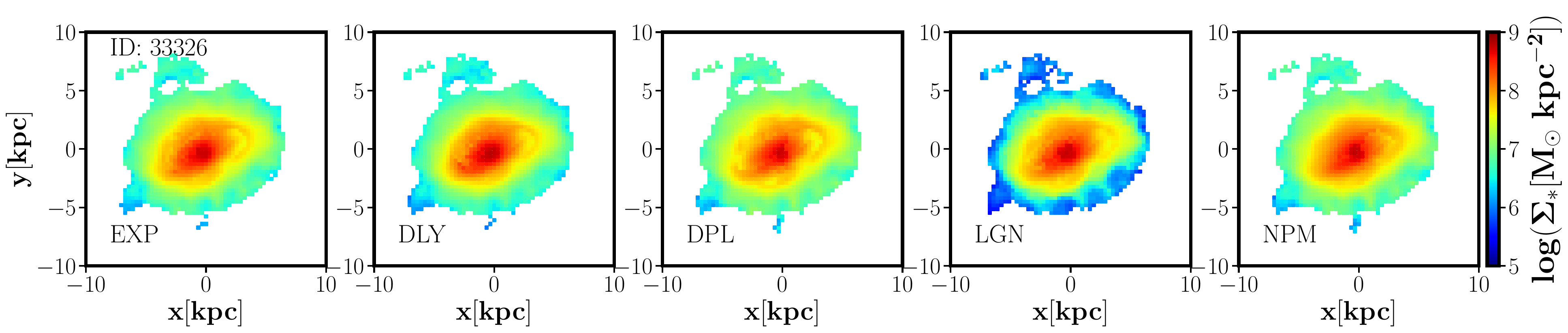}
    \caption{Comparison of stellar mass surface density maps derived from different SFH models for two observed galaxies.}
    \label{fig12}
\end{figure*}

\begin{figure*}
    \centering
    \includegraphics[width=1\linewidth ,height=0.19\linewidth]{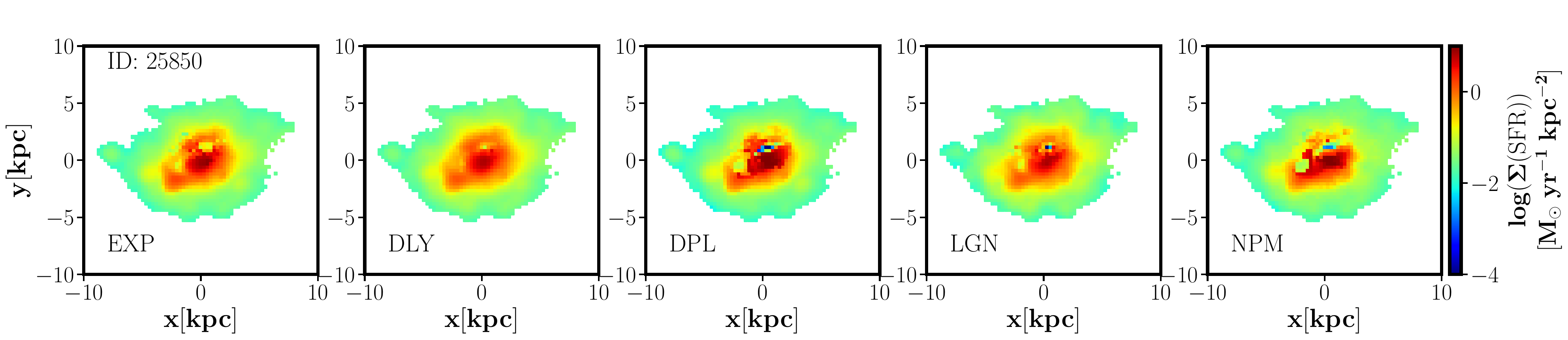}
    \includegraphics[width=1\linewidth ,height=0.19\linewidth]{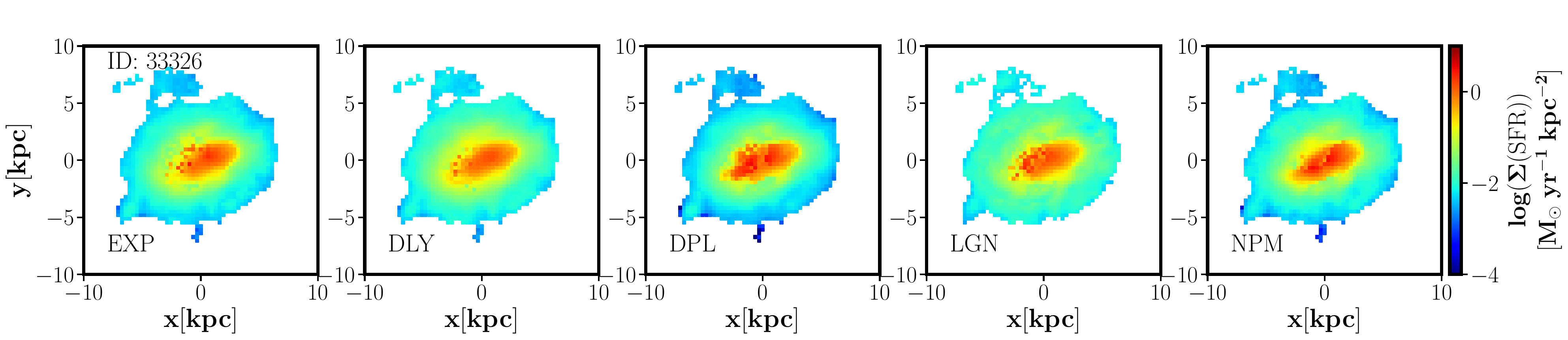}
    \caption{Same as Figure \ref{fig12}, but for SFR surface densities.}
    \label{fig13}
\end{figure*}

\begin{figure*}
    \centering
    \includegraphics[width=1\linewidth ,height=0.19\linewidth]{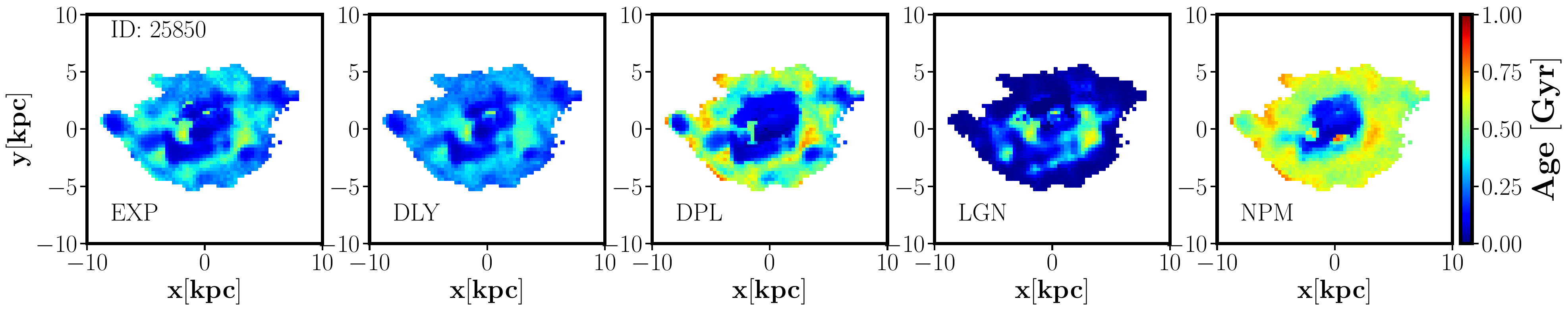}
    \includegraphics[width=1\linewidth ,height=0.19\linewidth]{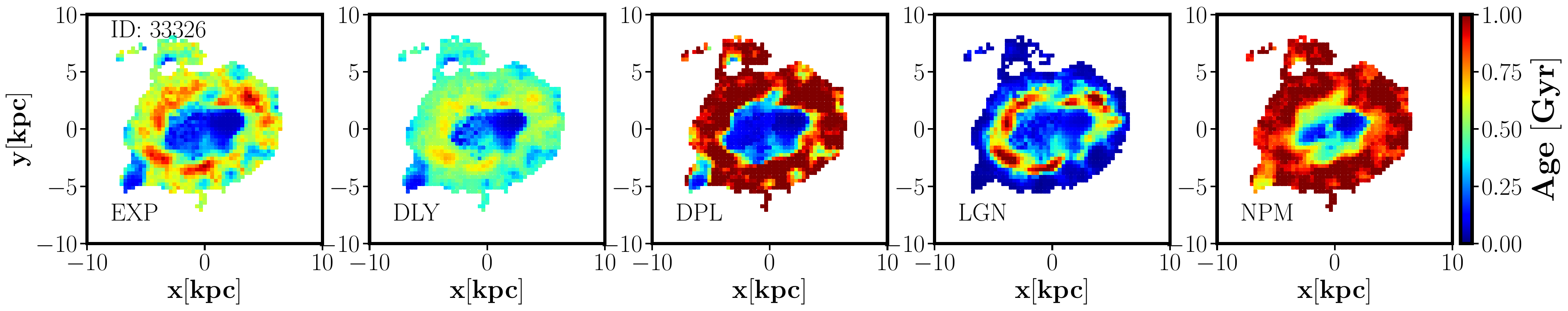}
    \caption{Same as Figure \ref{fig12}, but for stellar ages.}
    \label{fig14}
\end{figure*}

To further examine the spatial implications of these findings, we generated and compared surface density maps of stellar mass and SFR, as well as age maps, for two representative galaxies in our sample (Figures \ref{fig12}, \ref{fig13}, and \ref{fig14}). These maps, produced using different SFH assumptions, reveal the spatial distribution of these key properties. The stellar mass surface densities are relatively consistent across different SFH models, with only minor variations attributable to the specific SFH model used. Although the overall mass distribution remains stable, further investigation is needed to assess the impact of SFH assumptions on the galaxies' structural parameters, which will be explored in future studies using a larger sample.

The SFR surface density maps derived from the DPL and LGN SFH models are closely aligned and effectively highlight regions of recent star formation. However, the differences between SFH models are more clearly revealed in the stellar age maps. Here, we observe notable variations between each model, with the resolved DPL model as a reference, which could have provided the most accurate age estimates, based on the findings from our simulations. These spatially resolved maps offer valuable insights into how different SFH models perform in capturing the spatial distribution of key properties, with implications for understanding galaxy formation and evolution.

\begin{deluxetable}{ccccc}
\tablenum{7}
\tablecaption{Median deviation and scatter of stellar masses, star formation rates, and formation ages ($t_{50}$ and $t_5$) from the DPL model for each SFH model applied to the observed galaxies in the spatially resolved (pixel-by-pixel) fitting.\label{tb7}}
\tablewidth{0pt}
\tablehead{
\colhead{Model} & \colhead{$\Delta \mathrm{\log}(\mstar/\msun)$} & \colhead{$\Delta \mathrm{log(SFR)}$} & \colhead{$\Delta t_{50}$} & \colhead{$\Delta t_{5}$}\\
\colhead{} & \colhead{} & \colhead{[$\msun$ yr$^{-1}$]} & \colhead{[Gyr]} & \colhead{[Gyr]}
}
\startdata
EXP & 0.01 $\pm$ 0.01 & -0.01 $\pm$ 0.18 & 0.01 $\pm$ 0.04 & 0.30 $\pm$ 0.31 \\
DLY & 0.01 $\pm$ 0.04 & -0.03 $\pm$ 0.22 & 0.04 $\pm$ 0.07 & 0.18 $\pm$ 0.24 \\
LGN & -0.02 $\pm$ 0.05 & -0.09 $\pm$ 0.11 & 0.00 $\pm$ 0.04 & 0.22 $\pm$ 0.32 \\
NPM & 0.03 $\pm$ 0.01 & -0.05 $\pm$ 0.09 & 0.06 $\pm$ 0.06 & -0.49 $\pm$ 0.25 \\
NPM$^{\dagger}$ &  0.03 $\pm$ 0.06 & -0.21 $\pm$ 0.27 &  0.07 $\pm$ 0.14 & 0.00 $\pm$ 0.25\\
PSP$^{\dagger}$ & 0.07 $\pm$ 0.11 & -0.16 $\pm$ 0.38 & 0.12 $\pm$ 0.29 & -0.60 $\pm$ 0.28 
\enddata
\end{deluxetable}

\begin{figure*}
    \centering
    \includegraphics[width=0.49\linewidth]{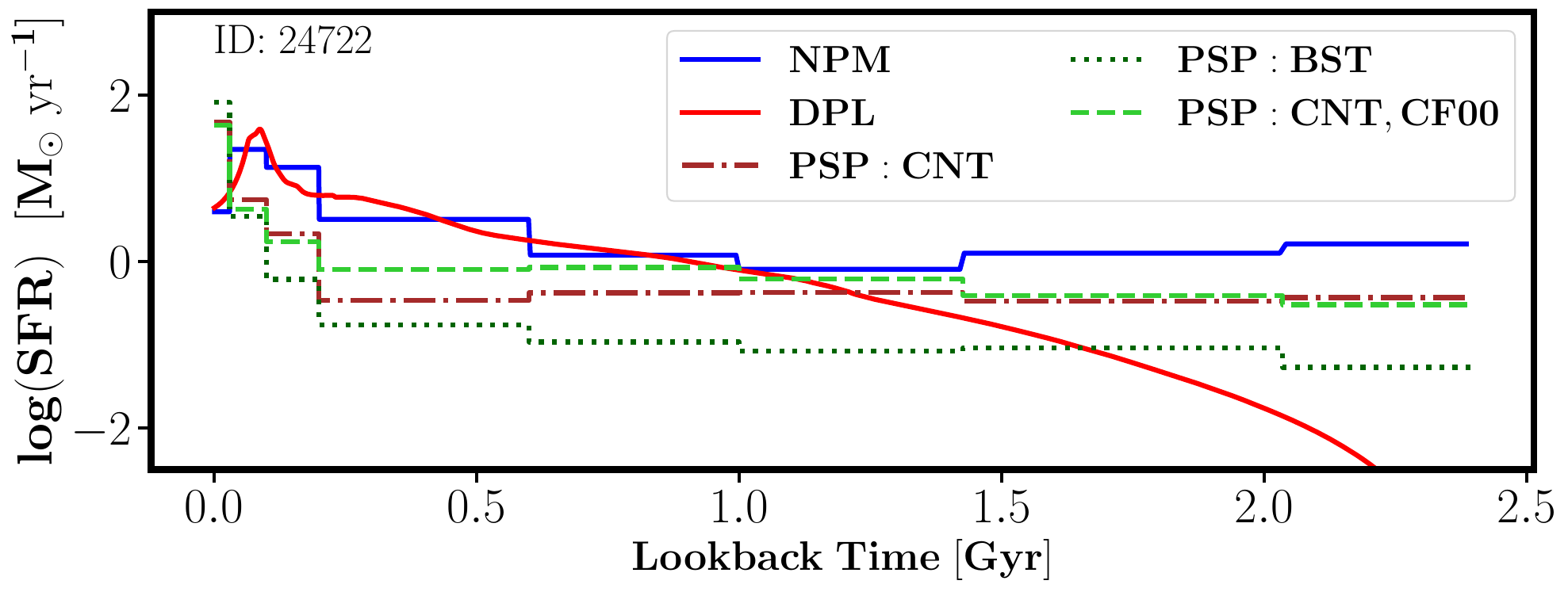}
    \includegraphics[width=0.49\linewidth]{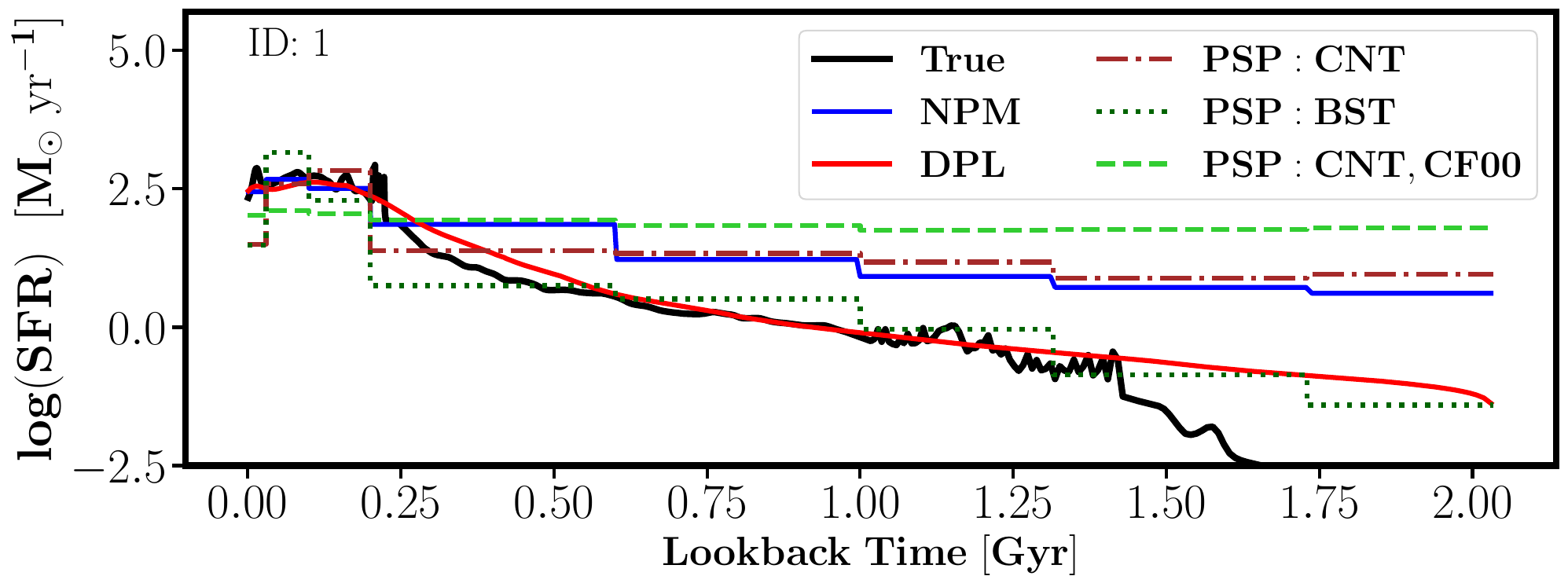}
    \includegraphics[width=0.49\linewidth]{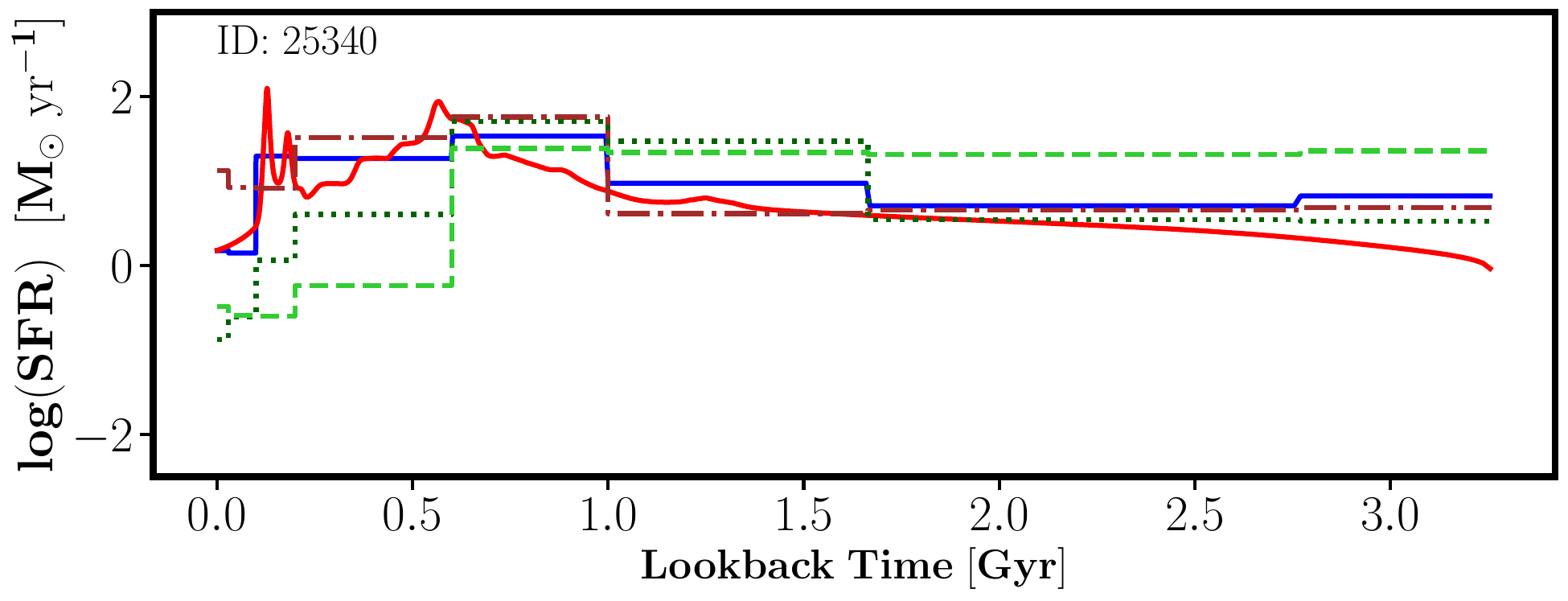}
    \includegraphics[width=0.49\linewidth]{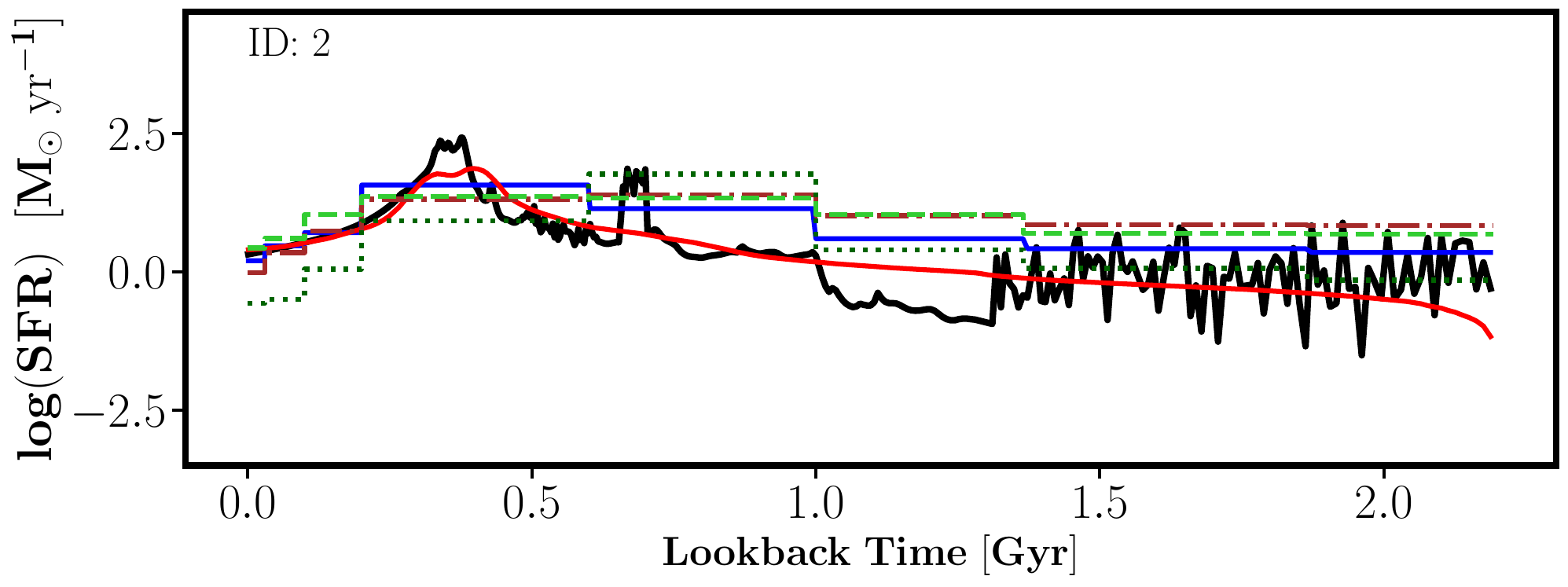}
    \includegraphics[width=0.49\linewidth]{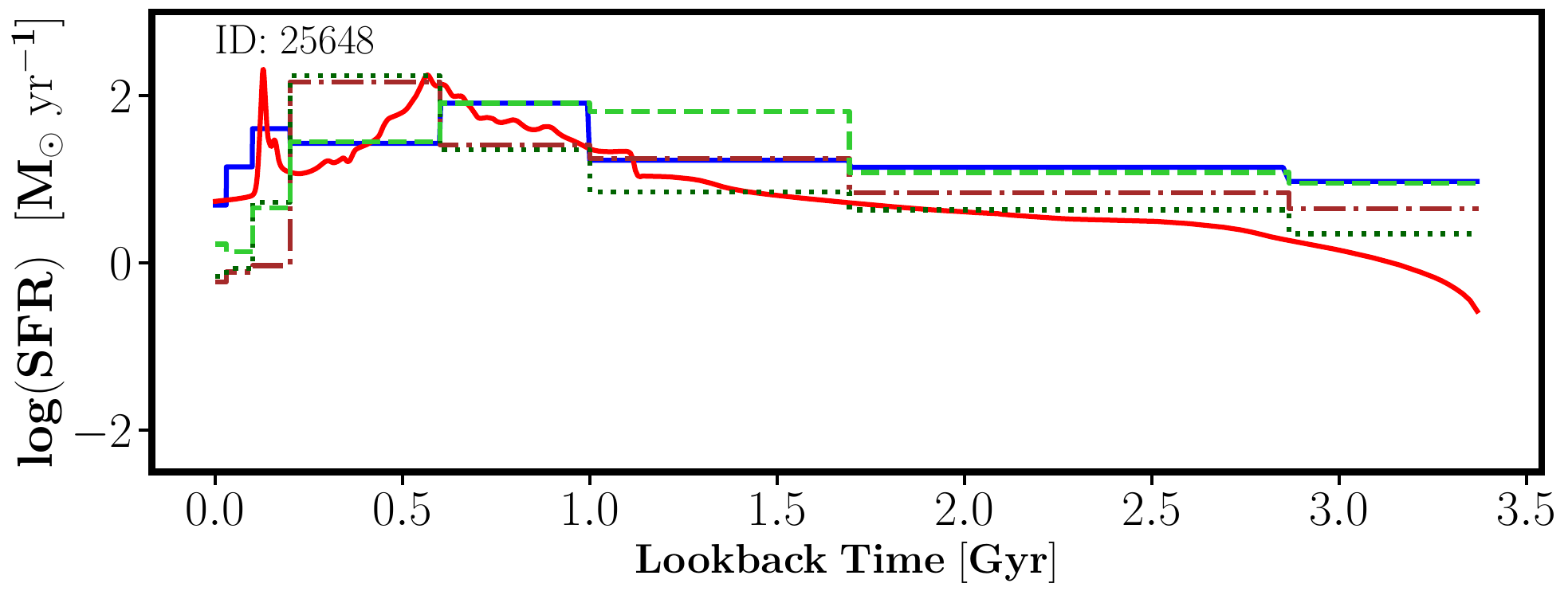}
    \includegraphics[width=0.49\linewidth]{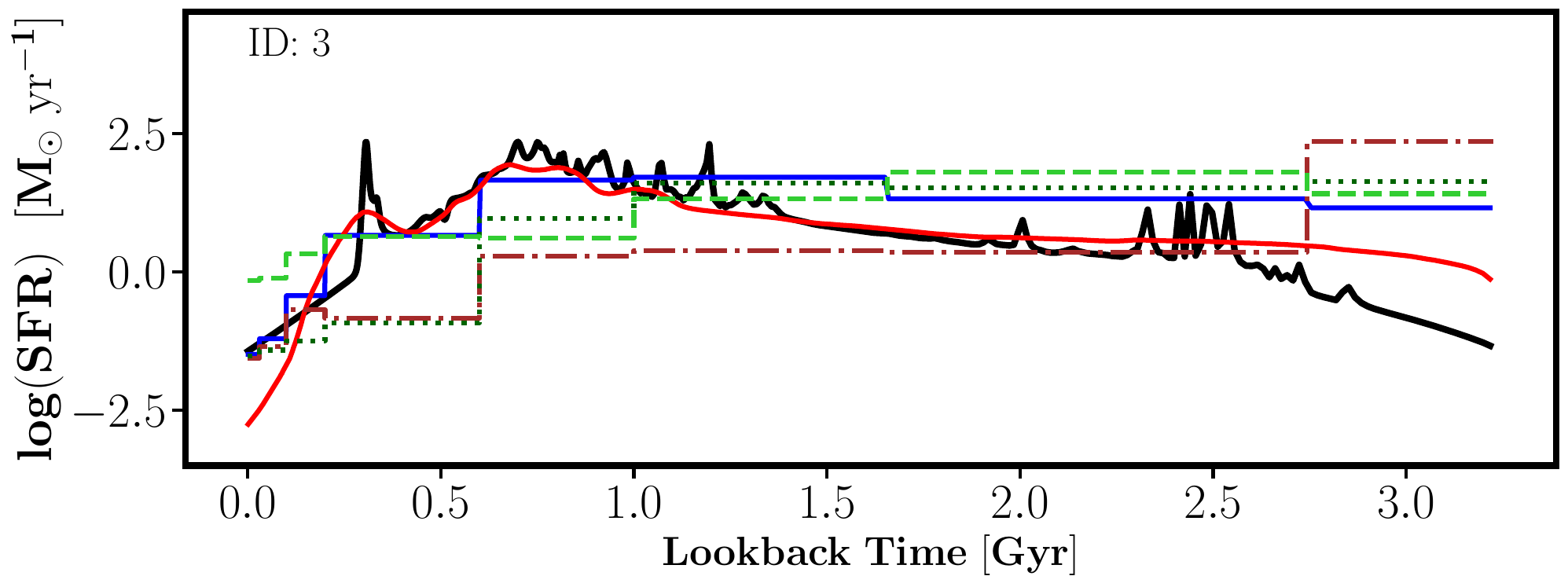}
    \includegraphics[width=0.49\linewidth]{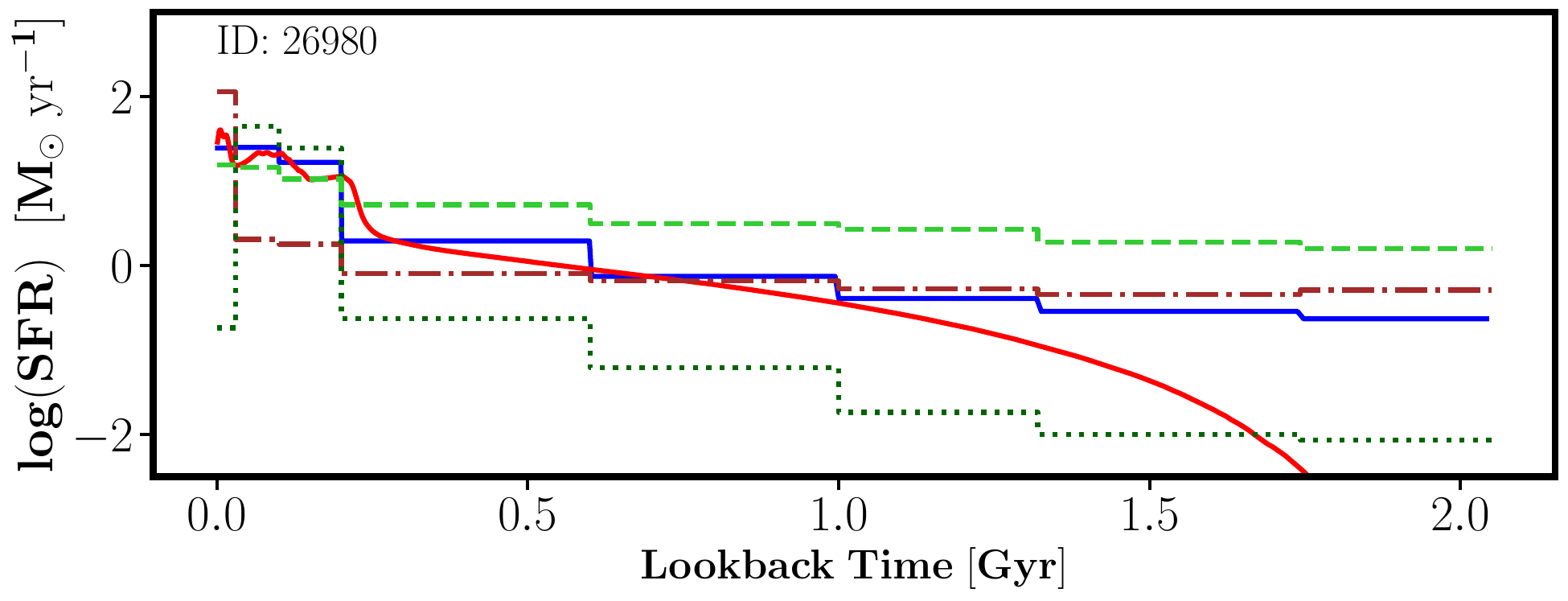}
    \includegraphics[width=0.49\linewidth]{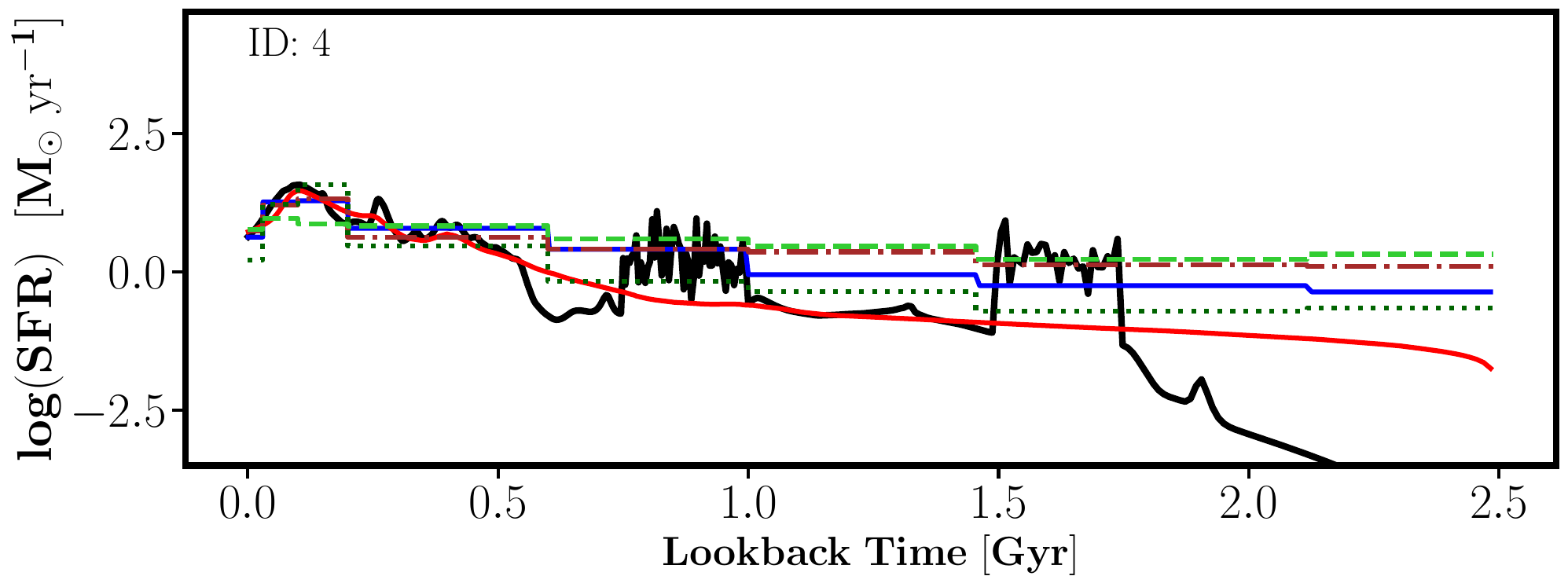}
    \includegraphics[width=0.49\linewidth]{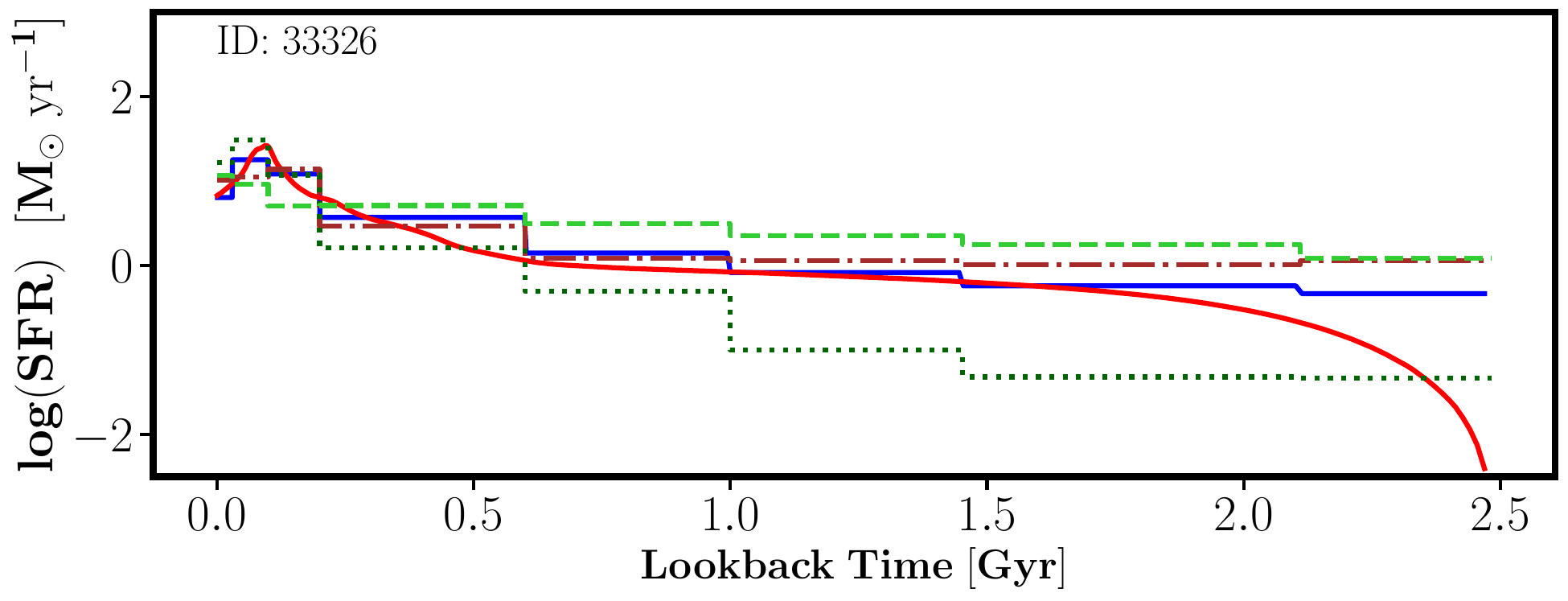}
    \includegraphics[width=0.49\linewidth]{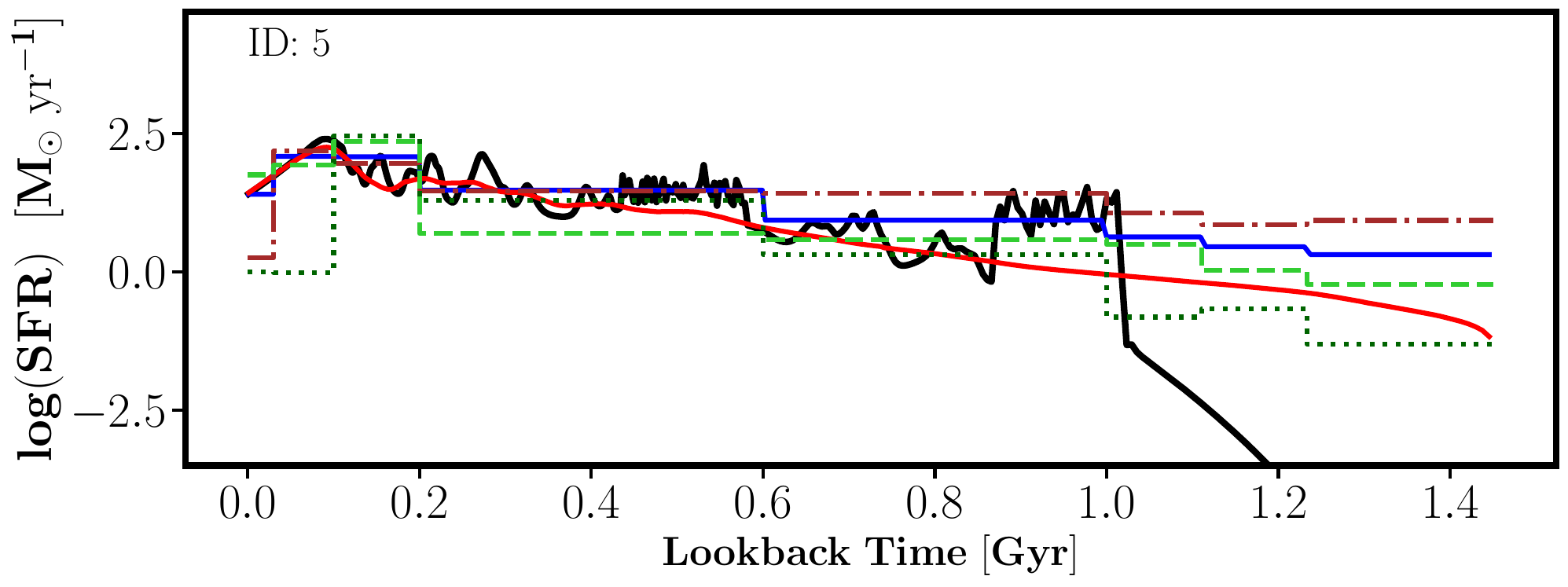}
    \caption{Comparison of different non-parametric methods applied to a sample of observed galaxies (left panels) and mock galaxies (right panels). The sensitivity of non-parametric models to prior assumptions is demonstrated.}
    \label{fig15}
\end{figure*}

\section{Discussion} 
\subsection{Performance of SFH Models}

Our analysis of the mock and observed galaxies demonstrates the crucial role that spatially resolved SED fitting plays in accurately recovering star formation histories (SFHs) and key galaxy properties. While unresolved methods, particularly those relying on parametric SFHs, tend to underestimate critical parameters like ages \citep{carnall2019}, spatially resolved approaches substantially reduce these biases. The consistency observed across different SFH models within the spatially resolved framework highlights the robustness of this method \citep{abdurrouf2023}. Notably, parametric models with greater flexibility, such as DPL and LGN, are more adept at capturing the inherent variations in galaxies' SFHs.

While non-parametric models can trace the general evolution of SFHs \citep{leja2019a}, they often introduce biases in estimating specific ages, particularly during the early star formation epochs. In the spatially resolved analysis, the non-parametric (NPM) SFH model tends to overestimate the formation ages by approximately $0.12 \pm 0.08$ Gyr (for mock galaxies). The unresolved non-parametric models (e.g., PSP$\dagger$) exhibit even larger deviations, with \textbf{$\Delta\log(\mstar/\msun) = 0.11 \pm 0.11$} and $\Delta$ Age = $0.28 \pm 0.35$ Gyr \citep[see also e.g.,][]{vanmierlo2023}. This overestimation stems from non-parametric models often assigning more weight to earlier time bins in the SFH reconstruction, emphasizing the need for caution when interpreting these results. We note that using different SED modeling codes with varying assumptions about stellar populations can also impact the age estimates, as discussed in \citet{whitler2023}. \citet{turner2024} also showed that an increasing SFH prior can lead to more physically reasonable results.

To explore this further, we compared the non-parametric continuity model with the bursty model for our observed galaxies (Figure \ref{fig15}). The left panels are assigned to a sample of the observed galaxies and the right panels are for the mock galaxies.  The brown dot-dashed lines represent the continuity PSP model (PSP: CNT), while the green dot-dashed lines depict the bursty PSP model (PSP: BST). As shown, the increased flexibility of the bursty PSP model leads to a decrease in average SFRs at earlier times. However, in some cases, such as galaxies 26980, the change in non-parametric model assumptions has a notable effect on recent SFR estimates. This is not observed for the resolved parametric models, where SFHs consistently converge to similar values. For further exploration, we used the double component dust model of the \cite{charlot2000} with the continuity assumption (PSP: CNT, CF00) and illustrated as the dashed green lines. Comparing this model with the PSP: CNT model (brown dot-dashed lines), depicts some notable differences in some cases (e.g., 25340). These results illustrate the challenges unresolved non-parametric models face, as they are highly sensitive to assumed priors and the distribution of star formation events across time (age bins) \citep[see][]{leja2019a}. Modifying the fiducial SFH prior in non-parametric models could help address some of the identified issues. For example, adopting priors that allow for increasing SFH bases over time may more accurately reflect realistic galaxy formation histories and mitigate the biases highlighted in this study \citep[see e.g.,][]{turner2024}.

Using more flexible non-parametric methods, such as Dense-Basis Gaussian Processes \citep{iyer2019}, reduces the reliance on parametric forms for age bins. However, this approach remains sensitive to the choice of the kernel, which can influence the recovery of rapid changes in star formation activity. Non-parametric models are also computationally demanding, particularly in pixel-by-pixel analysis applied to large samples of galaxies. On average, fitting a single galaxy required 20 minutes per CPU per pixel, with about 650 pixels per galaxy, leading to 216 hours per galaxy for non-parametric models, significantly longer than parametric models like DPL, which reduced runtime by a factor of 10-15. While this added flexibility is valuable, it comes at the cost of interpretability, sensitivity to noise, and higher computational expense.

In contrast, flexible parametric models like the DPL strike a balance between simplicity and tractability, especially for pixel-by-pixel analysis of large datasets. The DPL model provides sufficient flexibility to capture a broad range of star formation histories (SFHs), including both rising and declining star formation rates, as well as episodic bursts and quenching events. Additionally, the DPL model helps mitigate overfitting risks that highly flexible non-parametric models might introduce, especially in noisy regions.

SFH modeling based on simulation-motivated approaches is often constrained by the physical assumptions within the simulations, such as feedback mechanisms and cosmological parameters. This can limit the flexibility needed to match pixel-by-pixel observations, though future studies will be necessary to fully test this.

\begin{figure}
    \centering
    \includegraphics[width=\linewidth]{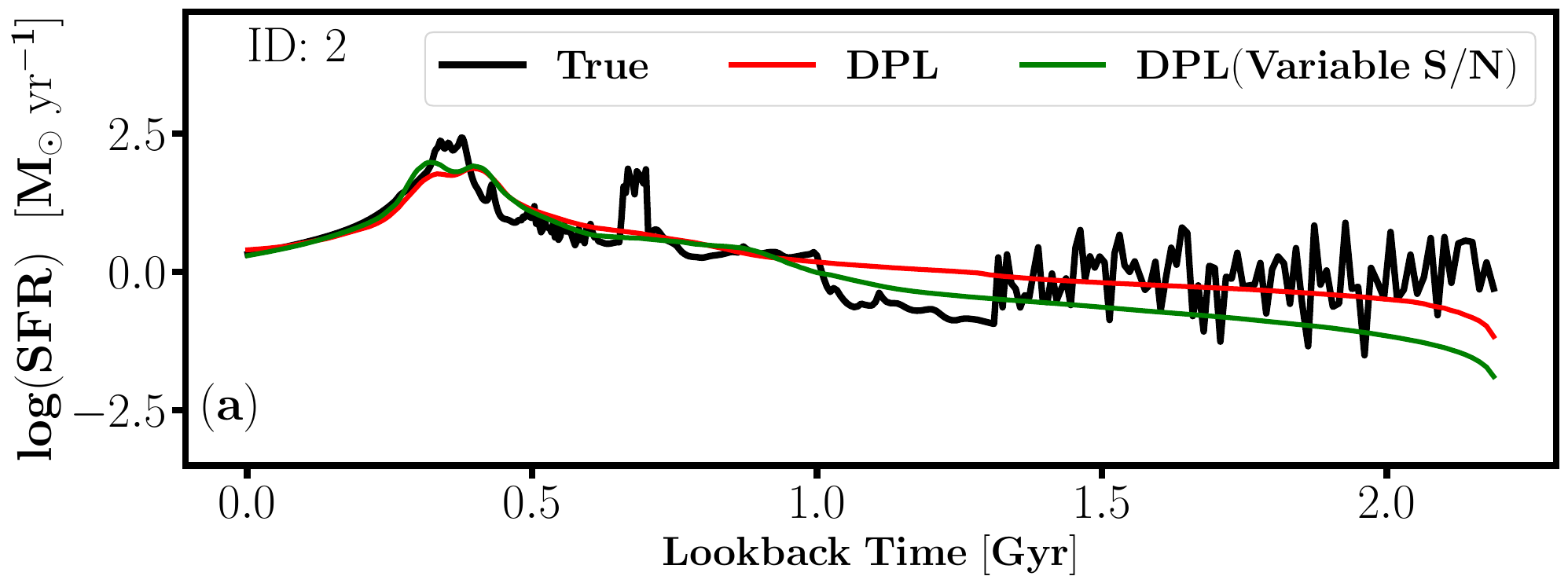}
    \includegraphics[width=\linewidth]{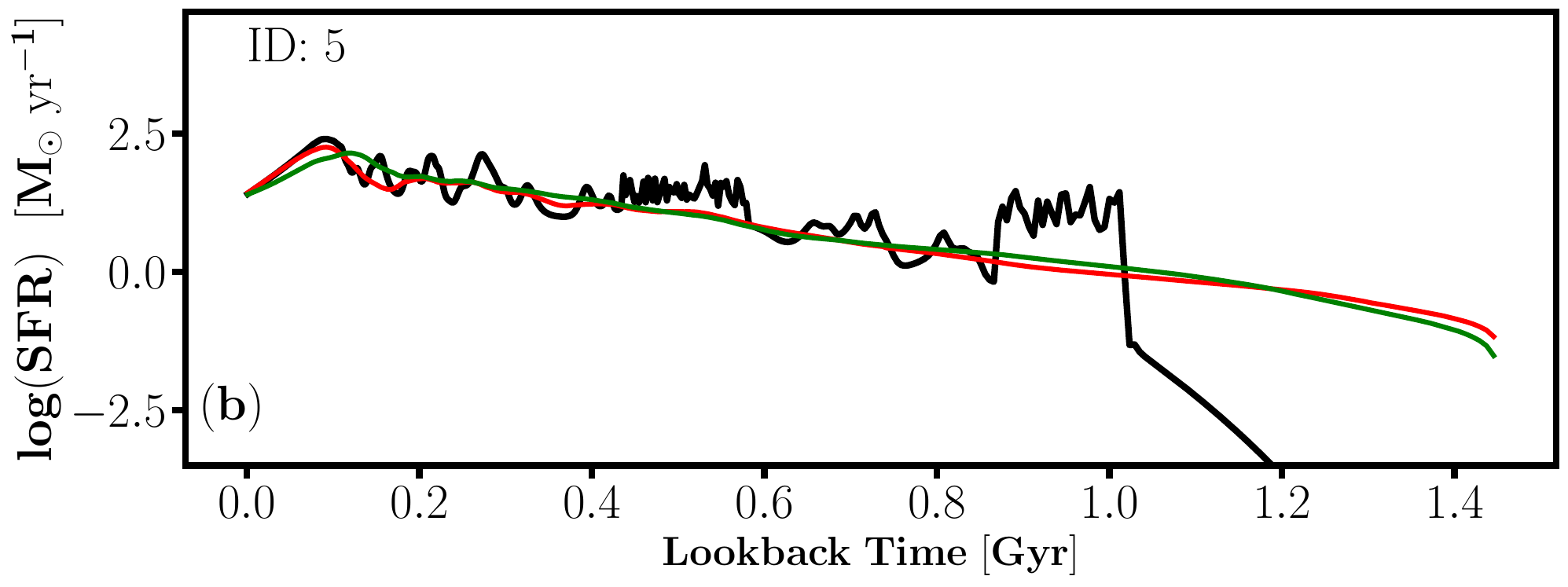}
    \includegraphics[width=\linewidth]{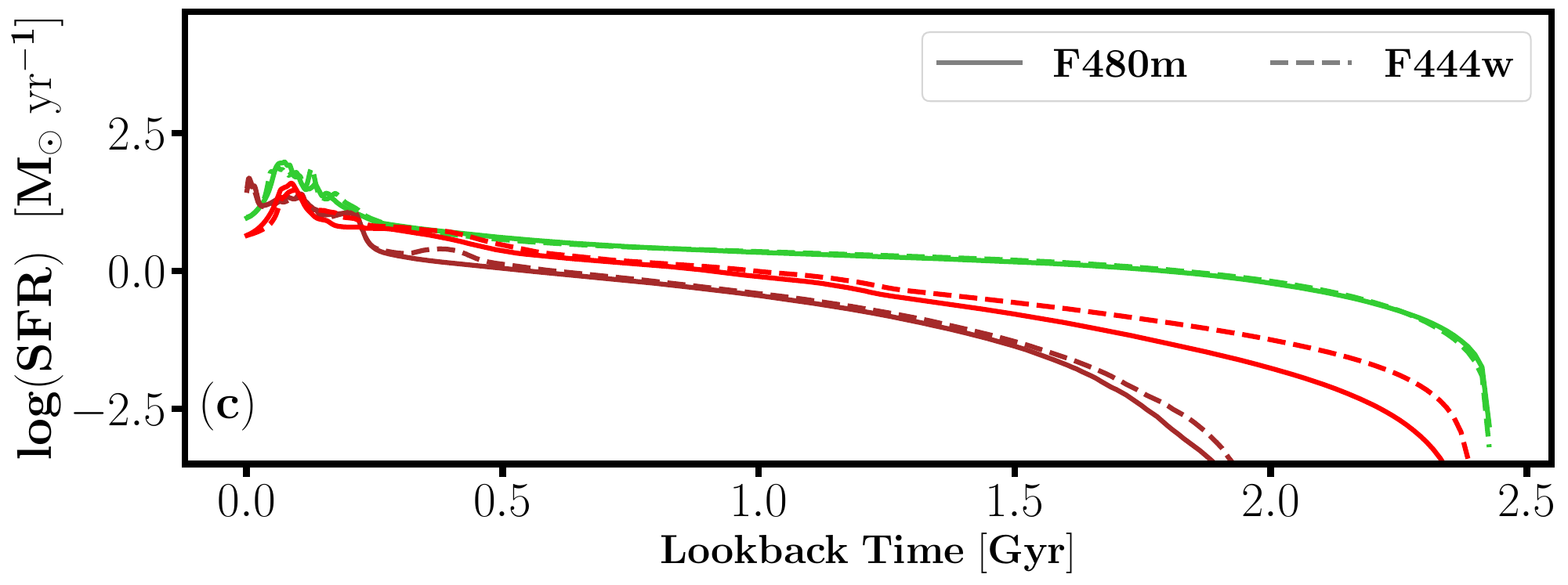}
    \caption{Panels (a) and (b) show the recovered SFHs for two mock galaxies (IDs 2 and 5) before and after reducing the S/N, illustrating that while lower S/N affects individual pixels, the overall SFH trends remain robust. Panel (c) compares the SFHs derived from spatially resolved analysis at varying image resolutions for three observed galaxies. The images are convolved to match the F444W and F480M filter responses. The resulting SFHs show subtle differences between resolutions. The brown, red, and green curves represent the SFHs of galaxy IDs 24722, 26980, and 27864, respectively.} 
    \label{fig16}
\end{figure}

\subsection{Caveats and Limitations}

The pixel-by-pixel analysis presents a major advantage by providing resolved maps of a galaxy's key physical properties, allowing for a detailed examination of their spatial distributions \citep{abdurrouf2023}. Additionally, resolved SFH analysis benefits from simpler treatments of dust and metallicity variations across the galaxy. However, this method also poses challenges, especially in the outer regions of galaxies, where the low signal-to-noise ratio (S/N) and potential light loss can introduce uncertainties. Accurately estimating these uncertainties is essential to reduce bias, and both our simulations and observations suggest that these errors are appropriately handled. Nevertheless, the method does not account for potential missing light beyond the detected boundaries, which must be considered when deriving total stellar masses.

We note that for our pixel-by-pixel analysis of simulated and observed galaxies, Voronoi binning \cite{cappellari2003} was not applied due to the high signal-to-noise (S/N) threshold in both datasets, particularly with an S/N$>10$ for simulated galaxies and careful noise handling in the observed ones. For the mock galaxies, we assumed a 5\% flux uncertainty, resulting in uniformly high S/N values, ensuring that pixel-level fitting was feasible without the need for spatial binning. The robustness of the derived maps was validated by direct comparisons with the true input maps, showing consistent and accurate recovery of parameters without needing to aggregate pixels. To assess the role of S/N further on the recovered SFHs, we artificially reduced the S/N in selected mock galaxies (IDs 2 and 5). The results showed minor differences in the recovered SFHs (see panels a and b of Figure \ref{fig16}), suggesting that while noise influences the precision of individual pixels, the overall SFH trends remain robust. For the observed galaxies, the consistency in 2D maps for mass, age, and SFR derived from different SFH models suggests that our results are stable against noise without the need for Voronoi binning.

Voronoi binning adaptively adjusts spatial resolution based on local S/N, preserving detail in high S/N regions while binning together low S/N pixels. However, binning also smooths out fine spatial structures, which are critical for studying structures of galaxies, particularly in low surface brightness regions where we aim to retain spatial gradients. Additionally, since low S/N pixels often recover broad posteriors that closely resemble the priors, averaging parameters after fitting does not necessarily lead to a more precise or accurate result. Instead, direct binning before SED fitting would be preferable if a significant improvement in constraints were required. While we opted for pixel-level fitting to maximize spatial resolution, alternative approaches such as local averaging in the outskirts could approximate the effects of Voronoi binning without compromising detail in high SNR regions. Overall, our methodology, supported by noise treatment and S/N-based tests, ensures robust parameter recovery while preserving the spatial information necessary for this study.

Resolution also plays a role when applying resolved models, as the point spread function (PSF) can cause fluxes to spread across adjacent pixels, correlating flux measurements and their errors. The PSF convolution was not applied to the mock galaxies in this study, hence making the results robust to resolution effects. We examined this for observed galaxies, by convolving the images of the observed galaxies to the resolution of the F444W filter using the DPL model. As shown in panel c of Figure \ref{fig16}, the SFHs from DPL models, convolved to the F444W and F480M resolutions, are consistent in terms of overall trends. Minor smoothing of details, such as the width and location of starburst peaks, is observed at longer wavelengths, but the general trends remain unchanged. Nevertheless, despite these subtle differences, the general trend indicates that spatially resolved analysis significantly reduces the biases that are common in unresolved photometry and is capable of predicting the SFHs. We note that we did not consider variations in pixel scales, which could potentially affect the final results \citep[see][for more discussion]{smith2018}.

In this study, we did not account for potential variations in the initial mass function (IMF) as a function of time, which could impact the derived star formation histories. The assumption of a constant IMF throughout cosmic time simplifies our analysis, but there is growing evidence suggesting that the IMF might evolve \citep{steinhardt2022}, particularly at higher redshifts or in specific environments such as starburst galaxies. Accounting for such variations could alter the stellar mass estimates and SFH reconstructions, particularly in the early phases of galaxy formation \cite{jermyn2018}. Additionally, we employed a simplified dust model when considering unresolved models. The simplicity of the dust model might introduce biases in the derived parameters, especially for galaxies with significant dust content. In future work, incorporating a more complex and time-varying IMF model, alongside more sophisticated dust treatments, will be crucial for obtaining a more accurate and nuanced understanding of the SFHs, particularly for high-redshift galaxies.

Another concern is the potential impact of active galactic nuclei (AGN) on our analysis. In our mock galaxies, AGNs were not included, so this was not an issue. For the observed galaxies, we accounted for the presence of emission lines in the SED fitting process, ensuring that they were modeled rather than significantly biasing the inferred stellar properties. While AGNs could influence the central regions, their overall contribution is expected to be minimal. Our dataset does not explicitly exclude AGN-hosting galaxies via emission line ratio diagnostics; however, the use of medium-band filters alongside broadband ones helps better characterize and mitigate the impact of emission lines on the SED fitting results.

We should note that an alternative approach to constructing mock galaxies is to leverage outputs from hydrodynamical zoom-in simulations. These simulations naturally incorporate self-consistent spatial distributions of stellar age, metallicity, and dust, which reflect the complex interplay of galaxy evolution processes. While such mock galaxies are invaluable for evaluating recovery methods in a cosmologically motivated context, their spatial resolution and noise characteristics often differ from those of real observations. In this study, we prioritized full control over the input parameters of the mock galaxies to ensure a systematic evaluation of our method and its sensitivity to key assumptions. Incorporating hydrodynamical simulation-based mocks would be a promising avenue for future work, offering a complementary assessment of our methodology under more realistic, complex conditions.

\subsection{Conclusions and Future Directions}
In conclusion, our analysis highlights key trends and discrepancies between spatially resolved and unresolved approaches, particularly in how they estimate fundamental galaxy properties like stellar mass, age, and star formation rates. Parametric SFH models, especially the DPL model, applied in spatially resolved frameworks, demonstrate high accuracy and consistency, making them a reliable tool for deriving physical parameters. Our findings further challenge the conclusions of \citet{narayanan2024}, who suggested that stellar masses of high-redshift galaxies are unconstrained, but align with some recent work suggesting that stellar masses are generally robust within a factor of about 3 for high-redshift galaxies \citep{coachrane2024}.

The DPL model strikes a balance between being flexible enough to capture diverse SFH behaviors and simple enough to avoid the overfitting or rigidity issues seen in non-parametric or simulation-motivated models. Its ability to adapt to each pixel's unique SFH gives it a distinct advantage in spatially resolved analysis, providing a detailed and accurate reconstruction of galaxy-wide SFHs. Compared to Gaussian Processes and simulation-motivated models, the DPL is likely more computationally efficient for pixel-by-pixel analysis, making it practical for large datasets. The methods presented in this study provide a foundation for applying similar analyses to larger samples of galaxies, offering deeper insights into galaxy formation and evolution.

\begin{acknowledgments}
We thank the anonymous referee for the thorough review and valuable comments, which improved this manuscript. The data products presented herein were retrieved from the Dawn JWST Archive (DJA). DJA is an initiative of the Cosmic Dawn Center (DAWN), which is funded by the Danish National Research Foundation under grant DNRF140. This research made use of Photutils, an Astropy package for detection and photometry of astronomical sources.
\end{acknowledgments}

\facilities{HST, JWST}



\software{BAGPIPES \citep{carnall2018}, Prospector \citep{leja2017, johnson2021}, Astropy \citep{astropy2013, astropy2018, astropy2022}, Pypher \citep{boucaud2016}, Cloudy \citep{ferland1998, ferland2013}, Photuils \cite{photutils2024}}

{}

\end{document}